\documentclass[aps,pra,twocolumn,groupedaddress,asmsymb,amsmath]{revtex4-1}

\usepackage{ulem}
\usepackage{stackrel}

\usepackage{graphicx}
\usepackage{color}
\usepackage{caption}
\usepackage[usenames,dvipsnames]{xcolor}

\def\utilde#1{\mathord{\vtop{\ialign{##\crcr
$\hfil\displaystyle{#1}\hfil$\crcr\noalign{\kern1.5pt\nointerlineskip}
$\hfil\widetilde{}\hfil$\crcr\noalign{\kern1.5pt}}}}}

\newcommand{\be}{\begin{equation}}
\newcommand{\ee}{\end{equation}}
\newcommand{\lb}{\label}

\newcommand{\bv}{{\bf v}}

\newcommand{\bx}{{\bf x}}
\newcommand{\br}{{\bf r}}

\newcommand{\bDelta}{{\mbox{\boldmath $\Delta$}}}
\newcommand{\grad}{{\mbox{\boldmath $\nabla$}}}
\newcommand{\bdot}{{\mbox{\boldmath $\cdot$}}}
\newcommand{\bdots}{{\mbox{\boldmath $:$}}}
\newcommand{\bzed}{{\mbox{\boldmath $0$}}}

\newcommand{\Dlim}{{{\mathcal D}\mbox{-}\lim}}
\newcommand{\us}{\underline{s}}

\newcommand{\up}{\underline{p}}
\newcommand{\ubeta}{\underline{\beta}}
\newcommand{\ulambda}{\underline{\lambda}}

\newcommand{\op}{\overline{p}}
\newcommand{\calQ}{{\mathcal Q}}
\newcommand{\calG}{{\mathcal G}}
\newcommand{\Dto}{{\stackrel{\mathcal{D}}{\longrightarrow}}}

\newcommand{\red}[1]{\textcolor{black}{#1}}

\begin{document}

%\preprint{}

\title{Cascades and Dissipative Anomalies in Relativistic Fluid Turbulence}
%and Dissipative Anomalies}


\author{Gregory L. Eyink${\,\!}^{1,2}$ and Theodore D. Drivas${\,\!}^1$}
%\thanks{}
\affiliation{${\,\!}^1$Department of Applied Mathematics \& Statistics, The Johns Hopkins University, Baltimore, MD, USA}
\affiliation{${\,\!}^2$Department of Physics \& Astronomy, The Johns Hopkins University, Baltimore, MD, USA}

\date{\today}

\begin{abstract}
We develop first-principles theory of relativistic fluid turbulence at high Reynolds and P\'eclet numbers. 
We follow an exact approach pioneered by Onsager, which we explain as a non-perturbative application 
of the principle of renormalization-group invariance. We obtain results very similar to those for 
non-relativistic turbulence, with hydrodynamic fields in the inertial-range described as distributional 
or ``coarse-grained'' solutions of the relativistic Euler equations. These solutions do not, however, 
satisfy the naive conservation-laws of smooth Euler solutions but are afflicted with dissipative anomalies
in the balance equations of internal energy and entropy. The anomalies are shown to be possible by 
exactly two mechanisms, local cascade and pressure-work defect. We derive ``4/5th-law''-type expressions 
for the anomalies, which allow us to characterize the singularities (structure-function scaling exponents) 
required for their non-vanishing. We also investigate the Lorentz covariance of the inertial-range fluxes,
which we find is broken by our coarse-graining regularization but which is restored in the limit that 
the regularization is removed, similar to relativistic lattice quantum field theory. In the formal limit as speed of  
light goes to infinity, we recover the results of previous non-relativistic theory. In particular, anomalous heat 
input to relativistic internal energy coincides in that limit with anomalous dissipation of non-relativistic kinetic energy.
\end{abstract}

\pacs{}
%\keywords{}

\maketitle

%\subsection{}
%\subsubsection{}

\section{Introduction}\label{sec:intro}

Relativistic hydrodynamics has a growing range of applications in current physics research, 
including energetic astrophysical objects such as gamma-ray bursts \cite{narayan2009turbulent} 
and pulsars \cite{bucciantini2014review},  high-energy physics of the early universe and heavy-ion collisions 
\cite{de2016hydrodynamic},  condensed matter physics of graphene \cite{fritz2008quantum,lucas2016transport}
and strange metals \cite{hoyos2013lifshitz,davison2014holographic}, and black-hole gravitational physics 
via the fluid-gravity correspondence in AdS/CFT \cite{baier2008relativistic,bhattacharyya2008nonlinear,bhattacharyya2008local,bhattacharyya2008conformal}. 
The ubiquity of relativistic hydrodynamics is natural, given that it represents a universal low-wavenumber 
description of relativistic quantum field-theories  at scales much larger than the mean-free path length.  
When the global length-scales of such relativistic fluid systems 
are even larger, as measured by the dimensionless Reynolds-number, then turbulent flow is likely. There is 
observational evidence for relativistic turbulence in high-energy astrophysical systems, e.g. gamma-ray 
bursts accelerate relativistic jets to Lorentz factors $\gamma \gtrsim 100$ and contain internal fluctuations with 
$\delta\gamma\sim 2$ \cite{piran2005physics}. Numerical simulations of relativistic fluid models have verified 
the occurrence of turbulence at high Reynolds numbers \cite{zrake2013magnetic,radice2013universality}.
Relativistic turbulence is also observed in numerical solutions of conformal hydrodynamic models  
\cite{carrasco2012turbulent,green2014holographic,westernacher2015scaling} and an analogous 
phenomenon is seen in their dual AdS black-hole solutions \cite{adams2014holographic}.  

Despite the importance of relativistic fluid turbulence at high Reynolds-number for many applications, 
there have been only a handful of theoretical efforts to elucidate the phenomenon 
\cite{fouxon2010exact,eling2011gravity,liu2011shocks}. Using a point-splitting 
approach, Fouxon \& Oz \cite{fouxon2010exact} derived statistical relations for relativistic turbulence that 
in the incompressible limit reduce to the famous Kolmogorov ``4/5th-law'' 
\cite{kolmogorov1941dissipation,Frisch95}. However, in the relativistic regime their  
relations have nothing to do with energy of the fluid. This seems to suggest a profound difference between 
relativistic and non-relativistic turbulence or, even more radically, an essential flaw in our current understanding 
of non-relativistic turbulence. As concluded by Fouxon \& Oz \cite{fouxon2010exact},  ``The interpretation of the Kolmogorov 
relation for the incompressible turbulence in terms of the energy cascade may be misleading.''  

We develop here the first-principles theory of relativistic fluid turbulence at high Reynolds and 
P\'eclet numbers, which reaches a very different conclusion.  We establish the existence of a 
relativistic energy cascade in the traditional sense and an even more fundamental entropy cascade. 
The appearance of thermodynamic entropy is not surprising, considering its central role in the theory 
of dissipative relativistic hydrodynamics \cite{israel1979transient,israel1979transientII,baier2008relativistic,bhattacharyya2008local,romatschke2010new}. 
Our analysis follows a pioneering work of Onsager \cite{onsager1945distribution,Onsager49} on incompressible 
fluid turbulence, who proposed that turbulent flows at very high Reynolds numbers are described by 
singular/distributional solutions of the incompressible Euler equations.  Onsager derived in 1945 
the first example of a conservation-law anomaly, showing by a point-splitting argument that the 
zero-viscosity limit of Navier-Stokes solutions can dissipate fluid kinetic energy for a critical $1/3$
H\"older singularity of the fluid velocity \cite{EyinkSreenivasan06}.  \red{Polyakov has pointed out the 
formal analogy of Kolmogorov's ``4/5th-law'' and its point-splitting derivation to axial anomalies in 
quantum gauge field theories \cite{polyakov1992conformal,polyakov1993theory}.} 
However, Onsager's analysis is deeper than the ensemble theory of Kolmogorov 
\cite{kolmogorov1941dissipation,kolmogorov1941local,kolmogorov1941degeneration} or ``K41'', because it 
applies to individual flow realizations. It is also formally 
exact and requires no statistical hypotheses, such as isotropic/homogeneous ensembles or mean-field arguments 
ignoring space-time intermittency.  Onsager's proposals were not understood at the time 
and he never published full proofs of his assertions.  Thus, the theory went ignored until 
Onsager's $1/3$ H\"older condition for anomalous energy dissipation was rederived  \cite{eyink1994energy}.  
This triggered a stream of work in the mathematical PDE community that has improved upon the analysis, 
notably \cite{CET1994,DuchonRobert2000,cheskidov2008energy}. More recently, 
concepts originating in the Nash-Kuiper theorem and Gromov's $h$-principle have been applied to 
mathematically construct dissipative Euler solutions of the type conjectured by Onsager 
\cite{de2012h,delellis2013continuous}. This new circle of ideas has led to a proof that Onsager's $1/3$ 
criticality condition for energy dissipation is sharp \cite{isett2016proof}.

Onsager's theory of dissipative Euler solutions and its application to fluid turbulence 
is still essentially unknown to the wider physics community, however.  This is unfortunate 
because it is the most comprehensive theoretical framework for high Reynolds turbulence
and generally applicable, not only to kinetic energy dissipation in incompressible fluid 
turbulence,  but also to cascades in magnetohydrodynamic turbulence 
\cite{caflisch1997remarks,aluie2017coarse},  to dissipative anomalies of Lagrangian invariants 
such as circulations \cite{Eyink06} and magnetic fluxes \cite{eyink2006breakdown}, and to 
cascades in compressible Navier-Stokes turbulence 
\cite{aluie2011compressible,aluie2013scale,eyink2017cascadesI,drivas2017onsager}. Furthermore, 
Onsager's analysis is based on very intuitive physical ideas. As we discussed in our earlier paper 
on non-relativistic compressible Navier-Stokes turbulence \cite{eyink2017cascadesI}
[hereafter, paper I] Onsager's argument is essentially a non-perturbative application of the principle of 
renormalization group invariance \cite{stueckelberg1951normalization,gell1954quantum,bogolyubov1955renormalization}. 
High-Reynolds turbulence is characterized by ultraviolet divergences  of gradients of the velocity and 
other thermodynamic fields, referred to as a ``violet catastrophe'' by Onsager \cite{onsager1945distribution}. 
Regularizing these divergences introduces a new arbitrary length-scale $\ell$ upon 
which objective physics cannot depend, and exploiting this invariance yields the main 
conclusions of the theory on fluid singularities, inertial range, local cascades, etc.  

Onsager's unpublished work in 1945 employed a point-splitting approach \cite{EyinkSreenivasan06}, but 
we exploit here a more powerful coarse-graining or ``block-spin'' regularization 
\cite{CET1994,eyink1995local} for relativistic fluid turbulence. 
Many essential steps were already taken in paper I on non-relativistic compressible turbulence, 
such as the identification of (neg)entropy as a key invariant and the development of appropriate 
non-perturbative tools of analysis, such as cumulant expansions for space-time coarse-graining
and mathematical distribution theory. Relativistic turbulence brings in some completely new difficulties, 
however. First, kinetic energy is usually given the central role in the theory of non-relativistic energy cascade, 
but kinetic energy is an unnatural quantity in relativity theory. We show here that internal energy 
is the appropriate basis for the theory of relativistic energy cascade. Another distinction  
of the relativistic theory is that our non-perturbative coarse-graining regularization preserves Galilean symmetry 
of non-relativistic fluid models but it breaks Lorentz-symmetry. This is reminiscent of the lattice 
regularization of relativistic quantum field-theories \cite{wilson1974confinement}, which breaks Lorentz 
symmetry for finite lattice spacing $a$ but recovers it in the continuum limit $a\rightarrow 0.$
The situation here is similar, as we show that Lorentz symmetry is restored as our 
regularization parameter $\ell\rightarrow 0,$ leading to a description by relativistic Euler equations. 
Further differences exist, such the unit normalization of relativistic velocity vectors, which leads 
to new terms in flux/anomaly formulas that do not appear non-relativistically. An important caveat
about the present work is that we consider only special-relativistic fluid turbulence in flat Minkowski 
space-time. General-relativistic (GR) fluids in curved space-times bring in additional technical difficulties.
These seem tractable but it makes sense to develop the theory first in Minkowski space-time, as
the simplest setting possible.  For remarks on full GR, see the conclusion. 

In this paper, we shall consider the $D=d+1$ dimensional Minkowski space-time for any space dimension $d\geq 1.$ This 
generality is motivated not only by the wider perspective it affords but also by fluid-gravity correspondence 
in AdS/CFT, which holds for general $D$ \cite{bhattacharyya2008conformal,haack2008nonlinear}.   We adopt 
signature $-+\cdots +$ of Minkowski metric $g^{\mu\nu}$. We shall follow standard relativistic notations, 
but we include explicit factors of speed of light $c$, e.g. space-time coordinates  $x^\mu=(x^0,\bx)=(ct,\bx),$ 
velocity vectors $V^\mu=\gamma (1,\bv/c),$ etc. rather than use natural units with $c=1.$ This facilitates taking the 
limit $c\to\infty$ for comparison with the results of paper I. 

\section{Relativistic Dissipative Fluid Models}\lb{fluidmodels}

We consider here a relativistic fluid with conserved stress-energy tensor $T^{\mu\nu}$
\be  \partial_\nu T^{\mu\nu}=0 \lb{eq1} \ee
and with one conserved current $J^\mu$ 
\be  \partial_\nu J^\nu =0. \lb{eq2} \ee
The latter may interpreted as a particle number current (e.g. baryon number) and the fluid models that we consider reduce 
in the limit $c\to\infty$ and at zeroth-order in gradients 
to the non-relativistic compressible Euler equations. This choice allows us to compare our results here to those derived in paper I 
for non-relativistic compressible turbulence. However, our analysis carries over straightforwardly to other fluid systems without the 
additional conserved current $J^\mu$ (e.g. zero chemical potential sectors, conformal fluids) and
to multicomponent systems with more than one conserved current (e.g. 2-fluid models of relativistic superfluids).    

Even with the restrictions to (\ref{eq1}),(\ref{eq2}), there are many possible fluid models. Unlike the non-relativistic case, where 
the compressible Navier-Stokes equations have a more canonical status and are employed almost universally 
in the fluid regime, there are still many dissipative relativistic fluid models competing as descriptions of the same 
physical system (e.g. see \cite{andersson2007relativistic}, section 14 or \cite{rezzolla2013relativistic}, Ch.6). We 
consider a broad class of dissipative relativistic fluid theories, which includes the traditional theories of Eckart 
\cite{eckart1940thermodynamics} and Landau-Lifschitz \cite{landau1959fluid} and the Israel-Stewart theory \cite{israel1979transient,israel1979transientII}, in which the number current and stress tensor
have the general form 
\be  J^\mu = n V^\mu + \sigma \hat{N}^\mu \lb{eq3} \ee
\begin{eqnarray}  T^{\mu\nu} &=& p\Delta^{\mu\nu} + \epsilon V^\mu V^\nu + \Pi^{\mu\nu}, \cr
\Pi^{\mu\nu}&=&\kappa \hat{Q}^{(\mu} V^{\nu)}
+\zeta\hat{\tau}\Delta^{\mu\nu} + 2\eta\hat{\tau}^{\mu\nu}. \lb{eq4} \end{eqnarray} 
Here $n$ is number density, $p=p(\epsilon,n)$ the pressure, and $\epsilon=u+\rho c^2$ the total energy density,
with $\rho=nm $ the rest-mass density for particle mass $m$ and $u$ the internal-energy density.  The 
velocity vector $V^\mu,$ to be specified below, is future time-like and $V_\mu V^\mu=-1$. The quantity 
$N^\mu=\sigma \hat{N}^\mu$ is a dissipative number current, $Q^\mu=\sigma \hat{Q}^\mu$
a dissipative heat current, and $\Pi^{\mu\nu}_{visc}=\tau\Delta^{\mu\nu} + \tau^{\mu\nu}$
a dissipative (viscous) stress tensor with $\tau=\zeta\hat{\tau}$ and $\tau^{\mu\nu}=2\eta\hat{\tau}^{\mu\nu}.$
Here we have defined $\Delta^{\mu\nu}=g^{\mu\nu}+V^\mu V^\nu$ as the projection onto the space direction
in the fluid rest-frame and the various dissipative terms satisfy 
\be  V_\mu \hat{N}^\mu = V_\mu \hat{Q}^\mu =V_\mu \hat{\tau}^{\mu\nu} =0 \lb{eq5} \ee
with $\hat{\tau}^{\mu\nu}$ also traceless and symmetric.  We have made an unconventional choice to 
factor out the overall dependences on particle conductivity $\sigma,$ thermal conductivity $\kappa,$
bulk viscosity $\zeta,$ and shear viscosity $\eta,$ in order to make clearer some of our arguments below. 
So-called particle- or Eckart-frame theories have $\hat{N}^\mu=0,$ so that $V^\mu$ is the time-like unit vector
in the $J^\mu$-direction and $n=-J^\mu V_\mu.$ On the other hand, energy- or Landau-Lifschitz-frame theories 
have $\hat{Q}^\mu=0,$ so that $V^\mu$ and $\epsilon$  are specified by the eigenvalue condition 
$T^{\mu\nu}V_\nu=-\epsilon V^\mu,$ with a time-like unit eigenvector. 
In the class of models that we will consider in detail,  there is also an entropy current $S^\mu$ (discussed further below) 
which satisfies a balance equation of the form 
\be  \partial_\mu S^\mu= \sigma\frac{\hat{N}^\mu\hat{N}_\mu}{T^2}+\kappa\frac{\hat{Q}^\mu\hat{Q}_\mu}{T^2}
+\zeta \frac{\hat{\tau}^2}{T} + 2\eta \frac{\hat{\tau}^{\mu\nu}{\hat{\tau}_{\mu\nu}}}{T}, \lb{eq6} \ee
whose righthand side, when all of the transport coefficients $\sigma,$ $\kappa,$ $\zeta,$
$\eta$ are positive, is nonnegative as required by the second law of thermodynamics. 
The specific assumptions made 
above are mostly to simplify our proof in the next section that effective coarse-grained equations 
obtained in the limit  $\sigma,$ $\kappa,$ $\zeta,$ $\eta\rightarrow 0$ correspond to distributional
Euler solutions. With some appropriate corresponding assumptions, our analysis will apply to any 
dissipative fluid model consistent with the thermodynamic second-law. In fact, our inertial-range analysis
is completely general and applies to any \red{distributional solution of the relativistic Euler equations, regardless
of the dissipative model limits used to obtain the particular solution (or to even solutions constructed by other means).} 

Defining the energy current 
\be {\mathcal E}^\mu=-T^{\mu\nu}V_\nu = \epsilon V^\mu + \kappa\hat{Q}^\mu, \lb{eq7} \ee
and internal energy current 
\be  U^\mu= {\mathcal E}^\mu-mc^2 J^\mu = u V^\mu + \kappa\hat{Q}^\mu, \lb{eq8} \ee 
it is straightforward to obtain from (\ref{eq1}),(\ref{eq2}) for all of the class of models we consider 
the balance equations of total and internal energy densities as 
\begin{eqnarray}
\partial_\mu {\mathcal E}^\mu &=& \partial_\mu U^\mu \cr 
&=& -(\partial_\mu V_\nu) T^{\mu\nu}=-p(\partial^\mu V_\mu) + {\mathcal Q}_{diss}\cr
&&  \lb{eq9} 
\end{eqnarray} 
with the dissipative ``heating'' of the fluid given by  
\be {\mathcal Q}_{diss} = -\kappa \hat{Q}^\mu A_\mu-\zeta \hat{\tau}\theta -2\eta  \hat{\tau}^{\mu\nu}\sigma_{\mu\nu}.  \lb{eq10} \ee
Here $A^\mu=\mathcal{D}V^\mu$ is the acceleration vector with $\mathcal{D}=V^\mu\partial_\mu$ the material derivative for an observer 
moving with 
the fluid, 
\be  \theta = \Delta^{\mu\nu} \partial_\nu V_\mu=\partial_\mu V^\mu \lb{eq11} \ee
is the relativistic dilatation, and 
\begin{eqnarray}  
\sigma_{\mu\nu} = \partial_{\left\langle\mu\right.} V_{\left.\nu\right\rangle} 
&\equiv&  \partial_{\left(\mu\right.}^\perp V_{\left.\nu\right)}-\frac{\theta}{d}\Delta_{\mu\nu} \cr
&=& \partial_{\left(\mu\right.} V_{\left.\nu\right)} + A_{\left(\mu\right.} V_{\left.\nu\right)}-\frac{\theta}{d}\Delta_{\mu\nu}
\lb{eq12} 
\end{eqnarray}
is the relativistic strain, for $\partial_\mu^\perp=\Delta^\alpha_\mu \partial_\alpha.$ We use here standard 
notations for relativistic fluids \cite{andersson2007relativistic,rezzolla2013relativistic}, in 
particular with $C_{(\mu\nu)}=\frac{1}{2}(C_{\mu\nu}+C_{\nu\mu})$ the symmetrization on $\mu,\nu,$ 
so that $\sigma_{\mu\nu}$ is symmetric, traceless, and $\sigma_{\mu\nu}V^\nu=0.$  
%with $C_{[\mu\nu]}=\frac{1}{2}(C_{\mu\nu}-C_{\nu\mu})$ the anti-symmetrization, 
%and $C_{\langle\mu\nu\rangle}=\frac{1}{2}(C_{\mu\nu}+C_{\nu\mu})$ 

The traditional theories of Eckart \cite{eckart1940thermodynamics} and Landau-Lifschitz 
\cite{landau1959fluid} have dissipative fluxes proportional to 
the following tensors: 
\be  \hat{N}^\mu = -T^2 \partial^\mu_\perp \lambda\lb{eq13} \ee
\be  \hat{Q}^\mu =-(\partial^\mu_\perp T + T A^\mu) \lb{eq14} \ee
\be  \hat{\tau} = -\theta, \quad  \hat{\tau}_{\mu\nu} = -\sigma_{\mu\nu} \lb{eq15} \ee
which are first-order in gradients, with particle-conductivity $\sigma=0$ for Eckart and thermal-conductivity 
$\kappa=0$ for Landau-Lifschitz, so that 
\be {\mathcal Q}_{diss} = \kappa A^\mu \partial_\mu T+\kappa A^\mu A_\mu +\zeta \theta^2 
+2\eta \sigma^{\mu\nu}\sigma_{\mu\nu}. \lb{eq16} \ee
In particular, ${\mathcal Q}_{diss}\geq 0$ for the Landau-Lifschitz theory. Above we have used the 
standard relativistic thermodynamic potentials, the temperature $T$ (or its inverse $\beta=1/T$) and 
$\lambda=\mu/T$ for the chemical potential $\mu.$ For reviews of relativistic thermodynamics, 
see \cite{israel1987covariant}, also \cite{andersson2007relativistic}, section 5 or 
\cite{rezzolla2013relativistic}, \S 2.3.7. Here we note only that the relativistic chemical potential differs from 
its Newtonian counterpart $\mu_N$ by a rest-mass contribution, $\mu=\mu_N + mc^2$. The 
entropy current of the Eckart and Landau-Lifschitz theories is defined in terms of the 
entropy density per volume $s(\epsilon,n)$ and the thermodynamic potentials as 
\be S^\mu =s V^\mu + \beta Q^\mu - \lambda N^\mu. \lb{eq17} \ee 
Using the thermodynamic second law $ds=\beta d\epsilon-\lambda dn$ and equations (\ref{eq2}) and (\ref{eq9}), it is 
then easy to check that the equation (\ref{eq6}) holds. However, as is well-known, the Eckart and Landau-Lifschitz 
theories are unstable, acausal and ill-posed in both linear \cite{hiscock1985generic} and nonlinear 
\cite{hiscock1988nonlinear} regimes. Thus, these theories are not useful as predictive evolutionary models
of relativistic fluids. 

The class of models that we consider also contain better-behaved models, however, such as the extended 
hydrodynamic theory of Israel-Stewart \cite{israel1979transient,israel1979transientII}. This is itself an entire class 
of models, each of which uses a different definition of the off-equilibrium fluid velocity. The particle-frame
and energy-frame versions have both been shown to be stable, causal, and hyperbolic in the linear 
\cite{hiscock1983stability,hiscock1988stability} and nonlinear \cite{hiscock1989effects,olsonphd1990,olson1990stability}
regimes, with somewhat better stability properties in the energy-frame. In these models the entropy current is
not given by (\ref{eq17}) but instead is modified by the addition of terms that are quadratic in the dissipative 
fluxes $N^\mu,$ $Q^\mu,$ $\tau,$ and $\tau^{\mu\nu}.$
The form of the entropy current may be illustrated by the expression that 
holds in the energy-frame Israel-Stewart theory \cite{olsonphd1990,olson1990stability}:
\begin{eqnarray}
&& S^\mu = s V^\mu-\lambda N^\mu -\frac{1}{2T} (\beta_0\tau^2+\beta_1 N_\alpha N^\alpha +\beta_2\tau_{\alpha\beta}
\tau^{\alpha\beta}) V^\mu\cr
&& \hspace{60pt} +\frac{\alpha_0}{T} \tau N^\mu + \frac{\alpha_1}{T} \tau^{\mu\nu}N_\nu. \lb{eq18} 
\end{eqnarray} 
The new term proportional to $V^\mu$ can be regarded as an off-equilibrium modification of the rest-frame entropy 
density $s,$ and thus the coefficients $\beta_i,\ i=1,2,3$ (not to be confused with $\beta=1/T!$) are required to be 
positive to ensure that non-vanishing gradients lower the entropy. The other two terms proportional to $\alpha_i,$ 
$i=1,2$ are purely spatial in the fluid rest-frame and describe second-order contributions to dissipative entropy transport.  
All of the $\alpha$ and $\beta$ coefficients are assumed to be smooth functions of $\epsilon,\ \rho.$
 Imposing the second law of thermodynamics
in the form of (\ref{eq6}) constrains the dissipative fluxes \footnote{The requirement imposed by the 
entropy balance (\ref{eq6}) far from fully characterizes the dissipative fluxes. As discussed by 
Israel \& Stewart \cite{israel1979transient} and Hiscock \& Lindblom \cite{hiscock1983stability}
(note added in proof) many additional second-order terms are possible in 
addition to those included in the standard Israel-Stewart model. Even additional first-order terms can be 
included under weakened symmetry assumptions and may be required by microscopic physics, 
e.g. \cite{son2009hydrodynamics}}. For example, for the energy-frame
Israel-Stewart theory one finds 
\be  \tau  = \zeta \hat{\tau} = -\zeta \left[\theta + \beta_0 \mathcal{D}\tau +\cdots \right], \lb{eq19} \ee
\be  N^\mu = \sigma \hat{N}^\mu = -\sigma T \left[ T\partial_\perp^\mu \lambda 
+ \beta_1 (\mathcal{D}N_\perp)^\mu+\cdots \right] \lb{eq20} \ee
%\be  Q^\mu =-(\partial^\mu_\perp T + T A^\mu) \lb{eq} \ee
\be \tau^{\mu\nu} = 2\eta \hat{\tau}^{\mu\nu} = -2\eta \left[\sigma^{\mu\nu} + \beta_2 (\mathcal{D}\tau)^{\langle\mu\nu\rangle}
+\cdots \right] 
\lb{eq21} \ee
with $(\mathcal{D}N_\perp)^\mu=\Delta^{\mu\nu} \mathcal{D}N_\nu$ and with 
$(\mathcal{D}\tau)^{\langle\mu\nu\rangle}=\mathcal{D}\tau^{\mu\nu}+\tau^{\mu\lambda} A_\lambda V^\nu +
\tau^{\nu\lambda} A_\lambda V^\mu$ the part of $\mathcal{D}\tau^{\mu\nu}$ symmetric, traceless, 
and orthogonal to $V^\mu.$ Here $(\cdots)$ indicates various terms that are second-order in gradients,
involving the fluxes $N^\mu,$ $\tau,$ $\tau^{\mu\nu}$ and the thermodynamic potentials. We note that 
in the case of the particle-frame Israel-Stewart model, nearly identical equations hold, but with $N^\mu\rightarrow Q^\mu$
and, in the middle equation (\ref{eq20}), $\sigma\rightarrow \kappa$ and $T\partial_\perp^\mu\lambda\rightarrow \partial_\perp^\mu T/T+A^\mu.$ 
Unlike the original Eckart-Landau-Lifshitz theories, the relations (\ref{eq19})-(\ref{eq21}) are not simple constitutive relations 
for the dissipative fluxes, but are instead evolutionary equations which must be solved in time together with the 
conservation laws (\ref{eq1}),(\ref{eq2}) in order to determine both the local thermodynamic variables and the dissipative fluxes. 

It is a curious fact that in the Israel-Stewart (IS) theories the ``energy dissipation" ${\mathcal Q}_{diss}$ in (\ref{eq9}) 
may possibly be negative and thus may not act to heat the fluid. Indeed, out of the entire class of models 
that we consider in this paper, only the (ill-posed) Landau-Lifschitz theory guarantees that ${\mathcal Q}_{diss}\geq 0.$
\red{It is generally argued that negative values of ${\mathcal Q}_{diss}$ cannot be realized within 
the physical regime of validity of a fluid description. Since the dissipative fluxes in the energy-frame Israel-Stewart (IS) model
differ from those in the Landau-Lifschitz (LL) theory only by terms second-order in gradients, it is plausible}
that for most circumstances the dissipative fluxes obtained by solving the IS model will be nearly the same 
as those given by the LL constitutive relations, when evaluated with the IS model solutions.  More generally, 
Geroch \cite{geroch1995relativistic} and Lindblom \cite{Lindblom96therelaxation} 
have argued that this close agreement with the Landau-Lifischitz/Eckart 
constitutive relations will hold in the energy/particle frame, respectively, for a wide set of extended dissipative relativistic 
fluid models that are hyperbolic, causal, and well-posed. We thus expect typically to have ${\mathcal Q}_{diss}\geq 0$ 
in energy-frame fluid models. Unfortunately, the arguments of \cite{geroch1995relativistic,Lindblom96therelaxation} fail in the presence 
of shocks with near-discontinuities extending down to lengths of the order of the mean-free-path. In fact, the IS fluid 
models and other broad classes of hyperbolic, causal, well-posed models of dissipative relativistic fluids do 
not even possess continuous solutions corresponding to strong shocks \cite{geroch1991causal,olsonphd1990}. 
Thus, perhaps even more than for the non-relativistic case, a better microscopic starting point for a 
theory of relativistic fluid turbulence might be relativistic kinetic theory or a relativistic quantum field-theory 
rather than a dissipative fluid model. Fortunately, our principal results do not depend upon any particular 
model of dissipation, but only require the general conservation laws (\ref{eq1}),(\ref{eq2}), a fluid description with 
variables given by local thermodynamic equilibrium, and the second law of thermodynamics.     

In this paper we examine the hypothesis that the entropy-production is anomalous in relativistic fluid 
turbulence. Thus, we assume in the ideal limit $\sigma,\kappa,\eta,\zeta\rightarrow 0$ that distributional 
limits of the entropy production exist: 
\begin{eqnarray}
\Sigma &=& \Dlim_{\sigma,\kappa,\eta,\zeta\rightarrow 0} \left[\frac{\sigma\hat{N}^\mu\hat{N}_\mu}{T^2}
+\frac{\kappa\hat{Q}^\mu\hat{Q}_\mu}{T^2} \right. \cr
&& \hspace{100pt} \left. +\frac{\zeta\hat{\tau}^2}{T} + \frac{2\eta\hat{\tau}^{\mu\nu}{\hat{\tau}_{\mu\nu}}}{T}\right] \cr
&=& \Sigma_{cond} + \Sigma_{therm} + \Sigma_{bulk} + \Sigma_{shear} >0 
\lb{eq22} \end{eqnarray}
We shall then show that any strong limits $\epsilon$, $\rho,$ $V^\mu$ of the local equilibrium fields
are weak solutions of the relativistic Euler equations, under very mild additional assumptions. The anomalous 
entropy production of these Euler solutions is shown to occur by a nonlinear cascade mechanism  and 
we characterize the type of singularities required for non-vanishing entropy cascade. As in the non-relativistic case, the ideal limit  
is really a limit of large Reynolds and P\'eclet numbers introduced by a non-dimensionalization of the 
fluid equations. Because the fluid velocity $V^\mu$ is already non-dimensional in natural units based on the speed
of light $c$ and is assumed to be of order unity, the Reynolds numbers are $Re_\eta=\rho_0 c^2 L_0/\eta$
and $Re_\zeta=\rho_0 c^2 L_0/\zeta$ as given by the shear and bulk viscosities \footnote{Note that the 
relativistic viscosities as defined in our paper are $c$ times their non-relativistic counterparts, because 
$\theta, \sigma^{\mu\nu}$ as $c\to\infty$ are $1/c$ times their non-relativistic analogues $\Theta, {\bf S}$ 
in paper I}, and the particle 
and thermal P\'eclet numbers are  $Pe_\sigma=\rho_0 L_0/\sigma T_0 (mc)^2$ and $Pe_\kappa
=\rho_0 c^3 L_0/\kappa T_0.$ Here $\rho_0 c^2$ is a typical energy density, $L_0$ a length
characterizing the injection scale of the flow (as well as the turnover time $L_0/c$ in natural units), 
and $T_0$ a temperature scale such as $T(\rho_0 c^2,\rho_0).$ 
There are additional dimensionless groups which multiply the terms 
of the dissipative fluxes that are second-order in gradients, but no assumption needs to be made 
in our analysis about their magnitudes. 

In addition to formulating a theory of the turbulent entropy 
balance, we shall also derive a turbulent internal energy balance and describe with precise 
formulas the relativistic energy cascade.  Conditions for the non-vanishing of the energy flux are very similar to those 
obtained in paper I for non-relativistic flow, and the relativistic 
energy flux reduces in the limit $c\rightarrow\infty$ to the non-relativistic kinetic energy flux. 
\red{An Onsager condition for non-vanishing energy-dissipation anomaly is obtained, 
assuming positivity of the dissipative heating. Our main result on entropy production anomaly 
requires no such additional assumption and the proof requires only modest changes to that for 
non-relativistic fluids, as we demonstrate in detail below.}

\section{Relativistic Coarse-Graining}\lb{coarse} 

We employ in our analysis a coarse-graining regularization very similar to that used in our 
non-relativistic study in the companion paper I. Just as in the non-relativistic case, non-vanishing
dissipative anomalies as in (\ref{eq22}) require that gradients $\partial^\perp_\mu V_\nu,$ $\partial_\mu^\perp T,$
$\partial_\mu^\perp\lambda$ must diverge as $\sigma,\kappa,\eta,\zeta\rightarrow 0$ and this 
makes it impossible to interpret the fluid dynamical equations in the naive sense in the ideal limit.
As in the non-relativistic problem, we can remove the ultraviolet divergences by space-time coarse-graining. 
An essential difference, however, is that coarse-graining with a spherically-symmetric filter kernel guarantees invariance of 
turbulent fluxes in non-relativistic flows under the full Galilean symmetry group, but there is no possible space-time coarse-graining
that can preserve Lorentz symmetry. For example, consider a general space-time filtering operation
of the velocity field
\be \bar{V}^\mu(x) = \int d^{D}r \ {\mathcal G}_\ell(r) \ V^\mu(x+r), \lb{eq23} \ee
with ${\mathcal G}_\ell(r)=\ell^{-D} {\mathcal G}(r/\ell).$ Then it is easy to check that Lorentz 
transformations $V^{\mu\prime}(x')=\Lambda^{\mu\nu} V_\nu(\Lambda^{-1}x')$
for $\Lambda\in SO(1,d)$ when applied to the coarse-grained field in (\ref{eq23}) correspond to a coarse-graining 
of the transformed field $V^{\mu\prime}(x')$, but with a different kernel
\be {\mathcal G}'(r')={\mathcal G}(\Lambda^{-1} r').  \lb{eq24} \ee
The kernels in the two frames are the same if and only if the coarse-graining kernel satisfies for all $r$
in Minkowski space and all $\Lambda\in SO(1,d)$ that 
\be {\mathcal G}(\Lambda r)={\mathcal G}(r).  \lb{eq25} \ee
This relation requires that ${\mathcal G}(r)$ depend upon the separation vector $r^\mu$ 
only through the relativistic proper-time interval $R^2=-r_\mu r^\mu.$ In that case, however,
the space-time integral of the kernel must diverge
\be \int d^{D}r\ {\mathcal G}(r) 
= \int_{-\infty}^{+\infty} dR^2\ {\mathcal G}(R^2) \int_{H_{R^2}} \frac{d^d\br}{|r^0|}=+\infty, \lb{eq26} \ee
because of the non-compactness of the hyperboloids $H_{R^2}=\{r:\ -r_\mu r^\mu=R^2\}.$
In contrast, in the non-relativistic case the orbits of the rotation group $SO(d)$ in its action
on space are the spheres $S_\rho=\{\br:\ |\br|=\rho\},$ which are compact and have finite area. 
Because of the divergence in (\ref{eq26}), it is impossible to define a coarse-graining operation which
commutes with Lorentz transformations and whose kernel satisfies the properties of positivity 
${\mathcal G}(r)\geq 0$ and normalization 
\be \int d^{D}r\ {\mathcal G}(r) = 1. \lb{eq27} \ee
Together with rapid decay and smoothness, these properties are necessary so that 
coarse-graining is a regularizing operation which represents a local space-time averaging. 
As we shall see below, this leads to a breaking of Lorentz-covariance of the coarse-grained 
fluid equations at finite $\ell$ and possible observer-dependence of quantities such as 
turbulent cascade rates. However, we shall see that there is restoration of Lorentz symmetry 
in the limit $\ell\rightarrow 0$ (similar to the restoration of Lorentz invariance in lattice field-theories
in the limit of lattice-spacing $a\rightarrow 0.$) 

The effect of Lorentz transformation on a filter kernel can be made more concrete by considering
a pure boost in the 1-direction, with rapidity $\varphi$ related to the relative velocity $w$ by $w=c\tanh\varphi.$ 
%and $\gamma=\cosh\varphi.$ 
Using standard light-front coordinates $x^\pm=(x^0\pm x^1)/\sqrt{2}$ 
in 0-1 planes \cite{dirac1949forms}, the boost transformation becomes
\be  x^{\prime\pm} = e^{\pm\varphi} x^\pm \lb{eq28} \ee 
with all other spatial variables $x^2,...,x^d$ remaining unchanged. A filter kernel ${\mathcal G}_\ell$
is thus transformed into 
\be {\mathcal G}_\ell^\prime(r') = {\mathcal G}_\ell(e^{-\varphi}r^{\prime+},e^{+\varphi}r^{\prime-},r^{\prime 2},...,r^{\prime d}). \lb{eq29} \ee 
Effectively, the coarse-graining scale is changed for the co-moving observer to $\ell_+=e^{+\varphi}\ell$
in the $+$ direction, to $\ell_-=e^{-\varphi}\ell$ in the $-$ direction, and unchanged in the remaining 
spatial directions $2,...,d.$ This discussion of the pure boost transformation underlines the fact that the 
notion of ``scale'' will be different for different observers. 

While any filter kernel that is smooth and rapidly decaying in space-time can be adopted, 
it is also possible to use more singular kernels that will still regularize the equations of motion.
For example, as in the non-relativistic case, it is possible to filter only spatially at 
fixed time instants \footnote{The fact that spatial coarse-graining alone regularizes time-derivatives
is due the the fact that the fields in question satisfy equations of motion that are (at least) 
first-order in time}:
\be {\mathcal G}_\ell(r) = G_\ell(\br) \delta(r^0), \lb{eq30} \ee 
where $G_\ell(\br)=\ell^{-d}G(\br/\ell)$ is a smooth kernel rapidly decaying in physical space. 
Such a coarse-graining does not, of course, remain instantaneous in other reference frames. For example,
for an observer moving with relative velocity $w$ in the 1-direction the kernel in (\ref{eq30}) transforms into 
 \begin{eqnarray} 
&& {\mathcal G}_\ell'(r^{\prime 0},r^{\prime 1},...,r^{\prime d}) \cr
&& \hspace{10pt} = \gamma^{-1}(w)G_\ell(r^{\prime 1}/\gamma(w),r^{\prime 2},...,r^{\prime d}) \delta(r^{\prime 0} +w r^{\prime 1}/c), \cr
&&
\lb{eq31} \end{eqnarray} 
with Lorentz-factor $\gamma(w)=(1-w^2/c^2)^{-1/2}$ according to the general transformation formula (\ref{eq25}). 
To the relatively moving observer the filtering kernel has become non-instantaneous and, furthermore, is 
elongated along the 1-direction with modified spatial scale $\ell'=\gamma(w) \ell$ in that direction. Such elongation 
corresponds to the well-known fact that a stationary blob of fluid at an instant in the original frame is 
length-contracted by the factor $1/\gamma(w)$ in the relatively moving frame but also sweeps through 
a distance larger by the factor $\gamma(w)$ as it moves in that frame. Once again, the notion of ``scale'' 
is seen to be different for different observers. 

Another singular kernel of some interest is a spatially-weighted average over the past light-cone:
\be {\mathcal G}_\ell(r) = G_\ell(\br) \delta(r^0+|\br|). \lb{eq32} \ee 
This is natural as an average that can be, in principle, computed at each point independently 
from incoming light-signals \cite{dunkel2009non}. It may also have some utility for numerical 
modelling of relativistic fluid turbulence by the Large-Eddy Simulation (LES) methodology 
\cite{meneveau2000scale,schmidt2015large,Radice2017} since such averages can be computed on arbitrary space-like 
Cauchy surfaces using only pre-computed (past) values of simulated fields.  
For the observer moving with relative velocity $w$ in the 1-direction, the light-cone average transforms 
into another light-cone average with a different spatial kernel:
\begin{eqnarray}
&& {\mathcal G}_\ell'(r') = \delta(r^{\prime 0}+|\br'|) \cr
&& \hspace{20pt} \times \left\{\begin{array}{ll}
                                                                               f G_\ell( f r^{\prime 1}, r^{\prime 2},...,r^{\prime d}) & \mbox{ for $r^{\prime 1}>0$} \cr
                                                                               f^{-1} G_\ell( f^{-1} r^{\prime 1}, r^{\prime 2},...,r^{\prime d}) & \mbox{ for $r^{\prime 1}<0$}
                                                                        \end{array}\right.
\lb{eq33} \end{eqnarray}
for $f=\sqrt{\frac{c-w}{c+w}}=e^{-\varphi}.$ In this particular case, the spatial kernel is elongated or contracted depending upon the relative 
signs of $w$ and $r^{\prime 1}$ and an initially reflection-symmetric kernel will not remain so in a boosted frame. 

A property of the space-time coarse-graining operation (\ref{eq23}) that must be kept in mind is that the 
coarse-grained fluid velocity vector $\bar{V}^\mu,$ while it remains future time-like, is not generally 
a unit vector with respect to the Minkowski pseudometric. Under coarse-graining 
\be V^\mu = \gamma(v)(1,\bv/c) \Longrightarrow \bar{V}^\mu=\overline{\gamma(v)}(1,\hat{\bv}/c), \lb{eq34} \ee 
where we introduced the $\gamma$-weighted space-time average 
\be \hat{\bv} =\overline{\gamma(v)\bv}/\overline{\gamma(v)}, \qquad \gamma(v) = (1-v^2/c^2)^{-1/2}. \lb{eq35} \ee
By convexity of the spatial Euclidean norm-square, 
\be |\hat{\bv}|^2 \leq \widehat{|\bv|^2} < c^2. \lb{eq36} \ee
Thus,
\be \bar{V}_\mu \bar{V}^\mu = - \overline{\gamma(v)}^2/\gamma({\hat{v}})^2<0  \lb{eq37} \ee
with $\hat{v}=|\hat{\bv}|$ and $\bar{V}^\mu$ remains future time-like.
However, generally $\overline{\gamma(v)}\neq \gamma(\hat{v})$ and thus $\bar{V}_\mu \bar{V}^\mu \neq -1.$ 
Non-unit normalization of $\bar{V}^\mu$ introduces new terms into the coarse-grained equations of motion 
in the relativistic case that have no counterpart non-relativistically. 

The most important feature of the space-time coarse-graining is that, for a fixed scale $\ell,$ 
all of the dissipative transport terms in the coarse-grained conservation laws 
\be   \partial_\nu \bar{T}^{\mu\nu}=0, \quad  \partial_\nu \bar{J}^\nu=0 \lb{eq38} \ee
become negligible in the ideal limit $\sigma,$ $\kappa,$ $\eta,$ $\zeta\rightarrow 0.$ As in the 
non-relativistic case, this negligible direct effect of dissipation leads to the crucial concept of the 
``inertial-range of scales''. Because of the key importance of this result, we give here a careful 
demonstration for the class of dissipative fluid theories treated in this paper.  For simplicity, we 
consider only filter kernels that are entirely smooth in space-time, as more singular kernels
(such as instantaneous or light-cone averages) would introduce additional purely technical complications.
See \cite{drivas2017onsager} for further discussion.  Furthermore, we assume that the kernel is $C^\infty,$
compactly supported in space-time and is thus a standard test function for space-time distributions,
which further simplifies the proofs. 

We illustrate the argument with the number conservation law,  which contains the single dissipative term
\be \partial_\mu \overline{\sigma \hat{N}^\mu}(x) = - \frac{1}{\ell} \int d^{D}r\ 
(\partial_\mu {\mathcal G})_\ell(r) \sigma(x+r) \hat{N}^\mu(x+r) \lb{eq39} \ee 
where an integration by parts has been performed. Introducing as a factor of unity $T\cdot (1/T)=1,$ 
the Cauchy-Schwartz inequality gives
\begin{eqnarray}
&& |\partial_\mu \overline{\sigma \hat{N}^\mu}(x)|\leq \frac{1}{\ell}
\sqrt{\int_{{\rm supp}({\mathcal G}_\ell)} d^{D}r \ (\sigma T^2)(x+r)} \cr
&& \hspace{30pt} \times \sqrt{\int d^{D}r  \ \frac{\sigma}{T^2}(x+r) |(\partial_\mu {\mathcal G})_\ell(r) \hat{N}^\mu(x+r)|^2 }\cr
&&
\lb{eq40} \end{eqnarray}
The first square-root factor vanishes in the ideal limit under mild assumptions (e.g. if $\sigma$
goes to zero uniformly in space-time and if the temperature $T$ remains locally square-integrable). 
If we can show that the second square-root factor remains bounded in the ideal limit, then the 
product will also go to zero. 

Because the projection tensor $\Delta^{\mu\nu}$ is symmetric and also non-negative (as seen by transforming 
into the fluid rest frame),  it defines an inner product for which another application of Cauchy-Schwartz gives
\be |(\partial_\mu {\mathcal G})_\ell(r) \hat{N}^\mu(x+r)|^2 \leq 
(\partial_\mu ^\perp G)_\ell(\partial^\mu_\perp G)_\ell(r) \cdot \hat{N}_\mu\hat{N}^\mu(x+r). \lb{eq41} \ee
The integral inside the second square-root in (\ref{eq40}) is thus bounded by 
\be \int d^{D}r  \ (\partial_\mu ^\perp G)_\ell(r)(\partial^\mu_\perp G)_\ell(r) \cdot \frac{\sigma \hat{N}_\mu\hat{N}^\mu}{T^2}(x+r) \lb{eq42} \ee
and the second factor in the integrand in (\ref{eq42}) above is just the entropy production due to particle conductivity. Because 
this entropy production is assumed to converge distributionally to a non-vanishing measure $\Sigma_{cond}$
in the ideal limit, this integral would remain bounded if the first factor were a smooth test function.
Unfortunately, the last statement is generally false, because $\partial^\mu_\perp G_\ell(r)=\Delta^{\mu\nu}(x+r)\partial_\nu G_\ell(r)$
acquires a ``rough'' dependence on space-time through the velocity $V^\mu(x+r)$ in the projection tensor. However, 
it is easy to show (see Appendix \ref{app:bound}) that
\be 0\leq (\partial_\mu ^\perp G)_\ell(r)(\partial^\mu_\perp G)_\ell(r) \leq 2\gamma^2(v) |(\partial G)_\ell(r)|^2_E \lb{eq43} \ee
where we introduced the factor $\gamma(v)=V^0(x+r)>0$ associated to the fluid velocity vector 
and also the Euclidean space-time norm 
\be  |(\partial G)_\ell(r)|^2_E =\sum_{\mu=0}^d | (\partial_\mu G)_\ell(r)|^2.  \lb{eq44} \ee 
The latter quantity no longer has any dependence on the fluid velocity vector $V^\mu(x+r)$ and 
it is a standard test function ($C^\infty$ and compactly supported) when the kernel ${\mathcal G}(r)$ has 
the same properties. Thus, the integral inside the second square root of (\ref{eq40}) is bounded by 
\be 2\|\gamma(v)\|^2_\infty \int d^{D}r  \ |(\partial^\mu G)_\ell(r)|_E^2 \cdot \frac{\sigma \hat{N}_\mu\hat{N}^\mu}{T^2}(x+r) \lb{eq45} \ee
and it remains finite in the limit, if we assume that $\|\gamma(v)\|_\infty=\sup_x |V^0(x)|<\infty.$  This requires 
an assumption that the fluid speed satisfies $v\leq c(1-\delta)$ for some fixed small $\delta\ll 1.$ 
Such a $\delta$ will obviously be observer-dependent.  Under these conditions we conclude that the coarse-grained dissipative number current 
term $\partial_\mu \overline{\sigma \hat{N}^\mu}(x) \rightarrow 0,$ vanishing pointwise in the ideal limit. 

The conclusion of this argument is that the coarse-grained particle conservation law for any fixed $\ell$
in the limit $\sigma,$ $\kappa,$ $\eta,$ $\zeta\rightarrow 0$ becomes 
\be  \partial_\mu \overline{J}^\mu=\partial_\mu \overline{nV^\mu}=0,  \lb{eq46} \ee
with the dissipative term tending to zero. We thus obtain the ideal particle conservation equation in a 
``coarse-grained sense''.  As shown in \cite{drivas2017onsager}, the validity of the ideal fluid equations in this 
coarse-grained sense for all $\ell>0$ is equivalent to their validity ``weakly'' or in the sense of distributions. 
We should emphasize the non-triviality of this result. Non-vanishing 
of the distributional limit $\Sigma_{cond}=\Dlim_{\sigma,\kappa,\eta,\zeta\rightarrow 0} \ \sigma \hat{N}_\mu\hat{N}^\mu/T^2$
requires that gradients of thermodynamic potentials must diverge, or $|\partial_\mu^\perp \lambda|\rightarrow +\infty,$
if the Landau-Lifschitz contribution (\ref{eq13}) to $\hat{N}^\mu$ is the dominant one. Nevertheless, even with 
such diverging gradients of fine-grained quantities (an ``ultraviolet divergence"), the coarse-grained equations 
are regularized and any limit fields $n,$ $V^\mu$ as $\sigma,$ $\kappa,$ $\eta,$ $\zeta\rightarrow 0$ will satisfy the ideal 
particle conservation law in the coarse-grained sense. For finite but very large values of the particle P\'eclet number 
$Pe_\sigma$ this means that there is a long ``inertial-range'' of scales $\ell$ where the coarse-grained ideal 
equation is valid. 

It is worth emphasizing that the dissipation length $\ell_\sigma$ where particle conductivity $\sigma$ 
becomes non-negligible is presumably observer-dependent at finite $Pe_\sigma,$ unlike the non-relativistic 
case where all observers in different Galilean frames will agree on the dissipation lengths. Notice that 
the upper bound in (\ref{eq45}) is not Lorentz-invariant, because the gamma factor $\gamma(v)$ and the Euclidean norm 
of the kernel gradient are both frame-dependent. In fact, consider the example of a coarse-graining 
average over a Euclidean ball in space-time with radius $\ell,$ as calculated by a certain observer. 
When $\ell\gg\ell_\sigma,$ then the dissipative contribution to the coarse-grained particle current  will be 
negligible to this observer. However, this same coarse-grained particle current to a co-moving observer 
with large relative velocity $w$ in the 1-direction will correspond to a filter kernel (\ref{eq29}) with dilated thickness $e^{+\varphi}\ell_\sigma$ 
in the + direction in the 0-1 plane and contracted thickness $e^{-\varphi}\ell$ in the - direction. As a consequence, 
$\partial_-$-gradients of coarse-grained fields become large for this observer. When $w$ is sufficiently 
close to $c$ so that $e^{-\varphi}\ell\simeq \ell_\sigma,$ then the co-moving observer may find that dissipative 
particle transport is non-negligible for the coarse-grained current in his frame of reference. Of course, in the ideal 
limit $\sigma\rightarrow 0$ with the scale $\ell$ of the filter kernel fixed, every observer will agree that dissipative 
transport has vanished in the coarse-grained particle current because $\ell_\sigma\rightarrow 0.$
  
All of the conclusions derived above hold also for the coarse-grained equations of energy-momentum conservation, 
where at fixed $\ell$ in the limit $\sigma,$ $\kappa,$ $\eta,$ $\zeta\rightarrow 0,$ any limiting fields
$\epsilon, \rho,$ and $V^\mu$ will satisfy 
\be \partial_\nu \overline{T}^{\mu\nu} = \partial_\nu (\overline{\epsilon g^{\mu\nu} + p\Delta^{\mu\nu}})=0, \lb{eq47} \ee 
with all dissipation terms tending to zero. Just as for particle-conservation, the range of $\ell$ 
over which these ideal equations are valid could be observer-dependent at finite Reynolds and P\'eclet numbers. 
The proof of these statements is very similar to that given above for particle-conservation, and we thus 
give complete details in Appendix  \ref{app:bound}. From the two equations (\ref{eq46}),(\ref{eq47}) we can conclude that 
the relativistic Euler equations hold in the coarse-grained sense at fixed scale $\ell$ in the limit 
of infinite Reynolds and P\'eclet numbers, that is, ideal relativistic Euler equations hold distributionally. 

%Here we just note that the thermal conductivity terms 
%\be \partial_\nu ( \overline{\kappa \hat{Q}^{(\mu}V^{\nu)}})=0 \lb{eq} \ee  
%are the most troublesome, because they are not orthogonal to the fluid velocity vector $V^\mu$ 
%in the Minkowski pseudo-norm. We then need to use estimates involving the Euclidean norm
%$|V|_E^2,$ which can become as large as $2\gamma^2(v)$ and diverge as $v\rightarrow c.$
%To show that the thermal-conductivity term is negligible as $\sigma,$ $\kappa,$ $\eta,$ $\zeta\rightarrow 0$
%we therefore need an additional assumption that $v\leq c(1-\delta)$ for some fixed small $\delta\ll 1.$
%Obviously such a $\delta$ will be observer-dependent. In general, we find that the particle-frame 
%formulation of dissipative relativistic fluid models leads to greater difficulties and requires 
%further special assumptions than does the energy-frame formulation in the treatment of relativistic 
%fluid turbulence. 

\textcolor{black}{There is one last remark on the coarse-graining regularization which has fundamental
importance in what follows. This coarse-graining is a purely passive operation which is applied
{\it a posteriori} to the fluid variables and which can effect no change whatsoever on any physical 
occurrence \cite{eyinkRec2015,eyink2017cascadesI}. In prosaic terms,  coarse-graining corresponds to 
``removing one's spectacles'' and observing the physical evolution at a reduced space-time resolution 
$\ell.$ The effective dynamical description is changed by regularization, of course, with coarse-grained variables 
satisfying  much more complex equations than fine-grained fields. This is not unexpected 
because the coarse-grained variables are like ``block-spins'' in the renormalization-group theory 
of critical phenomena \cite{kadanoff1966scaling,wilson1971renormalization}, and such Wilson-Kadanoff RG procedures
typically lead to very complicated effective descriptions. In fact, an implementation of the 
coarse-graining by integrating out unresolved fields in a path-integral formulation yields 
an effective dynamics with higher-order nonlinearity, long-time memory, and induced stochasticity 
\cite{eyink1996turbulence}. This is a manifestation of the ``closure problem'', in which coarse-grained variables 
like $\bar{V}^\mu(x)$ no longer satisfy simple closed equations of motion. As we shall see below, Onsager's 
method does not solve this problem, but instead bypasses it by exploiting  ``4/5th-law''-type 
expressions for new, unclosed expressions. The essential idea is then to invoke the independence 
of the physics on the arbitrary coarse-graining scale $\ell.$ This simple invariance principle turns 
out to yield non-trivial consequences.}

\section{Energy Cascade}\lb{energy}

It is reasonable to expect that relativistic fluids at very high Reynolds and P\'eclet numbers should exhibit a 
turbulent energy cascade, just as do non-relativistic incompressible and compressible fluids. 
However, the familiar notion of kinetic energy cascade is not appropriate for relativistic turbulence, because
kinetic energy is not a natural concept within relativity theory. On the other hand, we have seen in 
our discussion of non-relativistic compressible fluids in paper I that energy cascade can be understood 
from the coarse-grained dynamics of the internal energy. Because the concept of internal energy 
remains valid in relativistic thermodynamics, it provides a good basis for the theory of relativistic 
energy cascade. 

A resolved energy current is defined most simply as 
\be \underline{{\mathcal E}}^\mu_\ell = -\overline{T}^{\mu\nu} \overline{V}_\nu, \lb{eq48} \ee 
which (like resolved kinetic energy in non-relativistic turbulence) is a nonlinear function of coarse-grained
quantities. As we have shown in some detail in Appendix \ref{app:bound},  for length-scales $\ell$ in the inertial-range, 
or for all fixed $\ell$ in the ideal limit $\sigma,$ $\kappa,$ $\eta,$ $\zeta\rightarrow 0,$
\begin{eqnarray}
\overline{T}^{\mu\nu}
&=& \overline{\epsilon V^\mu V^\nu + p \Delta^{\mu\nu}} \cr
&=& \bar{p}g^{\mu\nu} +\overline{hV^\mu V^\nu}, \lb{eq49} 
\end{eqnarray} 
where $h=\epsilon+p$ is the {\it relativistic enthalpy}. If subsequent to the ideal limit 
$\sigma,$ $\kappa,$ $\eta,$ $\zeta\rightarrow 0,$ one considers the limit of regularization length-scale $\ell\to 0,$ 
one finds that 
\be \Dlim_{\ell\rightarrow 0}\underline{{\mathcal E}}^\mu_\ell = \epsilon V^\mu, \lb{eq50} \ee
For this to hold, one needs only some modest regularity of the limiting variables $\epsilon,$ $p,$ $V^\mu,$ 
such as finite (absolute) 4th-order moments in local space-time averages.  Thus, the resolved energy current converges 
distributionally in the ``continuum limit'' $\ell\rightarrow 0$ to the fine-grained energy current 
of the Euler fluid. The naive energy balance obtained by setting ${\mathcal Q}_{diss}=0$ in (\ref{eq9}) does not
follow, however. To obtain the correct result, we can use the balance equation for the resolved energy current 
\begin{eqnarray}
\partial_\mu \underline{{\mathcal E}}^\mu_\ell &=& \partial_\mu\left(-\bar{V}_\nu \bar{T}^{\mu\nu}\right)\cr
 &=& -(\partial_\mu\bar{V}_\nu) \bar{T}^{\mu\nu} \cr
&=& -\bar{p} (\partial^\nu\bar{V}_\nu) - (\partial_\mu\bar{V}_\nu) \overline{hV^\mu V^\nu}. 
\lb{eq51} \end{eqnarray} 
obtained from $\partial_\mu \bar{T}^{\mu\nu}=0$ and (\ref{eq49}). 
The last term in equation (\ref{eq51}) would not be present in the fine-grained internal energy balance for a smooth
Euler solution, because of the orthogonality condition $(\partial_\mu V_\nu) V^\nu=0.$ This term 
is the source of possible energy dissipation anomalies in relativistic fluid turbulence and it gives the simplest 
representation of turbulent energy flux. 

Despite the simplicity of the above formulation, we shall follow here an alternative approach 
based upon a {\it relativistic Favre-averaging}, similar to that employed in paper I 
for non-relativistic compressible turbulence. It should be emphasized that the entire theory 
presented below could be developed just as easily using the equation (\ref{eq51}). However, 
the relativistic Favre-averaging approach is convenient to compare with results of I in the limit $c\to 0.$ 
The proper relativistic generalization of Favre-averaging is motivated by the appearance 
of the enthalpy in (\ref{eq49}). Note that the ``null energy condition'' $h\geq 0$ is a condition for stability 
of thermodynamic equilibrium \cite{hiscock1983stability} and in the strict 
form $h>0$ is required for causality of the relativistic Euler fluid \cite{geroch1990dissipative}.   
%Note that $h=(\partial\epsilon/\partial n)_{\hat{s}}$ with $\hat{s}=s/n$ the entropy-per-particle. Thus, $h>0$ when $\epsilon$ increases 
%with $n$ for fixed $\hat{s}.$ One may also use a Favr\'e average based upon $n,$ with slightly more complicated expressions resulting. 
We thus define the Favre-average coarse-graining for a relativistic fluid by
\be \widetilde{f} = \overline{h f}/\overline{h} \lb{eq52} \ee 
with enthalpy-weighting. With this definition, (\ref{eq49}) becomes 
\begin{eqnarray} 
\bar{T}^{\mu\nu}&=&\bar{p}g^{\mu\nu} +\bar{h} \widetilde{V^\mu V^\nu} \cr
&= & \bar{p}g^{\mu\nu} +\bar{h} \tilde{V}^\mu \tilde{V}^\nu +\bar{h} \widetilde{\tau}(V^\mu,V^\nu) 
\lb{eq53} \end{eqnarray}  
As in the non-relativistic theory, expanding in the $p$th-order cumulants $\tilde{\tau}(f_1,...,f_p)$ 
of the Favre-average produces only a single ``unclosed'' term in the 
coarse-grained stress-energy tensor, whereas expanding in $p$th-order cumulants $\bar{\tau}(f_1,...,f_p)$ 
of the unweighted space-time coarse-graining would produce more such unclosed terms. This is a 
significant advantage of the Favre-average for potential applications to ``large-eddy simulation'' (LES) 
modeling of relativistic fluid turbulence. Within the Favre-averaging approach, it is convenient to 
define the resolved energy current by 
\be \utilde{{\mathcal E}}^{\mu}= -\widetilde{V}_\nu \bar{T}^{\mu\nu} +\frac{\bar{p}}{\bar{h}}\bar{\tau}(h,V^\mu)
- \frac{1}{2}\tilde{\tau}(V_\nu,V^\nu) \overline{h V^\mu}  \lb{eq54} \ee 
Alternative expressions for this current follow from (\ref{eq49}),  the relation  
\be \tilde{V}^\mu=\overline{V}^\mu+(1/\overline{h})\tau(h,V^\mu), \lb{eq55} \ee
and $V^\nu V_\nu=-1,$ from which one can easily derive 
\be -\widetilde{V}_\nu \bar{T}^{\mu\nu} +\frac{\bar{p}}{\bar{h}}\bar{\tau}(h,V^\mu)
 = \overline{hV^\mu} + \overline{h V_\nu V^\nu V^\mu} -\widetilde{V^\mu V^\nu} \overline{h V_\nu} 
-\overline{p}\,\overline{V}^\mu\lb{eq56}. \ee
Thus, one obtains that \footnote{The corresponding expression for the resolved energy current defined 
in (\ref{eq48}) is ${{\mathcal E}^\mu}_{{\!\!\!\!\!\!}^{-}}$ $=\overline{\epsilon}$ $\overline{V}^\mu+\overline{\tau}(h,V^\mu)
+\overline{h}\ {\tau}{\!\!\!}^{\sim}(V^\nu,V_\nu) V^{{\!\!\!\!\!}^{\sim\mu}}
-\overline{h}\ {\tau}{\!\!\!}^{\sim}(V^\mu,V_\nu) V^{{\!\!\!\!\!}^{\sim\nu}}$}
\begin{eqnarray}
&& \utilde{{\mathcal E}}^{\mu} 
= \overline{\epsilon V^\mu} + \bar{\tau}(p,V^\mu) +\overline{h}\Big[ \frac{1}{2}\tilde{\tau}(V_\nu,V^\nu)\tilde{V}^\mu \cr
&& \hspace{50pt} + \tilde{\tau}(V_\nu,V^\mu)\tilde{V}^\nu + \tilde{\tau}(V_\nu,V^\nu,V^\mu)\Big]  \cr
&& \hspace{17pt} = \overline{\epsilon}\overline{V}^\mu + \bar{\tau}(h,V^\mu) +\overline{h}\Big[ \frac{1}{2}\tilde{\tau}(V_\nu,V^\nu)\tilde{V}^\mu \cr
&& \hspace{50pt} + \tilde{\tau}(V_\nu,V^\mu)\tilde{V}^\nu + \tilde{\tau}(V_\nu,V^\nu,V^\mu)\Big].
\lb{eq57} \end{eqnarray}
Either from this expression or directly from the definition (\ref{eq54}) one can see that 
\be \Dlim_{\ell\rightarrow 0}\utilde{{\mathcal E}_\ell^\mu} = \epsilon V^\mu, \lb{eq58} \ee

\vspace{-10pt} 
\noindent under the same assumptions as (\ref{eq50}). Once again, however, the naive energy balance (\ref{eq9}) 
for the limiting current ${\mathcal E}^\mu=\epsilon V^\mu$ need not hold with $\calQ_{diss}=0$, but instead 
\textcolor{black}{may be modified by a turbulent dissipative anomaly if 
\be {\mathcal Q}_{diss} := \Dlim_{\eta,\zeta,\kappa,\sigma\rightarrow 0} {\mathcal Q}_{diss}^{\eta,\zeta,\kappa,\sigma} 
\neq 0
\lb{eq72} \ee
for ${\mathcal Q}_{diss}^{\eta,\zeta,\kappa,\sigma}$ given by (\ref{eq10}). We emphasize 
that the vanishing or not of the limit in (\ref{eq72}) is an objective physical fact, which cannot 
depend upon any coarse-graining.}  

To obtain the correct equation, one can use the inertial-range balance equation for the energy current 
defined in (\ref{eq54}).  Using $\partial_\mu \bar{T}^{\mu\nu}=0,$ one gets after some straightforward 
calculations that 
\be
\partial_\mu \utilde{{\mathcal E}}^{\mu}=-\bar{p} (\partial^\nu\bar{V}_\nu) + {\mathcal Q}^{flux}_\ell 
\lb{eq59} \ee

\vspace{-10pt} 
\noindent with relativistic energy flux defined by 
\begin{eqnarray} 
&& {\mathcal Q}^{flux}_\ell\equiv  \frac{1}{\bar{h}} (\partial_\nu\bar{p})\bar{\tau}(h,V^\nu) \cr 
&&  \vspace{30pt} -\bar{h} (\partial_\mu\widetilde{V}_\nu) \tilde{\tau}(V^\mu,V^\nu) 
-\frac{1}{2}\partial_\nu\overline{hV^\nu}\ \tilde{\tau}(V_\mu,V^\mu).  
\lb{eq60} \end{eqnarray} 
The energy flux ${\mathcal Q}_\ell^{flux}$ can be interpreted as the ``apparent dissipative heating''
in the large-scales based only on measurements resolved at that scale.
The first two terms in the energy flux (\ref{eq60}) are relativistic generalizations of the baropycnal work 
and the deformation work as defined by Aluie \cite{aluie2011compressible,aluie2013scale} for non-relativistic 
compressible fluids, whereas the third term has no non-relativistic analogue. The balance equation (\ref{eq59}) is 
formally very similar to the non-relativistic balance equation (I;57) for the ``intrinsic large-scale internal energy'',  
defined in (I;58). Not only does (\ref{eq59}) resemble the non-relativistic balance equation 
derived in paper I, but we show in Appendix \ref{Nonrel-Lim} that it reduces to it in the formal limit $c\to 0.$ 
In particular, the relativistic energy flux that we defined in (\ref{eq60}) converges as $c\rightarrow\infty$
to the non-relativistic expressions in \cite{aluie2011compressible,aluie2013scale} and in (I;43). 

Now \textcolor{black}{let us exploit the fact that a non-zero energy dissipation anomaly in the 
ideal limit (\ref{eq72}) cannot depend upon any particular choice of the regularization scale $\ell$. 
Subsequent to the limit $\eta,\zeta,\kappa,\sigma\rightarrow 0$ one can thus }
consider the limit $\ell\rightarrow 0$ of the coarse-grained internal energy balance (\ref{eq59}) 
for the relativistic Euler fluid. It follows from (\ref{eq58}) that the lefthand side converges 
distributionally to $\partial_\mu(\epsilon V^\mu),$  because the overall derivative $\partial_\mu$ 
can be transferred to a test function. We also define
the distributional product of the dilatation $\theta=\partial_\mu V^\mu$ and the pressure $p$ by  
a standard procedure \cite{Oberguggenberger92}
\be p \circ \theta = \Dlim_{\ell\rightarrow 0}\ \overline{p}\cdot \overline{\theta}, \lb{eq61} \ee
just as in the non-relativistic case in paper I. Although all of the cumulant factors appearing in the 
energy flux (\ref{eq60}) vanish as $\ell\rightarrow 0,$ the flux ${\mathcal Q}^{flux}_\ell$ itself need 
not vanish because the space-time gradients multiplying them diverge in the same limit. 
By taking the limit $\ell\rightarrow 0$ of (\ref{eq59}), one thus obtains for the relativistic Euler solutions 
the distributional energy balance 
\be \partial_\mu\left(\epsilon V^\mu \right)= -p \circ \theta + {\mathcal Q}_{flux}, 
\lb{eq62} \ee
with a possible anomaly due to energy cascade given by 
\begin{eqnarray} 
{\mathcal Q}_{flux} &=& \Dlim_{\ell\rightarrow 0} {\mathcal Q}_{flux,\ell} \cr 
&=&  \Dlim_{\ell\rightarrow 0}  - (\partial_\nu\bar{V}_\mu) \overline{hV^\mu V^\nu}. 
\lb{eq63} \end{eqnarray} 
Note that the second expression in the equation above arises from the corresponding 
$\ell\rightarrow 0$ limit of (\ref{eq51}). 

A condition for the non-vanishing of the anomaly ${\mathcal Q}^{flux}$ can be obtained just as 
in the non-relativistic case (see \cite{aluie2013scale}  and section V of paper I), by deriving 
``4/5th-law''-type expressions for the turbulent energy flux. The key point 
is that the cumulants of fields with respect to the space-time coarse-graining can be written 
instead as cumulants of their space-time increments with respect to an  average over 
displacement vectors $r^\mu$ weighted by the filter kernel. That is,
\be \bar{\tau}_\ell(f_1,...,f_p)(x+a) = \langle (\delta f_1)\cdots (\delta f_p)\rangle^{cum}_{\ell,a}, \lb{eq64} \ee
where 
\be  \delta f_i(x;r) = f_i(x+r)-f_i(x), \lb{eq65} \ee
are space-time increments and where, for any function $h(r),$
\be  \langle h\rangle_{\ell,a}=\int d^{D}r\ {\mathcal G}_\ell(r-a)\ h(r). \lb{eq66} \ee 
The superscript $cum$ in (\ref{eq64}) denotes the $p$th-order cumulant part of any $p$th-order moment.
The details of the proof are given in Appendix B of \cite{eyinkRec2015}, but the essential point is 
that cumulants are invariant under shifts of variables by constants and the increment 
$\delta f_i(x;r)$ is the shift of $f_i(x+r)$ by the quantity $-f_i(x)$ which is ``constant'', i.e. 
 independent of $r^\mu.$ The translation by the spacetime vector $a^\mu$ in (\ref{eq64}) 
 is useful to derive expressions for all space-time gradients of coarse-graining cumulants
 in terms of increments, by differentiating with respect to $a^\mu$ and then setting $a^\mu=0.$ 
For example, for $p=1$ one obtains with $\bar{f}=\tau(f)$ that  
\be \partial_\mu \bar{f}(x)= -\frac{1}{\ell} \int d^{D}r \ (\partial_\mu {\mathcal G})_\ell(r)\ \delta f(x;r),  \lb{eq67} \ee 
and analogous expressions for all $p>1$ and all orders of derivatives (\cite{eyinkRec2015}, Appendix B).
Expanding the Favr\'e-average cumulants into cumulants of the unweighted coarse-graining, one
thus obtains expressions for all of the contributions to the energy flux in terms of space-time increments 
of the thermodynamic fields. 

From these expressions in terms of space-time increments, we can derive necessary conditions 
for turbulent energy dissipation anomalies. Let us defined scaling exponents of space-time structure 
functions by
\be \zeta_q^f =\liminf_{|r|_E\rightarrow 0}\frac{\log \|\delta f(r)\|_q^q}{\log|r|_E}, \lb{eq68} \ee
where $\|\delta f(r)\|_q$ is the space-time $L_q$ norm of the increment and $S_q^f(r)=\|\delta f(r)\|_q^q$
is thus the $q$th-order (absolute) structure-function of $f.$ From the expressions 
in (\ref{eq64}),(\ref{eq66}) one can see that the baropycnal work term in (\ref{eq60}) vanishes as 
$\ell\rightarrow 0,$ unless for every $q\geq 3$
\be \zeta_q^p +\zeta_q^h+\zeta_q^v\leq q. \lb{eq69} \ee
Likewise, the deformation work and 
the third term in (\ref{eq60}) vanish as $\ell\rightarrow 0$ unless for every $q\geq 3$ either 
\be \zeta_q^h+2\zeta_q^v\leq q, \lb{eq70} \ee
or 
\be 3 \zeta_q^v\leq q.  \lb{eq71} \ee 
The arguments here closely parallel those in paper I for the non-relativistic case. In deriving these results 
we have assumed that the enthalpy $h$ is bounded away from both zero 
and infinity. The inequalities (\ref{eq69})-(\ref{eq71}) demonstrate that singularities of the fluid variables 
$\epsilon,$ $\rho,$ and $V^\mu$ are required in the ideal limit $\sigma,$ $\kappa,$ $\eta,$ $\zeta\rightarrow 0$
in order to obtain a non-vanishing energy dissipation anomaly from turbulent cascade. This is a scale-local 
cascade process as long as all of the structure-function exponents satisfy $0<\zeta_q^f<q$ for $f=p,$ $h,$ $v$ 
\cite{eyink2005locality}. 

The internal energy balance (\ref{eq62}) of limiting Euler solutions can also be obtained from the fine-grained
internal energy balance (\ref{eq9}) of the dissipative fluid model, by taking directly the limit 
$\sigma,$ $\kappa,$ $\eta,$ $\zeta\rightarrow 0.$ In particle-frame fluid models, the 
dissipative heat current contribution $\partial_\mu (\kappa \hat{Q}^\mu)$ can be shown to vanish
by arguments similar to those applied to the dissipative terms in the coarse-grained conservation laws.
The details are presented in Appendix \ref{fine-balance}. \textcolor{black}{Defining 
${\mathcal Q}_{diss}$ as in (\ref{eq72})} and defining also  
\be p * \theta = \Dlim_{\sigma,\kappa,\eta,\zeta\rightarrow 0}\ p\cdot \theta, \lb{eq73} \ee 
we then obtain the distributional balance equation
\be \partial_\mu\left(\epsilon V^\mu \right)= -p * \theta + {\mathcal Q}_{diss}. \lb{eq74} \ee
As in the non-relativistic case discussed in I, one must expect that the limit $p * \theta$ in (\ref{eq73}) is generally 
distinct from $p \circ \theta$ in  (\ref{eq61}), that is, the double limits of $\bar{p}_\ell\bar{\theta}_\ell$ for  
$\eta,\zeta,\kappa,\sigma\rightarrow 0$ and for $\ell\rightarrow 0$ do not commute. In fact,
the quantities $p * \theta$ and ${\mathcal Q}_{diss}$ are presumably not completely universal and 
may depend upon the particular sequence $\eta_k,\zeta_k,\kappa_k,\sigma_k\rightarrow 0$ 
used to reach infinite Reynolds and P\'eclet numbers. This is known to be true in the non-relativistic 
limit, as verified in paper I. 
%For example, in relativistic shock solutions these two quantities are found to be Prandtl-number 
%dependent, following  arguments similar to those in Appendix A for the non-relativistic case. 
However, it is a consequence of (\ref{eq74}) that the particular combination $-p * \theta + {\mathcal Q}_{diss}$
depends only upon the limiting weak solution and not upon the particular sequence of transport 
coefficients used to obtain it. 

A comparison of (\ref{eq62}) and (\ref{eq74}) shows that the two 
balance equations can be simultaneously valid only if $-p\circ \theta+{\mathcal Q}_{flux} 
=-p*\theta +{\mathcal Q}_{diss}.$ In that case, by introducing the 
{\it relativistic pressure-work defect}  
\be \tau(p,\theta)\equiv p*\theta-p\circ \theta \lb{eq75} \ee
we can then rewrite the inertial-range balance (\ref{eq62}) as 
\be \partial_\mu\left(\epsilon V^\mu \right)= -p * \theta + {\mathcal Q}_{inert}, \lb{eq76} \ee
where the total {\it inertial energy dissipation} is defined by 
\be {\mathcal Q}_{inert}\equiv \tau(p,\theta) + {\mathcal Q}_{flux} = {\mathcal Q}_{diss}, \lb{eq77} \ee 
As in the non-relativistic case considered in paper I, the inertial-range energy dissipation can arise not only from 
energy cascade but also from pressure-work defect. Relativistic shock solutions provide explicit 
examples with $\tau(p,\theta)\neq 0$ (Appendix \ref{shocks}).  Unlike the non-relativistic case, it is not known 
rigorously that ${\mathcal Q}_{diss}\geq 0.$ 

Another important distinction of the relativistic situation is that neither the energy flux ${\mathcal Q}^{flux}_\ell$
nor the pressure-work $\bar{p}_\ell\bar{\theta}_\ell$ at finite $\ell$ are Lorentz-invariant scalars, 
whereas the corresponding quantities are Galilei-invariant in non-relativistic compressible turbulence.
Although $\bar{p}_\ell\bar{\theta}_\ell$ and the expression (\ref{eq60}) for ${\mathcal Q}^{flux}_\ell$ appear 
to define invariant scalars, they involve the kernel ${\mathcal G}_\ell(r),$ which is not 
frame-invariant. Thus, the coarse-graining regularization breaks Lorentz-symmetry, somewhat 
similar to lattice-regularizations in relativistic quantum field-theory with finite lattice constant $a$.  
In contrast, the fine-grained dissipation ${\mathcal Q}_{diss}^{\eta,\zeta,\kappa,\sigma}$ and 
the fine-grained pressure-work $p\cdot\theta$ are both Lorentz-scalars, and thus their ideal limits 
${\mathcal Q}_{diss}$ and $p*\theta$ as $\eta,\zeta,\kappa,\sigma\rightarrow 0$ must be invariant 
as well. It may appear somewhat unsatisfactory that the energy flux ${\mathcal Q}^{flux}_\ell$
and the resolved pressure-work $\bar{p}_\ell\bar{\theta}_\ell$ at finite $\ell$ are observer-dependent. 
However, Lorentz-invariance is restored in the $\ell\rightarrow 0$ limit, as easily proved for the 
combinations $-p\circ\theta+{\mathcal Q}_{flux}$ and, in particular, ${\mathcal Q}_{inert}=\tau(p,\theta)+{\mathcal Q}_{flux}.$
The invariance of $-p\circ\theta+{\mathcal Q}_{flux}$ can be seen from its equality with both 
$\partial_\mu(\epsilon V^\mu)$ and $-p*\theta+{\mathcal Q}_{diss},$ which are Lorentz-scalars.
Likewise, ${\mathcal Q}_{inert}={\mathcal Q}_{diss}$ which is an invariant scalar.  
It is reassuring that the net inertial-range dissipation is observer-independent for the limit $\ell\rightarrow 0.$ 

This invariance must hold, within some limits, also for $\ell$ finite but very small, at large Reynolds and 
P\'eclet numbers.  The reason is that the only effect of a change of inertial frame is to change 
the filter kernel from ${\mathcal G}_\ell$ to ${\mathcal G}'_\ell$ as in (\ref{eq24}), but the $\ell\rightarrow 0$ limits 
of $\bar{p}_\ell\bar{\theta}_\ell$ and ${\mathcal Q}^{flux}_\ell$ as distributions, when they exist at all, 
must be independent of the specific filter kernel adopted \cite{Oberguggenberger92}. This argument 
implies that the two distributions $p\circ\theta$ and ${\mathcal Q}_{flux}$ are, in fact,  
Lorentz-invariant scalars separately and not only in combination
 \footnote{\red{In somewhat more detail, the pressure-work term transformed to a new Lorentz frame 
 is $((p\star {\mathcal G}_\ell)(\theta\star{\mathcal G}_\ell))'=
 (p'\star {\mathcal G}_\ell')(\theta'\star {\mathcal G}_\ell'),$  where ``$\star$'' 
 denotes space-time convolution. Because of the independence of the distributional product on ${\mathcal G},$  
 one recovers $(p\circ\theta)'=p'\circ\theta'$ in the limit $\ell\to 0.$ The distributional limit 
 ${\mathcal Q}_{flux}$ is then also independent of the filter kernel and a Lorentz scalar, because 
$p\circ \theta$ and $\partial_\mu(\epsilon V^\mu)$ separately possess those properties}}. For sufficiently small $\ell$ inside 
a long inertial range at large $Re$ and $Pe$, this invariance of the $\ell\rightarrow 0$  limiting distributions 
must hold approximately. On the other hand, some observer dependence presumably arises for $\ell$ small but 
non-zero. For example, two observers moving at sufficiently high relative velocities may disagree about the 
negligibility of the microscopic dissipation for the same coarse-grained fields. For one observer 
${\mathcal Q}^{inert}_\ell$ may account for all of the dissipation of resolved fields, while for the other 
the combination ${\mathcal Q}^{inert\prime}_\ell+ \underline{{\mathcal Q}}_{diss}^{\eta,\zeta,\kappa,\sigma\prime}$ 
is necessary to account for all of the dissipation in resolved fields, where $\underline{{\mathcal Q}}_{diss}^{\eta,\zeta,\kappa,\sigma}$
is the resolved viscous and conductive dissipation \footnote{Precisely,  
$\uline{{\mathcal Q}}_{diss}^{\eta,\zeta,\kappa,\sigma}=-(\partial_\mu\overline{V}_\nu)\overline{\kappa \hat{Q}^{(\mu}V^{\nu)}+\zeta\hat{\tau}\Delta^{\mu\nu}+2\eta{\hat{\tau}}^{\mu\nu}}$}.
The observed flux contributions will then be distinct.

We have focused in this section on the large-scale/resolved internal energy balance, but there is as well 
a complementary budget for the unresolved/subscale energy current. In the case of an unweighted 
space-time coarse-graining,  
the unresolved current can be naturally defined  
by $\underline{K}^\mu=-\bar{\tau}(T^{\mu\nu},V_\nu),$ so that its sum with the resolved current $
\underline{{\mathcal E}}^\mu=-\bar{T}^{\mu\nu}\bar{V}_\nu$ 
accounts for the total energy current.  Likewise within the Favre-average coarse-graining approach, the 
subscale internal energy current can be defined as $\utilde{K}^\mu= \overline{{\mathcal E}}^\mu
-\utilde{{\mathcal E}}^{\mu},$ which with (\ref{eq54}) gives

\vspace{-10pt}
\be \utilde{K}^{\mu} = -\bar{\tau}(V_\nu,T^{\mu\nu}) +\widetilde{V^\mu V^\nu}\bar{\tau}(h,V_\nu) 
+ \frac{1}{2}\tilde{\tau}(V_\nu,V^\nu) \overline{hV^\mu}. \lb{eq78} \ee
From the separate balance equations for $\overline{{\mathcal E}}^\mu$ and $\utilde{{\mathcal E}}^{\mu}$ it 

\vspace{-7pt}
\noindent easily follows that  
\be \partial_\mu \utilde{K}^{\mu}= -\bar{\tau}(p,\theta) + \bar{\calQ}_{diss}-\calQ_{flux,\ell}. 
\lb{eq79} \ee 

\vspace{-10pt} 
\noindent The source term on the righthand side is the difference between the true dissipative heating $\bar{\calQ}_{diss}$ and 
the ``apparent dissipation'' $\calQ_{flux,\ell}$ based on measurements at scales $>\ell,$ together with the 
pressure-work defect $\bar{\tau}(p,\theta)$ which represents the difference between the true pressure-work
$\overline{p*\theta}$ and the apparent pressure-work $\overline{p}\cdot\overline{\theta}$  based on fields
resolved also down to scales $\ell.$  
%In the limit $\ell\rightarrow 0$ of perfect resolution, the unresolved internal 
%energy obviously vanishes, with $-\bar{\tau}(T^{\mu\nu},V_\nu),$ converging distributionally to zero. 
%Both sides of the budget equation (???) also vanish in that limit (the right side as a consequence of (???)), and 
%the equation reduces to a tautology. However, we shall see in the following section that it provides crucial
%information for $\ell$ positive but tending to zero.  
Using the expression (\ref{eq57}) for $\utilde{{\mathcal E}}^{\mu},$ the subscale internal energy current can be 

\vspace{-7pt}
\noindent rewritten in terms of relativistic Favre-average cumulants of the velocity.
In particular, its negative becomes 
\begin{eqnarray}
-\utilde{K}^{\mu}=
&=& \overline{h}\Big(\frac{1}{2}\tilde{\tau}(V_\nu,V^\nu) \widetilde{V}^\mu +\tilde{\tau}(V_\nu,V^\mu) \widetilde{V}^\nu\cr
&& \hspace{10pt} +\tilde{\tau}(V_\nu,V^\nu,V^\mu)\Big) + \bar{\tau}(p,V^\mu). 
\lb{eq80} \end{eqnarray} 
Substituting this expression into (\ref{eq79}) yields a balance equation very similar in form to 
the non-relativistic subscale kinetic energy balance obtained in (I;64), and in fact formally reducing to the latter in the limit $c\to \infty$
(Appendix \ref{Nonrel-Lim}). This identity will prove very important for the discussion in the following section.  

\section{Entropy Cascade}\lb{entropy}

\textcolor{black}{Hydrodynamic turbulence, as any other macroscopic irreversible process, must be consistent with 
the second law of thermodynamics. In the relativistic case, in particular, positive entropy production is a 
primary constraint on dissipative fluid models \cite{israel1979transient,israel1979transientII,baier2008relativistic,bhattacharyya2008local,romatschke2010new}}. 
For non-relativistic compressible turbulence we have argued in paper I that there is a cascade of (neg)entropy, 
which is in addition to energy cascade and which is even more fundamental. All of these arguments carry 
over to relativistic fluid turbulence. The resolved pressure-work in the balance equations (\ref{eq51}) 
or (\ref{eq59}) for the large-scale internal energy current is a space-time structured source of 
internal energy. In relativistic thermodynamics, as in the non-relativistic case, the entropy per volume 
$s(\epsilon,\rho)$ is a concave function of $\epsilon$ and $\rho,$ so that the creation of large-scale structure in $\epsilon$
corresponds to a decrease of entropy at large-scales. To balance this destruction, one can then expect that there 
will be an inverse cascade of the entropy which is injected by microscopic dissipation/entropy production.  
As in the non-relativistic case, we may define a ``resolved entropy''
\be \underline{s} = s(\overline{\epsilon},\overline{\rho}) \lb{eq81} \ee
and an ``unresolved/subscale entropy''
\be \triangle s=\overline{s(\epsilon,\rho)}-s(\overline{\epsilon},\overline{\rho})\leq 0, \lb{eq82} \ee 
whose non-positivity follows from the concavity of the entropy. It is somewhat more natural to consider the
{\it negentropy} or {\it information} density $\iota(\epsilon,\rho)=-s(\epsilon,\rho),$ which is convex and whose 
unresolved/subscale contribution $\Delta\iota=-\triangle s$ is non-negative. In this equivalent picture, 
the pressure-work injects negentropy at large scales, which should cascade forward to small scales 
where it can be efficiently destroyed by dissipative transport.  In order to formalize such notions, one
must derive a balance equation for the large-scale entropy.

This balance is straightforward to derive after taking the limit $\eta$, $\zeta,$ $\sigma,$ $\kappa\rightarrow 0$
for fixed positive $\ell.$ Using the first law of thermodynamics $ds=\beta d\epsilon-\lambda dn$ and $\bar{{\mathcal D}}
=\bar{V}_\mu\partial^\mu,$ one gets 
\be \bar{{\mathcal D}}\us = \ubeta\bar{{\mathcal D}}\bar{\epsilon} - \ulambda\bar{{\mathcal D}}\bar{n} \lb{eq83} \ee
where we employ the notation $\underline{\phi}=\phi(\overline{\epsilon},\overline{\rho})$ for arbitrary 
smooth functions $\phi$ of $\epsilon,$ $\rho.$ The equations
\be \bar{{\mathcal D}}\bar{n} = -\bar{n}\bar{\theta} -\partial_\mu \bar{\tau}(n,V^\mu) \lb{eq84} \ee
%and 
\be \bar{{\mathcal D}}\bar{\epsilon} = -\bar{\epsilon}\bar{\theta} -\overline{p*\theta}
- \partial_\mu \bar{\tau}(\epsilon,V^\mu) + \overline{{\mathcal Q}}_{diss} \lb{eq85} \ee
are direct consequences of (\ref{eq46}) and (\ref{eq74}). Using the Gibbs 
homogeneous relation $(\epsilon+p)/T=s+ \lambda n,$
one obtains after some straightforward calculations a balance equation of the following form:  
\red{
\be 
\partial_\mu\underline{S}^\mu  = 
\Sigma^{inert}_\ell. \lb{eq86} \ee}
The vector whose divergence appears on the left 
\be \underline{S}^\mu= \us\bar{V}^\mu +\ubeta\bar{\tau}(\epsilon,V^\mu) -\ulambda \bar{\tau}(n,V^\mu)  \lb{eq87} \ee 
is a natural expression for the resolved entropy current, with
$\us\bar{V}^\mu$ describing the entropy transport by large-scale advection, $\ubeta\bar{\tau}(\epsilon,V^\mu)$ 
the entropy transport due to sub-scale internal energy current, and $\ulambda \bar{\tau}(n,V^\mu)$ the entropy transport 
due to subscale number current. It should be noted that entropy current due to such turbulent subscale transport
will not generally be othogonal to $\bar{V}^\mu$ in the Minkowski pseudometric, and thus 
not purely spatial in the rest-frame of the coarse-grained fluid velocity. 

\red{The source on the righthand side of (\ref{eq86}) is the 
inertial-range entropy production 
\be \Sigma^{inert}_\ell = -I^{mech}_\ell+\ubeta \overline{{\mathcal Q}}^{diss} + \Sigma^{flux}_\ell,  \lb{eq89} \ee  
where anomalous input of negentropy from pressure work is defined by 
 \be  I^{mech}_\ell =  \ubeta \left( \overline{p*\theta}-\up\ \bar{\theta}\right) \lb{eq88} \ee
and (forward) negentropy flux is by 
\be  \Sigma^{flux}_\ell=(\partial_\mu \ubeta) \bar{\tau}(\epsilon,V^\mu)-(\partial_\mu\ulambda)\bar{\tau}(n,V^\mu) . \lb{eq90} \ee
The latter expression} is also natural, as it represents entropy production due to subscale transport of internal
energy and particle number acting against large-scale gradients of the (entropically) conjugate thermodynamic potentials. 
In particular, $\Sigma^{flux}_\ell>0$ when the subscale transport vectors are ``down-gradient'', or opposite to the 
gradients of $T$ and $\lambda$.  Finally, note that one can further decompose the anomalous negentropy input as 
 \be  I^{mech}_\ell =   \ubeta \bar{\tau}(p,\theta) + I^{flux}_\ell \lb{eq91} \ee
 where the first term is the contribution from the pressure-dilatation defect and the second term 
 \be I^{flux}_\ell=\ubeta (\op-\up) \bar{\theta} \lb{eq92} \ee
 is ``flux-like'', representing work of subscale pressure fluctuations against large-scale dilatation. These 
 expressions are exactly analogous to those derived in section VI of paper I for the turbulent entropy balance of 
 non-relativistic compressible fluid flows. In fact,  as we show in Appendix \ref{Nonrel-Lim}, the formal limit 
 $c\rightarrow\infty$ recovers the previously derived non-relativistic expressions. 

\textcolor{black}{Now consider the case that there is a non-vanishing entropy production anomaly as in 
(\ref{eq22}). If such an anomaly exists, it cannot depend upon the arbitrary coarse-graining scale $\ell$.
Thus, } for ideal turbulence at infinite Reynolds and P\'eclet numbers, we may consider the subsequent limit
$\ell\rightarrow 0$ of the inertial-range entropy balance, with the coarse-graining regularization removed. 
This yields a fine-grained entropy balance for the relevant weak solutions of the relativistic Euler equations:
\red{
\be  \partial_\mu(s V^\mu)= \Sigma_{inert}, \lb{eq93} \ee
Because all coarse-graining cumulants vanish distributionally as $\ell\rightarrow 0,$ the resolved 
entropy current must converge in the sense of distributions to $s V^\mu$ under relatively mild 
assumptions (e.g. when $\epsilon$ and $\rho$ are bounded in space-time). The limit $\Sigma_{inert}$ of the 
source (\ref{eq89}) is $\Sigma_{inert}=-I_{mech}+\Sigma_{flux}+\beta\circ {\mathcal Q}_{diss},$ 
where $I_{mech}=I_{flux}+\beta\circ \tau(p,\theta)$ with 
\be  \beta\circ \tau(p,\theta) = {\mathcal D}\mbox{-}\lim_{\ell\rightarrow 0} \ubeta \bar{\tau}(p,\theta).\lb{eq94} \ee
and where
\be  \beta\circ {\mathcal Q}_{diss} = {\mathcal D}\mbox{-}\lim_{\ell\rightarrow 0} \ubeta \bar{{\mathcal Q}}_{diss}.  \lb{eq95} \ee
The limit source need not vanish. Although entropy is conserved for smooth solutions of relativistic Euler equations,  
there may be anomalous entropy production} for weak solutions. Relativistic shock solutions with 
discontinuities in the fluid variables are, of course, a well-known example of such dissipative weak solutions
(Appendix \ref{shocks}).  We shall see below, however, that even continuous solutions may exhibit anomalous entropy production.  

Precisely the same balance equation can be obtained by taking the limit $\eta,\ \zeta,\ \sigma,\ \kappa\rightarrow 0$ limit 
of the fine-grained entropy balance (\ref{eq6}) for the dissipative fluid model. The limit of the dissipative entropy production 
is, of course, obtained directly from our fundamental hypothesis (\ref{eq22}). The fine-grained entropy current for the 
dissipative fluid model also converges to $s V^\mu$ in the limit $\eta,\ \zeta,\ \sigma,\ \kappa\rightarrow 0.$ This can be 
verified without great difficulty for models of the Israel-Stewart class. Recall that in such models the entropy current does not 
have the naive form (\ref{eq17}) which it assumes in the Eckart-Landau-Lifschitz models, but is instead modified 
as in (\ref{eq18}) by terms that are second-order in gradients. Taking the latter energy-frame expression as a 
concrete example, we factor out the dependence upon the transport coefficients $\eta,\ \zeta,\ \sigma$ and introduce the 
rescaled variables $\hat{\tau}^{\mu\nu},$ $\hat{\tau},$ $\hat{N}^\mu.$ This yields the representation   
\begin{eqnarray}
&& S^\mu = s V^\mu- \sigma \lambda \hat{N}^\mu -\frac{1}{2} (\zeta\beta_0\Sigma_\zeta+ \sigma\beta_1\Sigma_\sigma +2\eta\beta_2\Sigma_\eta) V^\mu\cr
&& \hspace{60pt} + \zeta\sigma\frac{\alpha_0}{T}\hat{\tau}\hat{N}^\mu + \eta\sigma\frac{\alpha_1}{T} \hat{\tau}^{\mu\nu}\hat{N}_\nu 
\lb{eq96} \end{eqnarray}
Here we have denoted as $\Sigma_\zeta$, $\Sigma_\sigma,$ $\Sigma_\eta$ the three terms in the fine-grained 
entropy production (\ref{eq6}) that are proportional to $\zeta,$ $\sigma,$ $\eta,$ respectively. According to our fundamental
hypothesis (\ref{eq22}), these converge to positive distributions $\Sigma_{bulk},$ $\Sigma_{cond},$ $\Sigma_{shear}$ in the 
limit $\eta,\ \zeta,\ \sigma \rightarrow 0.$ Because of the remaining factors of $\zeta,$ $\sigma,$ $\eta,$ appearing 
in (\ref{eq96}), however, one should expect that the $\beta$-terms will all vanish in that limit. Likewise, the $\alpha$-terms 
should vanish because they are quadratic in the transport coefficients $\zeta,$ $\sigma,$ $\eta.$ These arguments 
are not rigorous because the factors involving $\epsilon,$ $\rho,$ $V^\mu$ in those terms do not remain smooth in the limit. 
It is possible nevertheless to show by simple inequalities that these terms do vanish in the sense of distributions and, thus,
$\Dlim_{\eta,\zeta,\sigma,\kappa\rightarrow 0} S^\mu_{\eta,\zeta,\sigma,\kappa}= s V^\mu.$ For details, see Appendix 
\ref{fine-balance}. One thus obtains finally the entropy-balance for the limiting Euler solution 
\be  \partial_\mu(sV^\mu) = \Sigma_{diss},  \lb{eq97} \ee
with $\Sigma_{diss}>0$ given by the limit in (\ref{eq22}).  The equality 
\red{\be \Sigma_{inert}=\Sigma_{diss} \lb{eq98} \ee}
is demanded by consistency with the inertial-range limiting balance (\ref{eq93}), just as in the non-relativistic theory.

Given that anomalous entropy production is possible for weak solutions, what degree of singularity of the fluid variables is 
required for a non-vanishing anomaly? To answer this question, we can prove an Onsager-type singularity theorem which 
gives necessary conditions for an anomaly. The basic idea is the same as in the non-relativistic case 
\cite{drivas2017onsager} and is easy to explain. We first rewrite the resolved entropy balance (\ref{eq86}) as
\be
\partial_\mu \underline{S}^\mu 
= \ubeta \left(\overline{{\mathcal Q}}_{diss} - \bar{\tau}(p,\theta)\right)  -I^{flux}_\ell + \Sigma^{flux}_\ell 
\lb{eq99}\ee
The flux-terms $I^{flux}_\ell$ and $\Sigma^{flux}_\ell$ may be readily expressed in terms of space-time increments 
of the fluid variables, using the cumulant-expansion methods described in section \ref{energy}. The term which is
difficult to estimate directly is the one involving $\overline{{\mathcal Q}}_{diss} - \bar{\tau}(p,\theta)$. Note that 
these two quantities separately may be non-universal and may depend upon the particular sequence 
$\zeta_k,$ $\sigma_k,$ $\eta_k\rightarrow 0$ used to obtain the limiting Euler solution. Fortunately, exactly 
the same combination appears in the balance equation (\ref{eq79}) for the subscale internal-energy current 
$\utilde{K}^{\mu}.$ \ Thus, 

\vspace{-7pt}
\noindent one can define an {\it intrinsic resolved entropy current} in the Favre-averaging approach as
\begin{eqnarray}
&& \utilde{{\mathcal S}}^{*\mu} = \underline{{\mathcal S}}^{\mu} - \ubeta \utilde{K}^{\mu} \cr
&& \hspace{17pt} = \us\bar{V}^\mu +\ubeta\bar{\tau}(h,V^\mu) -\ulambda \bar{\tau}(n,V^\mu) \cr
&& + \ubeta  \overline{h}\Big(\frac{1}{2}\tilde{\tau}(V_\nu,V^\nu) \widetilde{V}^\mu +\tilde{\tau}(V_\nu,V^\mu) \widetilde{V}^\nu
+\tilde{\tau}(V_\nu,V^\nu,V^\mu)\Big), \cr 
&&
\lb{eq100} \end{eqnarray} 
where the second equality uses (\ref{eq80}). It follows from the two balance equations (\ref{eq79}) and (\ref{eq99}) that this 
intrinsic entropy current satisfies the following balance:

\vspace{-10pt}
\red{\begin{equation}
\partial_\mu \utilde{{\mathcal S}}^{*\mu} = \Sigma^{inert*}_\ell \lb{eq101}
\end{equation}}

\vspace{-10pt} 
\noindent \red{where net inertial-range entropy production is defined by 
\be \Sigma_\ell^{inert*} = -I^{flux}_\ell + \Sigma^{flux*}_\ell \lb{eq126} \ee} 

\vspace{-10pt}
\noindent with the intrinsic negentropy flux 
\begin{eqnarray}
&& \Sigma_\ell^{flux *} = \Sigma_\ell^{flux} -(\partial_\mu\ubeta) \utilde{K}^{\mu}  + \ubeta \calQ^{flux}_\ell \cr
%\vspace{5pt} 
&& \hspace{32pt} = (\partial_\mu \ubeta) \bar{\tau}(h,V^\mu)-(\partial_\mu\ulambda)\bar{\tau}(n,V^\mu) 
 + \ubeta \calQ^{flux}_\ell \cr
&& +\overline{h}(\partial_\mu \ubeta)  \Big(\frac{1}{2}\tilde{\tau}(V_\nu,V^\nu) \widetilde{V}^\mu +\tilde{\tau}(V_\nu,V^\mu) \widetilde{V}^\nu
+\tilde{\tau}(V_\nu,V^\nu,V^\mu)\Big). \cr
&&  \lb{eq102}  \end{eqnarray}   
Just as for the naive version of the resolved entropy current, \ $\Dlim_{\ell\rightarrow 0} \utilde{{\mathcal S}}^{*\mu} = s V^\mu,$ \
since all of the additional 

\vspace{-7pt}
\noindent cumulant terms vanish in the limit. Furthermore, and crucially, all source terms on the 
righthand side of (\ref{eq101}) are ``flux-like'' and are products of sub-scale cumulant terms and gradients of resolved 
fields, which allows us to express them in terms of space-time increments.  There is a rough analogy of our entropy 
current modification with the Israel-Stewart correction, in that our current modification is a higher-order moment 
of the coarse-graining average: whereas the naive entropy current in the first line of (\ref{eq100}) involves at most 2nd-order 
moments of $\epsilon,$ $n,$ and $V^\mu,$ the correction on the second line involves 3rd-order moments. 
Note, however, that our correction term does not have to be small relative to the naive term. 

A fundamental observation is that all individual terms in the intrinsic entropy balance (\ref{eq101}) depend only upon the 
limiting Euler solution and not on the sequence used to obtain it. In fact, the same equation can be 
obtained from the distributional Euler solution directly, without considering the underlying 
microscopic model (dissipative fluid dynamics, kinetic equation, quantum field-theory, etc.) 
To see this, one can use the homogeneous Gibbs relation $\us=\ubeta(\overline{\epsilon}+\underline{p})-\ulambda\overline{n}$
and the definition $\utilde{K}^{\mu}=\overline{{\mathcal E}}^\mu-\utilde{{\mathcal E}}^{\mu}$ to rewrite 
intrinsic entropy current as

\vspace{-8pt}
\be \utilde{{\mathcal S}}^{*\mu} = \ubeta\, \utilde{{\mathcal E}}^{\mu} +\ubeta\, \underline{p} \overline{V}^\mu -
\ulambda \overline{N}^\mu. \lb{eq103} \ee

\vspace{-10pt} 
\noindent One can then derive the intrinsic entropy balance (\ref{eq101}) directly from the inertial-range balance equation (\ref{eq59})
for $\utilde{{\mathcal E}}^{\mu}$, \ the particle conservation equation $\partial_\mu \overline{N}^\mu=0,$ \ and 

\vspace{-7pt} 
\noindent thermodynamic relation $\partial_\mu(\ubeta\, \underline{p})=\overline{n}(\partial_\mu\ulambda)
-\overline{\epsilon}(\partial_\mu\ubeta).$
This crucial observation implies that our results for anomalous entropy production are universal and apply 
to all distributional solutions of the relativistic Euler equations, not only those obtained as ideal limits 
of Israel-Stewart-type dissipative fluid models. 

The necessary conditions for anomalous entropy production follow directly from the 
intrinsic entropy balance (\ref{eq101}), exactly as for the non-relativistic case considered
in \cite{drivas2017onsager}. The conclusion is that the entropy anomaly can be non-zero
only if for every $q\geq 3$ at least one of the following three 
conditions is satisfied on the structure-function scaling exponents defined in (\ref{eq68}):
\be 2\min\{\zeta^\epsilon_q,\zeta_q^\rho\}+\zeta_q^v\leq q,  \lb{eq104} \ee 
\be \min\{\zeta^\epsilon_q,\zeta_q^\rho\}+2\zeta_q^v\leq q, \ \lb{eq105} \ee 
\be 3\zeta_q^v\leq q,  \lb{eq106} \ee
The first inequality (\ref{eq104}) is implied by (and thus replaces) the inequality (\ref{eq69}) shown 
earlier to be be necessary for non-vanishing of the baropycnal work as $\ell\rightarrow 0,$  
while the inequalities (\ref{eq105}),(\ref{eq106}) replace (\ref{eq70}),(\ref{eq71}) shown to be 
necessary for non-vanishing of the other two contributions to energy flux.  
The above inequalities would be equalities for a K41 dimensional 
scaling determined by mean energy flux, and the departure from the upper bound is a measure 
of the space-time intermittency of the solution fields \cite{Frisch95,eyink1995besov}. 
These upper bounds,
even if they hold as equalities, imply that $\epsilon,$ $\rho,$ $V^\mu$ must be non-smooth/singular
in spacetime for the ideal limit. %Indeed, if all the solution fields remained differentiable in the ideal limit, 
%then the exponents would have lower bounds $3q$ rather than upper bounds $q.$ 
Roughly speaking, limit solutions with anomalous entropy production can  have at most 
$1/3$ of a derivative in a space-time $L^q$-sense. 

For non-relativistic fluids, the conditions analogous to (\ref{eq104})-(\ref{eq106}) are known 
to be necessary also for an energy dissipation anomaly \cite{drivas2017onsager}. While 
${\mathcal Q}_{flux}=0$ if none of those conditions hold, it is in principle still possible that 
$\tau(p,\theta)={\mathcal Q}_{diss}>0.$ When the balance equation (\ref{eq59}) for resolved internal 
energy is rewritten as
\be
\partial_\mu \utilde{{\mathcal E}}^\mu =-\overline{p*\theta} + {\mathcal Q}^{inert}_\ell 
\lb{eq107} \ee

\vspace{-10pt} 
\noindent with 
\be {\mathcal Q}^{inert}_\ell = \bar{\tau}(p,\theta)+{\mathcal Q}^{flux}_\ell, \lb{eq108} \ee
then it differs strikingly 
from the balance equation (\ref{eq101}) for intrinsic resolved entropy, because the terms $\bar{\tau}(p,\theta)$ and 
${\mathcal Q}^{inert}_\ell$ are not determined uniquely as $\ell\rightarrow 0$ by the limiting weak Euler 
solution. Those terms in fact generally depend upon the the underlying dissipative fluid model sequence, 
as seen, for example, for the non-relativistic limit of shock solutions where a Prandtl-number dependence remains.
In \cite{drivas2017onsager}, vanishing energy dissipation anomaly is instead derived from the vanishing 
entropy production anomaly. \red{That proof carries over to relativistic fluids whenever the 
dissipative fluid model satisfies the bounds 
\be \Sigma_{diss}^{\zeta,\eta,\kappa,\sigma} \geq  {\mathcal  Q}_{diss}^{\zeta,\eta,\kappa,\sigma}/T \geq 0. \lb{eq109} \ee 
Amusingly, the only dissipative relativistic model in the class that we consider which guarantees (\ref{eq109})
is the classical energy-frame Landau-Lifschitz theory\footnote{Not even the classical particle-frame Eckart theory 
satisfies (\ref{eq109}), because of the presence of the sign-indefinite term $Q^\mu A_\mu$ in 
${\mathcal  Q}_{diss}^{\zeta,\eta,\kappa}$ for that model}, which is ill-posed and acausal! 
%Thus, there is no proof at this time for any viable relativistic fluid model that the conditions (\ref{eq104})-(\ref{eq106}) 
%are required for non-vanishing energy dissipation in the ideal limit. 
The result will be true,
nevertheless, if the viscous transport fields $\tau,$ $\tau^{\mu\nu}$ in the relativistic fluid model
are sufficiently well approximated by the constitutive relations of the Landau-Lifschitz 
theory.} Such results have been proved \cite{geroch1995relativistic,Lindblom96therelaxation}, but need to 
be extended to solutions with shocks or other milder turbulent singularities in order to show that 
conditions (\ref{eq104})-(\ref{eq106}) are necessary for anomalous energy dissipation. 

\section{Relations to Other Approaches}

We now briefly discuss the relation of our analysis with other approaches to relativistic
fluid turbulence that have been proposed in the literature.

\subsection{Barotropic fluid models}

In paper I we have criticized non-relativistic barotropic models as being physically inapplicable to 
fluid turbulence, since this is a strongly dissipative process. The same criticisms carry over to relativistic 
barotropic models, if those are defined as in \cite{gourgoulhon2006introduction,andersson2007relativistic}, 
for example. These authors take $\epsilon=\epsilon(\rho)$ as the condition for barotropicity, which implies
that $p=p(\epsilon,\rho)=p(\rho).$ As in the non-relativistic case, the internal energy per rest mass 
$e=u/\rho$ can be obtained from the integral 
\be e = \int \frac{p\, d\rho}{\rho^2} \lb{eq110} \ee
if and only if the fluid is isentropic with entropy per mass $s_m=s/\rho$ constant in space-time
(see \cite{rezzolla2013relativistic}, section 2.4.10). This is inconsistent with the irreversible 
production of entropy by turbulence. Furthermore, one obtains from (\ref{eq110}) and $\partial_\mu J^\mu=0$ that 
\be  \partial_\mu (u V^\mu) = - p\, \theta \lb{eq111} \ee 
which omits viscous heating. Barotropic equations of state together with formula (\ref{eq110}) for internal energy 
are thus physically inconsistent, as soon as one includes dissipative terms in $J^\mu$ and $T^{\mu\nu}$, 
and are unsuitable as fluid models of turbulence.  These remarks apply to the special case of polytropic equations of state 
with $p(\rho)=K \rho^\Gamma$ for exponent $\Gamma,$ whenever the internal energy density is determined from the relation 
$u=p/(\Gamma-1),$ as is very standard in numerical simulations with relativistic polytropic models. Such models 
cannot correctly represent the time-irreversible physics of relativistic fluid turbulence which is created by 
the spectrum of singularities that develop in the solutions. Note that barotropic fluid models in the sense of 
\cite{gourgoulhon2006introduction,andersson2007relativistic} are already known to be physically inadequate 
to describe the irreversible evolution of relativistic shocks (\cite{rezzolla2013relativistic}, section 2.4.10). 

These criticisms do not apply to relativistic barotropic equations of state if those are defined instead 
by the alternative condition $p=p(\epsilon),$ e.g. as in \cite{rezzolla2013relativistic}. 
Note that such a formulation of barotropicity is more general,
because it makes sense even when the constituent particles of the fluid have zero rest-mass 
and $\rho\equiv 0.$   There is no physical inconsistency of such an equation 
of state with irreversible entropy production by microscopic dissipation. For example, 
ultrarelativistic fluids with vanishingly small coldness $m c^2/k_BT\ll 1$ (\cite{rezzolla2013relativistic}, section 2.4.4) 
and models of hot, optically thick, radiation-pressure dominated plasmas  
(\cite{rezzolla2013relativistic}, section 2.4.8) both satisfy $p = \frac{1}{3}\epsilon$
and are thus barotropic in this second sense. Both of these models have a non-constant thermodynamic 
entropy, which can be made consistent with the second law of thermodynamics by addition of 
suitable dissipative terms to the ideal fluid equations. More generally, conformally invariant fluid 
models that describe low wavenumber dynamics of conformal quantum-field theories \cite{baier2008relativistic}
and non-conformal fluid models in the zero charge-density sector \cite{romatschke2010new} satisfy both 
$p=p(\epsilon)$ and dissipative second-order hydrodynamical equations similar to the Israel-Stewart
models consistent with the second law of thermodynamics. In fact, the exact shock solutions considered in 
Appendix \ref{shocks} are for conformal fluids \cite{liu2011shocks}. All of our conclusions apply 
to such models, with the simplification that hydrodynamics now reduces to the 
equation $\partial_\nu T^{\mu\nu}=0$ for the stress-energy tensor alone. 
 
\subsection{Point-Splitting and Statistical States}

In paper I we have argued that point-splitting regularizations are inadequate for non-relativistic compressible 
fluid turbulence and the same arguments hold for relativistic fluid turbulence. Previously, Fouxon \& Oz 
\cite{fouxon2010exact} have used a point-splitting technique in the setting of an externally forced relativistic fluid 
satisfying
\be \partial_\nu T^{\mu \nu} = F^\mu \lb{eq112} \ee 
for a Minkowski force $F^\mu.$ Assuming that a statistically homogeneous and stationary 
state exists, those authors derived an exact statistical relation 
\be \langle T_{0\mu}(\bzed,t) T_{i\mu}(\br,t) \rangle = \frac{1}{D} {\mathcal P}_\mu\, r_i  \quad \mbox{(no sum on $\mu$)}
\lb{eq113} \ee  
with $\langle\cdot\rangle$ denoting the ensemble-average and with ${\mathcal P}_\mu=\langle T_{0\mu}(\bzed,t) F_\mu(\bzed,t) \rangle$
a ``power input''. In the formal non-relativistic limit $c\to\infty$ , this relation reduces in conformal models with
sound speed $c_s=c/\sqrt{d}$ \cite{fouxon2008conformal} to the classical ``12/$d(d+2)$th-law''  
for $d$-dimensional incompressible fluid turbulence (e.g. \cite{gawedzki2002easy}), but for finite speeds 
of light the relation (\ref{eq113}) has nothing to do with energy of the fluid. As noted earlier, Fouxon \& Oz 
\cite{fouxon2010exact} concluded: ``Our analysis indicates 
that the interpretation of the Kolmogorov relation for the incompressible turbulence in terms of the energy cascade 
may be misleading.''

\textcolor{black}{
Needless to say, our analysis contradicts this conclusion. %There is no mathematical error 
%in the derivations of \cite{fouxon2010exact}, but their conclusions do not follow. 
We have already discussed the limitations of point-spitting regularizations in paper I and we shall not repeat that
discussion here. We only point out that the anomalies obtained by the point-splitting arguments of 
\cite{fouxon2010exact} are for quantities such as $T_{0\mu}^2(x),$ which are not conserved quantities 
even for smooth solutions of relativistic Euler equations and which have no obvious physical significance. 
The specific quantities are chosen in \cite{fouxon2010exact} simply so that a point-splitting 
regularization applies. One cannot conclude that energy cascade and energy-dissipation anomaly
must be absent in relativistic turbulence because a certain regularization method is insufficient to derive them.
The alternative coarse-graining regularization employed by us here shows that cascades and dissipative anomalies 
for both energy and entropy naturally arise in relativistic fluid turbulence. Furthermore, in conformal fluid models 
with $c_s=c/\sqrt{d},$ the relativistic energy flux $\mathcal{Q}_\ell^{flux}$ considered by us reduces in the 
non-relativistic limit to the standard kinetic energy flux for an incompressible fluid 
\be
\lim_{c\to \infty} c\mathcal{Q}_\ell^{flux} = -\rho_0\nabla\overline{\bv}: \tau(\bv,\bv)
\lb{incomp-flux} \ee
with constant mass density $\rho_0,$ \red{following the arguments in Appendix \ref{Nonrel-Lim}.
Aluie (private communication) has shown that the standard 4/5th-law of Kolmgorov, which is ordinarily 
derived by point-splitting, can also be obtained from (\ref{incomp-flux}) for incompressible Navier-Stokes
\cite{aluieunpub}}.  Thus, there is no unique way to extend the incompressible 4/5th law to relativistic turbulence, 
but our extension describes energy cascade in the relativistic regime.}

To underscore this point, we here briefly discuss the energy balance for forced statistical steady-states 
of relativistic fluid turbulence. This is a rather artificial setting quite distinct from most real-world relativistic 
turbulence, e.g. in astrophysics, in which there is no Minkowski force and no ensembles. We have therefore 
focussed in this paper on freely-evolving turbulence and individual flow realizations. However, our considerations
carry over directly to forced, steady-state ensembles. Note that the Minkowski force can quite generally be composed 
as 
\be F^\mu = \frac{1}{c^2} h A^\mu_{ext}-\frac{1}{c} Q_{cool} V^\mu, \lb{eq114} \ee   
with $V_\mu A^\mu_{ext}=0.$ Here $A^\mu_{ext}$ is an external acceleration field with units of 
(length)/(time)${\!\,}^2$ and $Q_{cool}$ is a cooling rate density with units of (energy)/(volume)(time). 
As usual, we include factors of $c$ to facilitate discussion of the non-relativistic limit.  The internal energy
balance in the presence of a Minkowksi force becomes 
\be \partial_\mu {\mathcal E}^\mu= -p\,\theta + \calQ_{diss} -\frac{1}{c}Q_{cool}. \lb{eq115} \ee
It follows that in a statistically homogeneous and stationary state, one has the fine-grained balance
\be \frac{1}{c}\langle Q_{cool}\rangle=\langle \calQ_{trans}\rangle + \langle \calQ_{diss}\rangle \lb{eq116} \ee
where $\calQ_{trans}=-p\theta$ is the mechanical production of internal energy by pressure-work. 
Our inertial-range internal-energy balance (\ref{eq59}) with the addition of the Minkowski force becomes
\be \partial_\mu \utilde{{\mathcal E}}^{\mu}=-\overline{p}_\ell\,\overline{\theta}_\ell + 
\calQ_\ell^{flux} - \tilde{V}_{\ell,\mu} \overline{F}^\mu_{ext,\ell}, \lb{eq117} \ee 

\vspace{-10pt}
\noindent including now the coarse-graining length-scale $\ell$ explicitly. One thus has 
\begin{eqnarray}
 \langle \tilde{V}_{\ell,\mu} \overline{F}^\mu_{ext,\ell}\rangle &=& -\langle \overline{p}_\ell\,\overline{\theta}_\ell\rangle
  + \langle \calQ_\ell^{flux}\rangle \cr
  &=& \langle \calQ_{trans} \rangle
  + \langle \calQ_\ell^{inert}\rangle,  
\lb{eq118} \end{eqnarray}   
where $\calQ^{inert}_\ell=\calQ^{flux}_\ell+\bar{\tau}_\ell(p,\theta)$ is the total inertial-range effective 
dissipation from both energy cascade and pressure-work defect and $Q_{trans}=-p*\theta.$
At length-scales scales much smaller
than the scale $L$ of the Minkowski force, $\langle \tilde{V}_{\ell,\mu} \overline{F}^\mu_{ext,\ell}\rangle \simeq 
(1/c)\langle Q_{cool}\rangle,$ and 
\be \langle \calQ_\ell^{inert}\rangle\simeq \frac{1}{c} \langle Q_{cool}\rangle - \langle \calQ_{trans}\rangle
=\langle \calQ_{diss}\rangle, 
\quad \ell\ll L.  \lb{eq119}\ee
We thus find that the ideal dissipation rate has constant ensemble-average for scales $\ell$ in the inertial-range, which 
equals the energy dissipation rate of the microscopic fluid model.   
This is formally identical to the statistical energy-balance relation that we obtained in the non-relativistic 
case, and reduces to it in the limit $c\to\infty.$

It is more traditional to expect that the effective energy dissipation rate at inertial-range 
lengths $\ell$ is set by the external input of kinetic energy by the large-scale forcing, but, of course, kinetic energy 
is not a natural relativistic quantity. Analogous constraints arise relativistically from the conditions
\be \langle F^\mu\rangle =0, \lb{eq120} \ee
which are necessary if a statistically homogeneous and stationary state is to exist for the forced 
fluid described by (\ref{eq112}). The $\mu=0$ condition gives that 
\be  \langle Q_{cool}\gamma \rangle = \frac{1}{c} \langle h A_{ext}^0\rangle. \lb{eq121} \ee
In the limit $c\to\infty$ this becomes 
\be  \langle Q_{cool} \rangle \simeq c \langle \rho A_{ext}^0\rangle = \langle \rho \bv\bdot {\bf A}_{ext}\rangle \lb{eq122} \ee 
Here we used orthogonality condition $A^0_{ext}=\bv\bdot {\bf A}_{ext}/c.$
Since the equation of motion projected orthogonal to $V^\mu$ takes the form ${\mathcal D}V^\mu= (1/c^2)A^\mu_{ext}+\cdots$ 
in the presence of a Minkowski force, the limit of the spatial components as $c\to\infty$ becomes $D\bv={\bf A}_{ext}+\cdots.$
Thus, (\ref{eq122}) is equivalent to the usual non-relativistic relation that $\langle Q_{cool}\rangle=
\langle Q_{in}\rangle,$ where $Q_{in}=\rho \bv\bdot {\bf A}_{ext}$ is the kinetic-energy injection rate per volume by the 
external forcing. We note in passing that the constraints $\langle F^i\rangle=0$ from the spatial components 
similarly reduce in the non-relativistic limit $c\to\infty$ to the condition $\langle\rho{\bf A}_{ext}\rangle=\bzed,$
or no net momentum injection by the external forcing. 

In addition to energy balance, there must also be an entropy balance for homogeneous and
stationary ensembles. In the presence of a Minkowski force, the fine-grained entropy balance 
(\ref{eq6}) is found using (\ref{eq115}) to be modified to 
\be \partial_\mu S^\mu = \Sigma_{diss}-\frac{1}{c}\beta Q_{cool}. \lb{eq123} \ee 
Thus, for a homogeneous and stationary ensemble
\be \langle  \Sigma_{diss}\rangle = \frac{1}{c}\langle \beta Q_{cool}\rangle \lb{eq124} \ee 
and microscopic entropy production is balanced by entropy removal by cooling. 
The inertial-range entropy balance (\ref{eq101}) is likewise modified by a Minkowski force,
with the divergence of (\ref{eq103}) using (\ref{eq117}) given by
\be \partial_\mu \utilde{S}^{*\mu} = \Sigma_\ell^{inert*} -\ubeta \tilde{V}_\mu \overline{F}^\mu. \lb{eq125} \ee 
When the Minkowski force is supported mainly at the large scale $L,$ one obtains the inertial-range mean balance 
\be \langle \Sigma_\ell^{inert*}\rangle = \langle \ubeta \tilde{V}_\mu \overline{F}^\mu\rangle\simeq 
 \frac{1}{c}\langle \beta Q_{cool}\rangle, \quad \ell\ll L.  \lb{eq127} \ee 
This mean entropy balance is formally the same as (I;104) for the non-relativistic case and reduces 
to it in the limit $c\to\infty.$ The physical picture is also the same as for non-relativistic compressible 
turbulence, with entropy produced at small scales inverse-cascading through the inertial range up to scales 
$\ell\simeq L$ where external cooling can remove the excess entropy. Equivalently (and perhaps more 
naturally), the negentropy injected by a large-scale cooling will forward cascade to small-scales where 
irreversible microscopic transport can destroy it. If one makes the distinction in (\ref{eq126}) between 
negentropy flux and anomalous negentropy input, then one can also write 
\be \langle \Sigma_\ell^{flux*}\rangle \simeq 
 \frac{1}{c}\langle \beta Q_{cool}\rangle + \langle I^{flux}_\ell\rangle, \quad \ell\ll L.  \lb{eq127} \ee 
where the negentropy flux proper is equal on average to the total negentropy input at large-scale,
both from external cooling and from anomalous negentropy input.  
 
\subsection{Linear Wave-Mode Decompositions} 

In paper I we have also called into question the validity of representing turbulent solutions by 
decompositions into linear wave modes. This is a very popular approach in non-relativistic 
plasma astrophysics and has recently been developed for Poynting-dominated relativistic MHD turbulence 
\cite{takamoto2016compressible}. We do not consider charged plasmas in the present paper but only 
fluids of electrically neutral particles, so that we shall just briefly discuss here the issues with 
decompositions into linear wave-modes. A basic problem is that thermodynamic relations such as 
$p=p(\epsilon,\rho)$ and $s=s(\epsilon,\rho)$ impose nonlinear constraints on solutions of the fluid 
equations, which thus live in nonlinear submanifolds  of function space.  Wave-modes $\epsilon',$ $\rho'$ 
obtained by linearization of the fluid equations around a uniform equilibrium background $\epsilon_0,$ $\rho_0$ 
only satisfy these thermodynamic constraints to linearized level. This may be an adequate representation 
when fluctuations are relatively small, satisfying $\epsilon'/\epsilon_0,$ $\rho'/\rho_0\ll 1.$ However turbulence 
generally produces fluctuations much larger than the means, where this linear approximation to the thermodynamic
relations is inadequate. Decomposition into linear wave-modes is thus clearly an approximation, with an unknown 
range of validity. We note that in conformal fluids with AdS gravity duals, the linear wave-mode decomposition
corresponds on the gravity side to the expansion in quasinormal modes about the uniform AdS black-hole.
Expansion in such quasinormal modes has recently been independently argued \cite{green2014holographic} 
to be inapplicable to the turbulent regime.

\section{Empirical Predictions and Evidence}

High-energy astrophysical plasma flows are probably the best candidates in Nature to exhibit 
relativistic fluid turbulence, but remote observations of such systems poorly constrain theory.
In order to confront theory with precise evidence, the only recourse at the moment is numerical 
simulations of turbulence for relativistic kinetic equations or dissipative fluid models. We shall
here briefly discuss the relations of our work to the existing body of numerical simulations.   
Confining attention to electrically neutral fluids, as considered in the present work, 
\red{the most relevant numerical studies have been motivated either by astrophysics 
\cite{zrake2013magnetic,radice2013universality} or by the fluid-gravity correspondence
\cite{carrasco2012turbulent,green2014holographic,westernacher2015scaling}.} Numerical 
codes exist for simulating the particle-frame Israel-Stewart model \cite{takamoto2011fast}, 
but we are aware of no turbulence simulations so far that exploit such codes. (The only exception is 
the study of \cite{green2014holographic} for a very similar second-order dissipative model of conformal 
fluids in $2+1$ space-time, discussed further below.) Instead, most 
studies have solved the relativistic Euler fluid equations using dissipative numerical schemes 
to remove the energy cascaded to small-scales rather than a physical viscosity.   

We first discuss the astrophysically motivated simulations in 3+1 space-times
with topology $T^3\times R$. Zrake \& MacFadyen \cite{zrake2013magnetic}
solved the stress-energy equation (\ref{eq112}) and the equation 
(\ref{eq2}) for conserved particle-number. They employed a relativistic ideal-gas equation 
of state $p=(\Gamma-1)u$ for $\Gamma=4/3,$ and adopted a Minkowski force 
\be  F^\mu = \rho A^\mu - \rho (u/u_0)^4 V^\mu, \lb{eq129} \ee 
with terms representing mechanical stirring and radiative cooling, respectively. The space-resolutions 
of their simulations were $256^3,$ $512^3,$ $1024^3,$ $2048^3$ and they had a mean relativistic Mach number 
of about $Ma=2.67$.  Radice \& Rezzolla \cite{radice2013universality} instead solved only the stress-energy 
equation (\ref{eq112}) for a radiation-pressure dominated fluid with  $p=(1/3)\epsilon$ and with a Minkowski force 
\be  F^\mu = F_0(t) (0,f^i) \lb{eq130} \ee 
for $f^i$ a zero space-average, solenoidal, random vector supported at low-wavenumbers. 
%Also, impose a minimum value $\epsilon_{\min},$ which corresponds to a heating of the fluid.  
They performed four runs with $F_0(t)=1,\ 2,\ 5,\ 10+(t/2),$ with space-resolutions 
$128^3,$ $256^3,$ $512^3,$ $1024^3,$ and with relativistic Mach numbers $Ma=0.362,$ 
$0.543$, $1.003$,  $1.759$.  The simulations of both groups are consistent with a forward energy 
cascade, although they had at their disposal no concrete formula such as our equation (\ref{eq60}) 
in order to make a precise measurement of relativistic energy flux. 

Both of these groups measured also the scaling exponents $\zeta_p^{\|v}$ of longitudinal 
velocity structure functions using the ESS procedure \cite{benzi1993extended}, and  
\cite{zrake2013magnetic} measured as well the exponents $\zeta_p^v$ for an 
absolute Minkowski-norm velocity structure-function. Both of these studies found  
$\zeta_p^{\|v}\leq p/3$ and $\zeta_p^v\leq p/3$ for $p\geq 3,$ consistent with our 
theoretical predictions. The phenomenological model of She-L\'ev\^eque \cite{she1994universal} was found to be a 
reasonable approximation to the ESS results for $\zeta_p^{\|v}$, but not for $\zeta_p^v$ 
in \cite{zrake2013magnetic}, which took on smaller values than $\zeta_p^{\|v}$ associated 
to greater space-time intermittency. When $p<3,$ our analysis makes no theoretical predictions 
for $\zeta_p^{\|v}$ or $\zeta_p^v,$ aside from the reasonable inference by concavity that $\zeta_p> p/3.$  
The direct (non-ESS) measurements of \cite{zrake2013magnetic} yielded 
$\zeta_2^{\|v}\doteq \zeta_2^v\doteq 1$ (Burgers-like), whereas \cite{radice2013universality} claimed 
consistency with $\zeta_2^{\|v}\doteq 2/3$ (K41). This discrepancy could be due to the larger Mach number
in the simulations of \cite{zrake2013magnetic} (see their Figure 1, which shows clear evidence 
of shocks). On the other hand, the spectra in Fig.2 of \cite{radice2013universality} at 
low wave-numbers are consistent with $\zeta_2^{\|v}>2/3$ and the higher wave-numbers are 
plausibly contaminated by bottleneck effects. In our opinion, neither of the simulations 
\cite{zrake2013magnetic,radice2013universality} achieved a long enough
inertial range to yield quantitatively reliable results for scaling exponents. 

Motivated by black-hole gravitational physics through the fluid-gravity correspondence 
\cite{baier2008relativistic,bhattacharyya2008conformal,bhattacharyya2008nonlinear}, 
there have also been simulations of relativistic fluid turbulence in $2+1$ space-time dimensions, 
both for free-decaying \cite{carrasco2012turbulent,green2014holographic} 
and externally-forced \cite{westernacher2015scaling} cases. Here, 
the evolution of low-wavenumber perturbations to black-holes in a $D+1$ dimensional, asymptotically
AdS space-time is expected to be equivalent to a relativistic hydrodynamics on the $D=d+1$ dimensional conformal
boundary of AdS space. Thus, 3+1 dimensional black-holes correspond to relativistic hydrodynamics 
in 2+1 dimensions. All of our considerations are independent of the space dimension $d$
and thus apply for $d=2,$ but this case is likely to be substantially more complex than $d>2.$
Even for incompressible fluid turbulence, $d=2$ is a much richer problem than 
$d>2.$ For example, freely-decaying and externally-forced incompressible turbulence appear 
substantially similar for $d>2,$ with both exhibiting an energy-dissipation anomaly.  
However, the enstrophy-dissipation anomaly predicted for $d=2$ incompressible turbulence 
\cite{kraichnan1967inertial,batchelor1969computation} 
appears only in forced turbulence, whereas there is no 
enstrophy-anomaly  for freely-decaying turbulence unless the initial data is very singular 
\cite{eyink2001dissipation,tran2006vanishing}. 
Viscous energy dissipation always tends to zero in $d=2$ incompressible turbulence, 
but the energy accumulates in large-scales by quite different mechanisms in the two cases: ``vortex merger'' 
\cite{Onsager49,mcwilliams_1984} for freely-decaying turbulence and ``inverse energy cascade'' 
\cite{kraichnan1967inertial} for forced turbulence. The previously-mentioned simulations of 2+1 relativistic 
turbulence also seem to indicate that there is no energy-dissipation anomaly there, and that vortex-merger 
and inverse-cascade processes occur. It should be kept in mind, however, that all of the discussed
simulations are at low relativistic Mach numbers. At higher Mach numbers, shocks 
will surely proliferate, leading to irreversible energy dissipation and entropy production.
Such behavior was observed in \cite{Kritsuk2017}  for simulations of $d=2$ non-relativistic 
compressible turbulence, motivated by large-scale dynamics of galactic 
accretion disks. We thus believe that the phenomenology of 2+1 relativistic turbulence will be 
quite non-universal, depending upon the relativistic Mach number, free-decay vs. forced, 
precise details of the initial-data, etc.   

The simulations cited above already largely support the present work, but our theory makes a rich array 
of further predictions for relativistic fluid turbulence that are easily subject to empirical test. Chief among 
these predictions are: (1) anomalous energy dissipation both by local energy cascade and by pressure-work defect; 
(2) anomalous input of negentropy into the inertial-range by pressure-work, in addition to any external input by 
large-scale cooling mechanisms; (3) negentropy cascade to small-scales through a flux of intrinsic inertial-range 
entropy; and (4) singularity or ``roughness'' of fluid fields to sustain cascades of energy and entropy, so that at least one of 
the exponent inequalities (\ref{eq104})-(\ref{eq106}) must hold. The explicit formulas (\ref{eq60}) for energy flux 
and (\ref{eq102}) for intrinsic entropy flux provide quantitative measures of cascades rates in relativistic 
turbulence. Furthermore, in order to provide mean fluxes of the predicted signs, the expressions (\ref{eq60}),(\ref{eq102}) 
require specific space-time correlations to develop, e.g. ``down-gradient turbulent transport'' with 
$\bar{\tau}(h,V^\mu),$ $\bar{\tau}(n,V^\mu)$ anti-correlated with the thermodynamic gradients 
$\partial_\mu T,$ $\partial_\mu \lambda,$ respectively. These many predictions provide an ample field
of study for future numerical investigation.  
    
\section{Summary and Future Directions}

The theory developed in this paper is based upon the hypothesis that relativistic 
fluid turbulence should exhibit dissipative anomalies of energy and entropy, similar to those 
observed for incompressible fluids. From this hypothesis alone, we have shown 
that the high Reynolds- and P\'eclet-number limit should be governed by distributional or ``coarse-grained''
solutions of the relativistic Euler equations. We have also demonstrated that precisely characterized 
singularities or ``roughness'' of the fluid fields is required to permit dissipative anomalies. 
The argument closely follows that of Onsager \cite{onsager1945distribution,Onsager49} for incompressible 
fluids, which we have explained as a non-perturbative application of the principle of renormalization-group invariance. 

One of the key open questions is certainly the extension of the present special-relativistic theory to 
general relativistic (GR) turbulence. There is reason to believe that much of the present theory will carry over 
straightforwardly to GR, since curved Lorentzian manifolds are locally diffeomorphic to Minkowski space.
However, new effects may arise if curvature scales become comparable to inertial-range 
turbulence scales. The main technical problem in extending our theory to GR is development  of 
suitable ``coarse-graining'' in curved space-times in order to regularize turbulent ultraviolet divergences.
Coarse-graining operations in GR have attracted recent interest also because of problems in cosmology 
and in the interpretation of cosmological observations, and much of this parallel work 
\cite{korzynski2010covariant,brannlund2010averaging} should carry over to general-relativistic turbulence.  
Here we may note that an Onsager singularity theorem has already been proved for incompressible fluid turbulence 
on general compact Riemannian manifolds, by exploiting a coarse-graining regularization defined with
a heat kernel smoothing \cite{isett2016heat}.  

Even in Minkowski space, there are important new directions of study opened by our work. Our quantitative 
formulas (\ref{eq60}) for energy flux and (\ref{eq102}) for entropy flux make possible an exploration 
of the physical mechanisms of relativistic turbulent cascades \cite{eyink2006multi,eyink2006turbulent}. 
The vortex-stretching mechanism 
of Taylor \cite{Taylor1937statistical} is widely believed to drive the $d=3$ incompressible energy cascade, 
but it is unclear whether such physics carries over to relativistic fluids. The equations 
of motion with the coarse-grained tensor (\ref{eq53}) derived in this paper also provide the mathematical foundations 
for Large-Eddy Simulation (LES) modeling of relativistic turbulence in Minkowski space
\cite{meneveau2000scale,schmidt2015large,Radice2017}. Such LES holds promise to be an important tool in numerical 
investigation of local turbulence in high-energy astrophysical events, such as gamma-ray bursts.  Finally, 
there are interesting implications of the present work for black-hole physics, because the fluid-gravity 
duality connects relativistic fluid-dynamics in $d+1$ Minkowski space-time to Einstein's equations in a 
Poincar\'e patch of a $(D)+1$ dimensional AdS black-hole solution. Thus, when high-Reynolds-number turbulence 
develops in a relativistic fluid in Minkowski space, our Onsager singularity theorem implies not only 
that the fluid fields must become ``rough'', but also that ``rough" metrics must develop in the turbulent 
solutions of the Einstein equations in the dual gravitational description.      

The ``roughness'' or H\"older-singularity of the turbulent velocity $V^\mu(x)$ in particular 
has profound implications for relativistic fluid turbulence. It was pointed out in a landmark work of Bernard 
et al. \cite{Bernardetal98} on non-relativistic incompressible turbulence that fluid velocities with H\"older exponent 
$h<1$ have non-unique Lagrangian particle trajectories. It was shown by those authors 
in a synthetic model of turbulence that the Lagrangian trajectories become ``spontaneously stochastic" 
in the high Reynolds-number limit, with randomness of trajectories persisting even when the initial particle 
location and the advecting velocity become deterministic and perfectly specified. It has subsequently been 
shown that such ``spontaneous stochasticity'' of Lagrangian particle trajectories holds at Burgers shocks 
\cite{EyinkDrivas14} and is necessary in incompressible Navier-Stokes turbulence for anomalous 
dissipation of passive scalars \cite{DrivasEyink2017ScalarA,DrivasEyink2017ScalarB}. These considerations 
carry over 
directly to relativistic fluid world-lines $X^\mu(X_0,\tau)$ defined by the equations 
\be dX^\mu/d\tau = V^\mu(X(\tau),\tau), \quad X^\mu(0)=X^\mu_0. \lb{eq131} \ee 
Because of the H\"older singularities of the turbulent velocity vector predicted by our analysis,
the fluid world-lines must become ``spontaneously stochastic'', with a random ensemble of world-lines passing 
through each fixed event $X_0$. This implies a turbulent breakdown of Lagrangian conservation laws 
that hold for smooth solutions of the relativistic Euler equations, such as the Kelvin Theorem 
\cite{greenberg1970general}; \cite{rezzolla2013relativistic}, section 3.7.5. Likewise, in relativistic 
astrophysical plasmas, the Alfv\'en Theorem on magnetic flux-conservation for ideal MHD solutions
\cite{lichnerowicz1967relativistic,bekenstein1978new}  must be fundamentally altered by spontaneous 
stochasticity effects. In non-relativistic theory, this fact leads to fast turbulent magnetic reconnection 
independent of collisional resistivity \cite{lazarian1999reconnection,eyink2006breakdown,eyink2013flux},
and our present work implies that the same turbulent mechanisms can act in relativistic 
magnetic reconnection.

\begin{acknowledgments}
We thank Hussein Aluie for sharing with us his unpublished work \cite{aluieunpub}. 
\end{acknowledgments}

\appendix

\section{Derivation of Coarse-Grained Relativistic Euler Equations}\label{app:bound} 

In this appendix we give key details of the proof of validity of the relativistic Euler equations 
in the coarse-grained  or ``weak'' sense for any ideal limits of thermodynamic fields 
$\epsilon,$ $\rho,$ $V^\mu$ as $Pe_\sigma,$ $Pe_\kappa,$ $Re_\eta,$ $Re_\zeta\rightarrow \infty.$

Most of the argument for the particle-conservation equation has been given 
in section \ref{coarse}.  One final estimate was left unproved, involving the Lorentz-invariant norm defined by 
$$ \Delta^{\mu\nu} A_\mu A_\nu $$
where $\Delta^{\mu\nu}=g^{\mu\nu}+V^\mu V^\nu$ projects perpendicular to the relativistic fluid velocity vector
$V^\mu$ with respect to the Minkowski pseudometric. Lorentz-transforming into the fluid rest frame $A_\mu\rightarrow A_\mu'$
$$ \Delta^{\mu\nu} A_\mu A_\nu = |{\bf A}'|^2, $$
coinciding with the standard Euclidean norm of the spatial part of the vector. The above norm is, in fact, only a semi-norm,
because $\Delta^{\mu\nu} V_\mu V_\nu=0.$ In deriving a bound on the dissipative terms in the coarse-grained 
conservation laws in section \ref{coarse}, we needed an estimate on this semi-norm from above in terms of the 
Euclidean norm. 

To obtain this, we note that 
$$ \Delta^{\mu\nu} A_\mu A_\nu = {\bf a}^\top \bDelta {\bf a}$$
where ${\bf a}$ is the $D$-dimensional vector with components $A_\mu$ of the covariant vector and $\bDelta$ is the 
$D \times D$-dimensional matrix with entries $\Delta^{\mu\nu}$ of the contravariant tensor. 
We then use the standard bound 
\be    | {\bf a}^\top \bDelta {\bf a}|  \leq \|\bDelta\|_2 \|{\bf a}\|_2^2 \lb{eqA1} \ee
where $\|\bDelta\|_2=\sqrt{\rho(\bDelta^\top\bDelta)}$ and $\rho({\bf M})$ is the spectral radius of the $D \times D$
dimensional matrix ${\bf M}$ (\cite{golub1996matrix}, section 2.3). Since $\bDelta$ is real, symmetric, one has 
furthermore $\|\bDelta\|_2=\rho(\bDelta).$ 
We thus must compute the eigenvalues of $\bDelta.$ This is simply done by an orthogonal transformation, which 
rotates the spatial part of the vector $V^\mu=\gamma(v)(1,\bv/c)$ into the 1-direction. Note that such a purely 
spatial rotation changes neither the Minkowski pseudonorm nor the Euclidean norm of $A_\mu.$ After this 
rotation, the matrix $\bDelta$ becomes block-diagonal, with a lower block which is the $(d-1)\times (d-1)$
identity matrix and an upper block which is the $2\times 2$ matrix 
$$ \bDelta_2 = \frac{1}{1-\beta_v^2}\left(\begin{array}{cc}
                                                             \beta_v^2 & \beta_v \cr
                                                             \beta_v & 1 
                                                             \end{array}\right), \qquad \beta_v=v/c. $$
The matrix $\bDelta_2$ has an eigenvalue $0$ with eigenvector $(-1,\beta_v)^\top$ (whose components are 
obviously those of the covariant vector $V_\mu=g_{\mu\nu}V^\nu$) and an eigenvalue 
$\frac{1+\beta_v^2}{1-\beta_v^2}$ greater than 1 with eigenvector $(\beta_v,1)^\top.$ It follows that 
$\rho(\bDelta)=\frac{1+\beta_v^2}{1-\beta_v^2}.$
Finally, noting that $\|{\bf a}\|_2^2=|A|_E^2,$ we obtain the bound 
$$ |\Delta^{\mu\nu} A_\mu A_\nu | \leq  \frac{1+\beta_v^2}{1-\beta_v^2} |A|_E^2. $$
This upper estimate is optimal, in that it can actually be achieved for a suitable space-like vector $A_\mu$
corresponding to the second eigenvector above. Since $1+\beta_v^2\leq 2,$ we obtain the bound 
stated in eq.(\ref{eq43}) in the main text. 

The dissipative terms in the coarse-grained energy-momentum conservation equation are estimated 
in a very similar fashion. Here we sketch briefly the bound for the shear-viscosity term, which can be  
written as 
\begin{eqnarray}
&& c_\mu \partial_\nu \Big(\overline{2\eta\hat{\tau}^{\mu\nu}}(x) \Big)\cr
&& \hspace{10pt} = - \frac{1}{\ell} \int d^{D}r\ 
c_\mu (\partial_\nu {\mathcal G})_\ell(r) \cdot 2\eta(x+r) \hat{\tau}^{\mu\nu}(x+r) \cr
&& \lb{eqA2} 
\end{eqnarray}  
and we have introduced a constant vector $c^\mu$ which can be set to 1 for any particular 
component of the equation and zero for the others, in order to select that component. 
Cauchy-Schwartz applied to this term gives
\begin{eqnarray}
&& \left|c_\mu \partial_\nu \Big(\overline{2\eta\hat{\tau}^{\mu\nu}}(x) \Big)\right|\leq \frac{1}{\ell}
\sqrt{\int_{{\rm supp}({\mathcal G}_\ell)} d^{D}r \ (2\eta T^2)(x+r)} \cr
&& \hspace{30pt} \times \sqrt{\int d^{D}r  \ \frac{2\eta}{T^2}(x+r) |c_\mu (\partial_\nu {\mathcal G})_\ell(r) 
\cdot \hat{\tau}^{\mu\nu}(x+r)|^2 }\cr
&&
\lb{eqA3} \end{eqnarray}
The first square-root factor goes to zero in the ideal limit under mild assumptions on $\eta$ and $T,$
as long as the second square-root factor remains bounded.  To estimate the second term we note 
that for any 2nd-rank covariant tensors $A_{\mu\nu},$ $B_{\mu\nu}$ the quantity 
$$ \Delta^{\mu\alpha}\Delta^{\nu\beta} A_{\mu\nu}B_{\alpha\beta}=\sum_{ij} A_{ij}' B_{ij}'$$
when the tensors are transformed to $A_{\mu\nu}',$ $B_{\mu\nu}'$ in the rest-frame of the fluid. 
The expression on the right is the standard Frobenius inner product of $d\times d$ matrices and 
thus the expression on the left is a degenerate inner product (vanishing whenever either tensor 
is a product of the form $V_\mu C_\nu$ or $C_\mu V_\nu$). Employing the Cauchy-Schwartz 
inequality for this degenerate inner-product gives
\begin{eqnarray}
 && |c_\mu(\partial_\nu {\mathcal G})_\ell(r) \cdot \hat{\tau}^{\mu\nu}(x+r)|^2 \cr
&& \leq  (c_\mu^\perp c_\perp^\mu) \cdot (\partial_\nu ^\perp G)_\ell(\partial^\nu_\perp G)_\ell(r) \cdot \hat{\tau}_{\mu\nu}\hat{\tau}^{\mu\nu}(x+r). 
\lb{eqA4} \end{eqnarray} 
The above inequality yields the following upper bound for the integral under the second square-root in (\ref{eqA3}): 
\be (c_\mu^\perp c_\perp^\mu) \int d^{D}r  \ (\partial_\nu ^\perp G)_\ell(r)(\partial^\nu_\perp G)_\ell(r) \cdot \frac{2\eta \hat{\tau}_{\mu\nu}\hat{\tau}^{\mu\nu}}{T^2}(x+r) \lb{eqA5} \ee
Now using equation (\ref{eq43}) in the main text and the similar inequality 
\be 0\leq (c_\mu^\perp c_\perp^\mu) \leq 2\gamma^2(v) |c|^2_E \lb{eqA6} \ee
we obtain our final estimate for the integral under the second square-root
\be 4|c|_E^2 \|\gamma(v)\|^4_\infty \int d^{D}r  \ |(\partial G)_\ell(r)|_E^2 \cdot \frac{2\eta \hat{\tau}_{\mu\nu}\hat{\tau}^{\mu\nu}}{T^2}(x+r) 
\lb{eqA7} \ee
This upper estimate converges in the ideal limit to 
\be 4|c|_E^2 \|\gamma(v)\|^4_\infty \int d^{D}r  \ |(\partial G)_\ell(r)|_E^2\ \Sigma_\eta(x+r) \lb{eqA8} \ee
and thus remains bounded. We conclude that the shear-viscosity term in the coarse-grained energy 
momentum equation vanishes in the ideal limit. 

Similar results are obtained for the bulk-viscosity term in the coarse-grained energy-momentum equation 
using the identity
\begin{eqnarray}
&& c_\mu \partial_\nu \Big(\overline{\zeta\hat{\tau}\Delta^{\mu\nu}}(x) \Big)\cr
&& \hspace{10pt} = - \frac{1}{\ell} \int d^{D}r\ 
c_\mu^\perp (\partial_\perp^\mu {\mathcal G})_\ell(r) \cdot \zeta(x+r) \hat{\tau}(x+r) \cr
&& \lb{eqA9} 
\end{eqnarray}  
and for the thermal-conductivity term using 
\begin{eqnarray}
&& c_\mu \partial_\nu \Big(\overline{\kappa\hat{Q}^\mu V^{\nu}+\kappa \hat{Q}^\nu V^\mu}(x) \Big)=\cr
&& %\hspace{10pt} 
- \frac{1}{\ell} \int d^{D}r\ \kappa(x+r) 
c_\mu^\perp\hat{Q}^\mu(x+r) \cdot  (\partial_\nu {\mathcal G})_\ell(r) V^\nu(x+r)  \cr
&& %\hspace{22pt}  
- \frac{1}{\ell} \int d^{D}r\ 
c_\mu V^\mu(x+r) \cdot \kappa(x+r) (\partial^\perp_\nu {\mathcal G})_\ell(r) \hat{Q}^\nu(x+r). \cr
&& \lb{eqA10} 
\end{eqnarray}  
The bulk-viscosity term is treated very similarly to the shear-viscosity term. For the thermal-conductivity term 
we need to use the standard Cauchy-Schwartz inequality $|c_\mu V^\mu|\leq |c|_E |V|_E$ and the following 
estimate for the Euclidean norm of the fluid velocity vector:
\be |V|_E^2=\gamma^2(v)(1+v^2/c^2) \leq 2 \gamma^2(v). \lb{eqA11} \ee 
The details are straightforward and left to the reader. 

\section{Non-Relativistic Limit}\lb{Nonrel-Lim}

Space-time coarse-graining kernels in relativistic theory $\calG(r)=\calG(r^0,\br)$ and in non-relativistic (Newtonian) 
theory $\calG_N(\br,\tau)$ are related by a simple change of dimensions through scaling with $c$: 
$$ \calG(r^0,\br) = (1/c)\calG_N(\br,r^0/c). $$
Thus,
\begin{eqnarray}
\overline{f}(x) &=& \int d^{D}r \ \calG_\ell(r) \ f(x+r)  \cr
                      &=& \int d^d\br \int d\tau \ \calG_{N,\ell}(\br,\tau) \ f(\bx+\br,t+\tau)\cr
                      &=& \overline{f}^N(\bx,t)   
\lb{eqB1} \end{eqnarray} 
and there is no need to distinguish between $\overline{f}$ and $\overline{f}^N$ as $c\rightarrow\infty.$ 
This is not true in general for more singular coarse-graining in space-time. Consider as an example 
the backward light-cone average with 
$$ \calG(r) = G(\br) \delta (r^0+|\br|). $$
In that case
$$ \overline{f}(x) = \int d^d\br\ G_\ell(\br) f(\bx+\br, t-|\br|/c) $$
Then in the limit $c\to \infty$ 
$$ \overline{f} = \overline{f}^N - \frac{1}{c} \int d^d\br\ G_\ell(\br) \ |\br| \dot{f}(\bx+\br,t)+ O\left(\frac{1}{c^2}\right)$$
where 
$$ \overline{f}^N(\bx,t) = \frac{1}{c} \int d^d\br\ G_\ell(\br)  f(\bx+\br,t) $$
is the non-relativistic instantaneous spatial coarse-graining. In this case, $\overline{f}$ and $\overline{f}^N$
are distinct. We shall assume hereafter a smooth space-time coarse-graining.  

Even with smooth space-time coarse-graining, the relativistic and non-relativistic Favre-averages
are distinct, because, respectively,  
$$ \tilde{f} := \overline{h f}/\overline{h}, \qquad \tilde{f}^N:=\overline{\rho f}/\overline{\rho} $$
where the first is weighted by $h=\rho c^2 + h_N,$ with $h_N=u+p$ the non-relativistic (Newtonian) 
enthalpy, and the second weighted by $\rho.$ Straightforward Taylor-expansion in $1/c^2$ gives 
$$ \tilde{f} = \tilde{f}^N  + \frac{1}{c^2\bar{\rho}^2}\left(\overline{f h_N}\ \overline{\rho^{\,\!}}-\overline{f\rho}\ \overline{h_N}\right) + O\left(\frac{1}{c^4}\right) $$
While the relativistic and non-relativistic Favre averages are distinct, they do agree to leading order in $1/c^2.$

With these preliminaries, we now consider the formal non-relativistic limit of $c\to\infty.$ We note the 
standard relations: 
\be \partial_\mu = (\frac{1}{c}\partial_t,\grad) \lb{eqB2} \ee
\be V^\mu = (1+\frac{1}{2c^2}|\bv|^2+ O(\frac{1}{c^4}), \frac{1}{c}\bv+ O(\frac{1}{c^3})\lb{eqB3} \ee 
\be {\mathcal D}=V^\mu\partial_\mu = \frac{1}{c} D+ O(\frac{1}{c^3}). \lb{eqB4}\ee 
with $D=\partial_t+\bv\bdot\grad,$ and 
\be \theta = \partial_\mu V^\mu =  \frac{1}{c} \Theta+ O(\frac{1}{c^3}), \lb{eqB5}\ee
with $\Theta=\grad\bdot\bv,$ or, more generally, 
\be \partial_\mu(f V^\mu) =  \frac{1}{c} [\partial_t f+ \grad\bdot(f\bv)]+ O(\frac{1}{c^3}). \lb{eqB6}\ee 
Furthermore, because cumulants of constants vanish, we have relations such as 
\be \bar{\tau}(V^0,f_2,f_3,...,f_n) = \frac{1}{2c^2}\bar{\tau}(|\bv|^2,f_2,f_3,...,f_n) \lb{eqB7}\ee 
\be \bar{\tau}(V^0,V^0,f_3,...,f_n) = \left(\frac{1}{2c^2}\right)^2\bar{\tau}(|\bv|^2,|\bv|^2,f_3,...,f_n) \lb{eqB8}\ee 
and so forth. The same relations hold also for Favre cumulants, just replacing $\overline{\tau}$ by $\widetilde{\tau}.$

\subsection{Inertial-Range Energy Balance} 

We consider first the energy balance (\ref{eq59}) or 
\be 
\partial_\mu \utilde{{\mathcal E}}^{\mu}=-\overline{p}\,\overline{\theta}+ 
\calQ^{flux}_\ell \lb{eqB9}\ee
Note that $\epsilon V^\mu = c^2 J^\mu + u V^\mu,$ so that 
\be \partial_\mu \overline{\epsilon V^\mu} = \partial_\mu \overline{u V^\mu} = 
\frac{1}{c}\left[\partial_t \overline{u} + \grad\bdot( \overline{u}\,\overline{\bv}+\overline{\tau}(u,\bv))\right] +  O(\frac{1}{c^3}). 
\lb{eqB10} \ee  
By the results (\ref{eqB3}),(\ref{eqB7}),(\ref{eqB8})
\be \overline{\tau}(p,V^\mu)=(O(\frac{1}{c^2}), \frac{1}{c}\overline{\tau}(p,\bv)+ O(\frac{1}{c^3})), 
\lb{eqB11}\ee
\begin{eqnarray}
&& \frac{1}{2}\tilde{\tau}(V_\nu,V^\nu)\tilde{V}^\mu= \cr 
&& \Big(\frac{1}{2c^2}\tilde{\tau}^N(v_i,v_i)+O(\frac{1}{c^4}), 
\frac{1}{2c^3}\tilde{\tau}^N(v_i,v_i)\bv+O(\frac{1}{c^5})\Big), \cr
&& \lb{eqB12}\end{eqnarray}  
\begin{eqnarray}
&& \tilde{\tau}(V_\nu,V^\nu,V^\mu)= 
\Big(O(\frac{1}{c^4}), \frac{1}{c^3}\tilde{\tau}^N(v_i,v_i,\bv)+O(\frac{1}{c^5})\Big), \cr
&& \lb{eqB13}\end{eqnarray}  
and 
\begin{eqnarray}
&& \tilde{\tau}(V_\nu,V^\mu)\tilde{V}^\nu= \cr 
&& \Big(O(\frac{1}{c^4}), 
-\frac{1}{2c^3}\tilde{\tau}^N(|\bv|^2,\bv)\cdot 1+\frac{1}{c^3}\tilde{\tau}^N(v_i,\bv)\tilde{v}^N_i+O(\frac{1}{c^5})\Big) \cr
&& \hspace{57pt} = \Big(O(\frac{1}{c^4}), 
-\frac{1}{2c^3}\tilde{\tau}^N(v_i,v_i,\bv) +O(\frac{1}{c^5})\Big). \cr
&& 
\lb{eqB14}\end{eqnarray}  
Putting all of these results together with the formula (\ref{eq57}) for $\utilde{{\mathcal E}}^{\mu}$ and 
$\overline{h}=\overline{\rho} c^2 + \overline{h}_N$ gives 

\vspace{-10pt} 
\begin{eqnarray}  
&& \partial_\mu \underline{{\mathcal E}}^{*\mu}\simeq \frac{1}{c}\partial_t \left(\overline{u}+\frac{1}{2}\overline{\rho}\tilde{\tau}^N(v_i,v_i)\right) \cr 
&& +\frac{1}{c}\grad\bdot\Big(\overline{u}\, \overline{\bv}+\overline{\tau}(h,\bv) + \frac{1}{2}\overline{\rho}\tilde{\tau}^N(v_i,v_i)\tilde{\bv}^N 
+\frac{1}{2}\overline{\rho}\tilde{\tau}^N(v_i,v_i,\bv)\Big). \cr
&& \lb{eqB15} 
\end{eqnarray}

Now consider relativistic energy flux given by 
\begin{eqnarray} 
&& {\mathcal Q}^{flux}_\ell=  \frac{1}{\bar{h}} (\partial_\nu\bar{p})\bar{\tau}(h,V^\nu) \cr 
&&  \vspace{30pt} -\bar{h} (\partial_\mu\widetilde{V}_\nu) \tilde{\tau}(V^\mu,V^\nu) 
-\frac{1}{2}\partial_\nu\overline{hV^\nu}\ \tilde{\tau}(V_\mu,V^\mu).  
\lb{eqB16} \end{eqnarray} 
Easily from previous estimates
\be \frac{1}{\bar{h}} (\partial_\nu\bar{p})\bar{\tau}(h,V^\nu)\simeq 
\frac{1}{c\bar{\rho}} \grad\bar{p}\bdot \bar{\tau}(\rho,\bv). \lb{eqB17} \ee
Next observe that 
\be 
\tilde{\tau}(V^\mu,V^\nu)=\left[
\begin{array}{c|c}
O(\frac{1}{c^4})  & O(\frac{1}{c^3}) \\ 
\hline
O(\frac{1}{c^3}) & \frac{1}{c^2}\tilde{\tau}^N(\bv,\bv)
\end{array}\right]
\lb{eqB18} \ee
and 
\be 
\partial_\mu\tilde{V}_\nu=\left[
\begin{array}{c|c}
O(\frac{1}{c^3})  & O(\frac{1}{c^3}) \\ 
\hline
O(\frac{1}{c^3}) & \frac{1}{c}\grad\tilde{\bv}^N
\end{array}\right], 
\lb{eqB19} \ee
so that 
\be \bar{h} (\partial_\mu\widetilde{V}_\nu) \tilde{\tau}(V^\mu,V^\nu) \simeq
\frac{1}{c}\,\bar{\rho} \,\grad\widetilde{\bv}^N\bdots \tilde{\tau}^N(\bv,\bv) \lb{eqB20} \ee 
For the last term, use $h V^\nu=c^2 J^\nu + h_N V^\nu$ to obtain
\be \partial_\nu\overline{hV^\nu}\simeq 
\frac{1}{c} [\partial_t h_N+ \grad\bdot(h_N \bv)]. \lb{eqB21} \ee 
Since also
\be \frac{1}{2}\tilde{\tau}(V_\nu,V^\nu)\simeq \frac{1}{2c^2}\tilde{\tau}^N(v_i,v_i), \lb{eqB22} \ee
we thus find 
\be  \frac{1}{2}\partial_\nu\overline{hV^\nu}\ \tilde{\tau}(V_\mu,V^\mu)=O\left(\frac{1}{c^3}\right). \lb{eqB23} \ee 
In conclusion, 
\begin{eqnarray}  
{\mathcal Q}^{flux}_\ell &\simeq& \frac{1}{c\bar{\rho}} \grad\bar{p}\bdot \bar{\tau}(\rho,\bv) - 
\frac{1}{c}\,\bar{\rho} \,\grad\widetilde{\bv}^N\bdots \tilde{\tau}^N(\bv,\bv) = \frac{1}{c} Q^{flux}_\ell\cr
&&  \lb{eqB24} \end{eqnarray}
where $Q^{flux}_\ell$ is the non-relativistic energy flux. 

From the results (\ref{eqB15}), (\ref{eqB24}), and $\overline{p}\,\overline{\theta}\simeq (1/c)\overline{p}\,\overline{\Theta},$ we 
thus obtain as the non-relativistic limit of the inertial-range  internal-energy balance for relativistic Euler that 
\begin{eqnarray}  
&& \partial_t \left(\overline{u}+\frac{1}{2}\overline{\rho}\tilde{\tau}^N(v_i,v_i)\right)
+\grad\bdot\Big(\overline{u}\, \overline{\bv}+\overline{\tau}(h,\bv) \cr
&& + \frac{1}{2}\overline{\rho}\tilde{\tau}^N(v_i,v_i)\tilde{\bv}^N  +\frac{1}{2}\overline{\rho}\tilde{\tau}^N(v_i,v_i,\bv)\Big)=
-\overline{p}\overline{\Theta}+Q^{flux}_\ell. \cr
&& \lb{eqB25}
\end{eqnarray}
This is nothing other than the non-relativistic balance equation for intrinsic large-scale internal energy,
obtained in equation (I;57) of paper I. 

There is no natural (covariant) relativistic analogue of the large-scale kinetic-energy balance 
(I;41) for non-relativistic  compressible turbulence. On the other hand, in any fixed inertial-frame, 
it is easy to see that the time-component of the relativistic Euler equation when coarse-grained 
\be \partial_\mu \overline{T}^{0\mu}=0, \lb{eqB26}\ee 
yields in the limit $c\to\infty$  the coarse-grained conservation of mass
\be  \partial\overline{\rho}+\grad\bdot \overline{\rho\,\bv}=0 \lb{eqB27}\ee 
to order $O(c)$ and the coarse-grained balance of total non-relativistic energy 
\be \partial_t\overline{\frac{1}{2}\rho|\bv|^2+u}+\grad\bdot \overline{\left(\frac{1}{2}\rho|\bv|^2+u+p\right)\bv}=0 \lb{eqB28} \ee 
to order $O(1/c).$ If one subtracts (\ref{eqB25}) from the latter equation (\ref{eqB28}), then one obtains 
\begin{eqnarray}
&& \partial_t(\frac{1}{2}\bar{\rho} |\tilde{\bv}^N|^2)+\grad\bdot\Big[(\bar{p}+\frac{1}{2}\bar{\rho} |\tilde{\bv}^N|^2)\tilde{\bv}^N
+\bar{\rho}\tilde{\tau}^N(\bv,\bv)\cdot\tilde{\bv}^N\cr
&& \hspace{80pt} - \frac{\bar{p}}{\bar{\rho}}\bar{\tau}(\rho,\bv) \Big] = \bar{p}\,\bar{\Theta}-Q_\ell^{flux}.
\lb{eqB29} \end{eqnarray} 
In this manner, the inertial-range kinetic energy balance equation (I; 41) of non-relativistic turbulence can be recovered
as $c\to\infty$ from the relativistic theory.

\subsection{Inertial-Range Entropy Balance}

We now consider the intrinsic inertial-range entropy current in the Favre formulation,
$\utilde{{\mathcal S}}^{*\mu} = \underline{{\mathcal S}}^{\mu} - \ubeta \utilde{K}^{\mu},$ and 

\vspace{-7pt}
\noindent its balance equation 
\begin{equation}
\partial_\mu \utilde{{\mathcal S}}^{*\mu} =  -I^{flux}_\ell + \Sigma^{flux*}_\ell
\lb{eqB30} \end{equation} 

\vspace{-10pt} 
\noindent with the intrinsic negentropy flux 
\be  \Sigma_\ell^{flux *} = \Sigma_\ell^{flux} -(\partial_\mu\ubeta) \utilde{K}^{\mu}  + \ubeta \calQ^{flux}_\ell.  
\lb{eqB31}\ee

\vspace{-7pt}
First we note a standard difference between relativistic and Newtonian thermodynamics, due to the distinction 
between rest-mass and energy in the latter:
\be \epsilon = \rho c^2 +u, \quad \lambda=\beta mc^2 +\lambda_N. \lb{eqB32}  \ee
See \cite{israel1987covariant} or \cite{rezzolla2013relativistic}, \S 2.3.6.  Using these relations, one easily finds that
\begin{eqnarray}
&& \underline{{\mathcal S}}^{\mu} = \us \overline{V}^\mu + \ubeta\bar{\tau}(u,V^\mu) -\ulambda_N\bar{\tau} (n,V^\mu) \cr 
&& = \left(\us+O(\frac{1}{c^2}), \ \frac{1}{c}[\us\bar{\bv}+\ubeta\bar{\tau}(u,\bv)-\ulambda_N\bar{\tau}(n,\bv)] +O(\frac{1}{c^3})\right) \cr
&& 
\lb{eqB33}\end{eqnarray} 
On the other hand, it follows directly from the formula (\ref{eq80}) for $\utilde{K}^{\mu}$ and the estimates in the previous 
subsection 

\vspace{-7pt} 
\noindent that 
\begin{eqnarray}
&& -\utilde{K}^{\mu} = \Big(\frac{1}{2}\bar{\rho}\tilde{\tau}^N(v_i,v_i) + O(\frac{1}{c^2}), \cr
&& \frac{1}{2c}\bar{\rho}\tilde{\tau}^N(v_i,v_i)\tilde{\bv}^N + \frac{1}{2c}\bar{\rho}\tilde{\tau}^N(v_i,v_i,\bv) 
+ \frac{1}{c} \bar{\tau}(p,\bv) + O(\frac{1}{c^3})\Big). \cr
&&
\lb{eqB34}\end{eqnarray} 
As an aside, we note that this last result implies that the balance equation (\ref{eq79}) for $\utilde{K}^{\mu}$
reduces in the limit $c\to\infty$

\vspace{-7pt} 
\noindent  to the non-relativistic balance equation (I;64) for the subscale 
kinetic-energy. We finally obtain that 
\be \partial_\mu \utilde{{\mathcal S}}^{*\mu} \simeq \frac{1}{c}[\partial_t\us^* + \grad\bdot {\underline{{\bf s}}}^*] 
\lb{eqB35} \ee 

\vspace{-10pt} 
\noindent where 
\be \us^*=\us+(1/2){{\underline{\beta}}} \overline{\rho}\ \tilde{\tau}^N(v_i,v_i)  \lb{eqB36}\ee
is the non-relativistic intrinsic inertial-range entropy and 
\begin{eqnarray} 
&& {\underline{{\bf s}}}^*=\us\overline{\bv}+{\underline{\beta}}{\overline{\tau}}(h_N,\bv)-{\underline{\lambda}}_N
{\overline{\tau}}(n,\bv)\cr
&& \hspace{20pt} +{\underline{\beta}}\Big[\frac{1}{2}\overline{\rho}\ \tilde{\tau}^N(v_i,v_i)\tilde{\bv}^N
+\frac{1}{2}\overline{\rho}\ \tilde{\tau}^N(v_i,v_i,\bv)\Big], \cr
&&
\lb{eqB37}\end{eqnarray} 
is the associated spatial-current. See (I;94),(I;96). 

On the other hand, using again the relation (\ref{eqB32}) between relativistic and Newtonian 
thermodynamic quantities, one finds that
\begin{eqnarray}
\Sigma_\ell^{flux}&=& (\partial_\mu\ubeta)\bar{\tau}(u,V^\mu) -(\partial_\mu \ulambda_N)\bar{\tau} (n,V^\mu) \cr
&\simeq & \frac{1}{c}\left[\grad\ubeta\bdot\bar{\tau}(u,\bv) - \grad\ulambda_N\bdot\bar{\tau} (n,\bv)\right] \cr
&=& \frac{1}{c} \Sigma_\ell^{flux,N}
\lb{eqB38}\end{eqnarray} 
where $\Sigma_\ell^{flux,N}$ is the (naive) entropy flux in non-relativistic compressible turbulence. 
Directly from (\ref{eqB34}) and the asymptotics for $\calQ_\ell^{flux}$ in the previous subsection, one finds that 
\begin{eqnarray} 
&& \ubeta \calQ_\ell^{flux}-(\partial_\mu \ubeta) \utilde{K}^{\mu} \simeq \frac{1}{c}\ubeta Q_\ell^{flux}+\frac{1}{2c}(\partial_t \ubeta) \bar{\rho}\tilde{\tau}^N(v_i,v_i) \cr
&& + \frac{1}{c}\grad\ubeta\bdot\left[\frac{1}{2}\bar{\rho}\tilde{\tau}^N(v_i,v_i)\tilde{\bv}^N + \frac{1}{2}\bar{\rho}\tilde{\tau}^N(v_i,v_i,\bv)+\bar{\tau}(p,\bv) \right]  \cr
&& \lb{eqB39} \end{eqnarray}
This corresponds exactly to eq.(I;90) in the non-relativistic theory. Finally,  the very simple equality
\be I_\ell^{flux} = \ubeta(\overline{p}-\underline{p})\overline{\theta}
\simeq \frac{1}{c} \ubeta(\overline{p}-\underline{p})\overline{\Theta}=\frac{1}{c} I_\ell^{flux,N} \lb{eqB40} \ee 
shows that the relativistic inertial-range entropy balance (\ref{eq101}) reduces in the limit $c\to\infty$ to the 
balance (I;95) of non-relativistic intrinsic inertial-range entropy. 

\section{Fine-Grained Balances of Internal Energy and Entropy in the Ideal Limit}\lb{fine-balance} 

In this appendix we derive the balances of internal energy and entropy for the relativistic Euler 
solutions by considering directly the ideal limit of the fine-grained balances from the dissipative fluid model. 

We begin by considering $\partial_\mu {\mathcal E}^\mu$ with the particle-frame energy current in Eq.(\ref{eq7}), 
or ${\mathcal E}^\mu=\epsilon V^\mu +\kappa \hat{Q}^\mu.$ We must show that the contribution
of the second term vanishes distributionally in the limit $\kappa,$ $\eta,$ $\zeta\to 0.$
After smearing with a general test function $\varphi,$   a straightforward estimate by a Cauchy-Schwartz 
inequality gives 
\begin{eqnarray}
&&  \left|\int d^{D}x\ (\partial_\mu \varphi) \kappa \hat{Q}^\mu\right| \cr
&& \hspace{10pt} \leq \sqrt{\int_{{\rm supp}(\varphi)} d^{D}x\ \kappa T^2 
\int d^{D}x\ (\partial_\mu^\perp \varphi \partial^\mu_\perp\varphi)  \frac{\kappa \hat{Q}_\mu \hat{Q}^\mu}{T^2}}, \cr
&& \hspace{10pt} \leq \sqrt{2\|\gamma(v)\|_\infty^2\int_{{\rm supp}(\varphi)} d^{D}x\ \kappa T^2 
\int d^{D}x\ |\partial\varphi|_E^2  \Sigma_\kappa}, \cr
&& \lb{eqC1} \end{eqnarray} 
using (\ref{eqA6}) and the definition $\Sigma_\kappa =\kappa \hat{Q}_\mu \hat{Q}^\mu/T^2$ to obtain
the last estimate.  The second integral inside the square root is bounded when 
$\Sigma_{therm}=\Dlim_{\eta,\zeta,\kappa\rightarrow 0} \Sigma_\kappa$ 
exists, while the first integral vanishes in the limit. We conclude that $\partial_\mu(\kappa \hat{Q}^\mu)\Dto 0$ 
as $\kappa,$ $\eta,$ $\zeta\to 0.$

We next consider $\partial_\mu S^\mu$ with the entropy current given by the energy-frame Israel-Stewart formula 
Eq.(\ref{eq96}).  We must show that only the term $\partial_\mu(sV^\mu)$ survives in the ideal limit and that all of the direct dissipative 
contributions vanish distributionally. The easiest to treat is the $\lambda\sigma \hat{N}^\mu$ term in $S^\mu,$
which gives a vanishing contribution by the same argument used above for $\kappa \hat{Q}^\mu.$

The terms $\eta\beta_2\Sigma_\eta V^\mu,$ $(1/2)\zeta\beta_0\Sigma_\zeta V^\mu$ $(1/2)\sigma\beta_1\Sigma_\sigma V^\mu$ 
all give contributions to $\partial_\mu S^\mu$ that are bounded in the same manner. We thus consider only the first.   
After smearing by a test function $\varphi,$ its contribution is bounded by 
\begin{eqnarray}
&&  \left|\int d^{D}x\ (\partial_\mu \varphi) V^\mu \eta\beta_2 \Sigma_\eta\right| \cr
&& \hspace{10pt} \leq \sqrt{2}\|\gamma(v)\|_\infty \int d^{D}x\ |\partial \varphi|_E \eta\beta_2 \Sigma_\eta \cr 
&& \hspace{10pt} \leq \sqrt{2}\|\gamma(v)\|_\infty  \sqrt{\int_{{\rm supp}(\varphi)} d^{D}x\ \eta^2\beta_2^2\Sigma_\eta} \cr 
&& \hspace{80pt} \times \sqrt{\int d^{D}x\ |\partial \varphi|^2_E \Sigma_\eta}, \cr
&& \lb{eqC2} \end{eqnarray} 
using $ |\partial_\mu\varphi V^\mu|\leq  |\partial\varphi|_E  |V|_E$ and (\ref{eqA11}) 
for the first inequality, and Cauchy-Schwartz for the second. Since $\Sigma_{shear}=
\Dlim_{\eta,\zeta,\sigma\rightarrow 0} \Sigma_\eta,$ the second square-root factor 
is bounded. For the first square-root factor note that 
\be \int_{{\rm supp}(\varphi)} d^{D}x\ \eta^2\beta_2^2\Sigma_\eta \leq \|\eta\beta_2\|_{L^\infty({\rm supp}(\varphi))}^2
\Sigma_\eta({\rm supp}(\varphi)), \lb{eqC3} \ee
with $\Sigma_\eta(K)=\int_{K} d^{D}x\ \Sigma_\eta$. 
For the compact set ${\rm supp}(\varphi),$ take $\psi\in C^\infty_c$ with $\psi\geq 0$ and $\psi\big|_{{\rm supp}(\varphi)}=1.$
Then $\Sigma_\eta({\rm supp}(\varphi))\leq \int d^{D}x\ \psi\ \Sigma_\eta$ so that
\be \limsup_{\sigma,\eta,\zeta\to 0} \Sigma_\eta({\rm supp}(\varphi))\leq  \int d^{D}x\ \psi\ \Sigma_{shear} \lb{eqC4} \ee 
follows also from $\Sigma_{shear}=\Dlim_{\eta,\zeta,\sigma\rightarrow 0} \Sigma_\eta.$ Finally, 
since $\|\eta\beta_2\|_{L^\infty({\rm supp}(\varphi))}\to 0$ as $\sigma,$ $\eta,$ $\zeta\to 0,$ the upper 
bounds (\ref{eqC2})-(\ref{eqC4}) show that the entire contribution vanishes in the ideal limit.

The terms 
$\eta\sigma \frac{\alpha_1}{T}\hat{\tau}^{\mu\nu}\hat{N}_\nu,$ 
$\zeta\sigma \frac{\alpha_0}{T}\hat{\tau} \hat{N}^\mu$ also give contributions 
that are both bounded similarly and we consider only the first. After smearing with a test function, 
\begin{eqnarray}
&&  \left|\int d^{D}x\ (\partial_\mu \varphi)  \eta\sigma \frac{\alpha_1}{T}\hat{\tau}^{\mu\nu}\hat{N}_\nu\right| \cr
&& \hspace{10pt} \leq \sqrt{\int_{{\rm supp}(\varphi)} d^{D}x\ \frac{1}{2}\sigma \eta \alpha_1^2 T \Sigma_\eta 
\int d^{D}x\ (\partial_\mu^\perp \varphi\partial^\mu_\perp\varphi) \Sigma_\sigma}, \cr
&& \lb{eqC5} \end{eqnarray} 
by Cauchy-Schwartz and the definitions of $\Sigma_\eta,$ $\Sigma_\sigma.$ Then
\be \int d^{D}x\ (\partial_\mu^\perp \varphi\partial^\mu_\perp\varphi) \Sigma_\sigma
\leq 2\|\gamma(v)\|_\infty^2 \int d^{D}x\ |\partial\varphi_E|^2 \Sigma_\sigma \lb{eqC6} \ee
using (A6) and 
\be  \int_{{\rm supp}(\varphi)} d^{D}x\ \sigma \eta \alpha_1^2 T \Sigma_\eta 
\leq \|\sigma \eta \alpha_1^2 T\|_{L^\infty({\rm supp}(\varphi))} \Sigma_\eta({\rm supp}(\varphi)). \lb{eqC7} \ee
The term $\Sigma_\eta({\rm supp}(\varphi))$ is bounded as in (\ref{eqC4}). In the ideal limit 
$\|\sigma \eta \alpha_1^2 T\|_{L^\infty({\rm supp}(\varphi))} \to 0$ and thus the bounds (\ref{eqC5})-(\ref{eqC7}) 
imply that the contribution to $\partial_\mu S^\mu$ vanishes distributionally as $\sigma,$ $\eta,$ $\zeta\to 0.$

We conclude that $\partial_\mu S^\mu\Dto \partial_\mu(s V^\mu)$ as $\sigma,$ $\eta,$ $\zeta\to 0$
when $S^\mu$ is given by the energy-frame formula Eq.(\ref{eq96}). The argument for the entropy 
current of the particle-frame Israel-Stewart theory is identical, with the replacements $\sigma\to\kappa,$
$\hat{N}^\mu\to \hat{Q}^\mu.$

\section{Relativistic Shock Solutions}\lb{shocks}  

\subsection{Reduced Conformal Model and Shock Solution}

We consider here an exact family of shock solutions for dissipative relativistic fluid models in 1+1 
space-time dimensions, which were obtained in the previous work of Liu \& Oz \cite{liu2011shocks}.
The 1+1 fluid models considered by those authors are reduced conformal fluids (RCF's) obtained 
from a $D=(d+1)$--dimensional conformal fluid (note that our $D$ is instead denoted  $2\sigma$ in \cite{liu2011shocks}) 
and have corresponding dimensionally-reduced gravity duals \cite{kanitscheider2009universal}. 
We recall that the equation of state for the pressure in $D$-dimensional conformal fluids is given by a 
power of the temperature
\begin{equation}\label{pressEqn}
p =\alpha T^D, 
\end{equation}
with a dimensionless constant $\alpha$. The tracelessness of the stress-energy tensor requires an energy density 
\be\label{energyForm}
\epsilon = (D-1) p = \alpha (D-1)T^D. 
\ee
There is no additional conserved current $J^\mu$ in the RCF's considered by \cite{liu2011shocks} and consequently 
$\lambda=0$. The resulting  first law of thermodynamics $d\epsilon = T ds$ as well as the homogenous Gibbs 
relation $h=\epsilon+p = sT$  imply that the entropy density is:
  \be\label{entrEqn}
  s = \alpha DT^{D-1}= D \alpha^{1/D}p^{(D-1)/D}.
  \ee
In the energy frame description, the non-ideal part of the stress-tensor (\ref{eq4}) is 
transverse to the velocity, $V_\mu \Pi^{\mu\nu}=0.$ As in \cite{liu2011shocks}, we consider 
only first-order terms in the gradient-expansion. Since bulk viscosity $\zeta=0$ for conformal fluids, 
the only transport coefficient at this order is shear viscosity $\eta$ with $\Pi^{\mu\nu}=-2\eta\sigma^{\mu\nu}.$
Upon reduction to $1+1$ dimensions, this appears as an effective bulk viscosity, so that 
\be \Pi^{\mu\nu}= -\zeta \theta \Delta^{\mu\nu}\ee
with 
$\zeta= \frac{1}{2\pi}\frac{D-2}{D-1}s.$ However, just as in \cite{liu2011shocks}, 
we take $\zeta:=\zeta(T)$ to be an arbitrary function, since none of our results depend upon any 
particular choice. 

Representing the two-velocity as $V^\mu= \gamma_v(1, \beta_v)$, any stationary solution 
of the 1+1 viscous model satisfies:
\begin{eqnarray}\label{cons1}
 \frac{d}{dx}\left[ \left( D p - \zeta \theta  \right)\gamma_v^2\beta_v \right]&=&0,\\ \label{cons2}
  \frac{d}{dx} \left[p(1+ D\gamma_v^2\beta_v^2) - \zeta \theta \gamma_v^2   \right]&=&0.
\end{eqnarray}
Equations (\ref{cons1}) and (\ref{cons2}) follow from 
$\nabla_\mu T^{\mu\nu}=0$ setting $\nu=0,1$ and they imply:
\begin{eqnarray}\label{feEqn}
 f_e &=& \left(D p- \zeta \theta\right)\gamma_v^2 \beta_v  \\
 f_p %= \frac{\epsilon}{D-1} \left(1+d\gamma_v^2\beta^2\right) -  (1+ \gamma_v^2 \beta^2) \zeta \theta 
& =& p \left(1+ D \gamma_v^2 \beta_v^2\right) -  \zeta \theta  \gamma_v^2 \label{fpEqn}
\end{eqnarray}
where $f_e\equiv T^{01}$ and $f_p \equiv  T^{11}$ are constant energy and momentum fluxes.  
Using (\ref{feEqn}) and (\ref{fpEqn}), \cite{liu2011shocks} obtained smooth 
viscous shock solutions by quadrature. We shall not employ these integral expressions, but 
only use the following important consequences of (\ref{feEqn}),  (\ref{fpEqn}): 
\be \label{PEqn}
\epsilon=(D-1)p = {f_e}/{\beta_v} - f_p.
\ee
\be \label{PEqn2}
p-\zeta\theta  =  f_p - {f_e}{\beta_v}. 
\ee
In particular,  the representation (\ref{PEqn}) of the pressure in terms of the velocity is analogous to the 
Bernoulli-type relation exploited by Becker to study shock solutions of the non-relativistic compressible 
Navier-Stokes equations for $Pr=3/4$ \cite{becker1922stosswelle}. Together, (\ref{PEqn}) and (\ref{PEqn2}) completely 
determine $\zeta \theta$ in terms of the velocity,  yielding identical results for any choice of viscosity $\zeta(T)$.   

The viscous model solutions of interest converge in the infinite Reynolds-number limit to stationary shock 
solutions of the relativistic Euler equations. These are piecewise constant, with a pre-shock velocity $\beta_0$
to the left, and post-shock value $\beta_1$ to the right. The possible values are obtained by equating the two 
expressions for the pressure from (\ref{PEqn}) and (\ref{PEqn2}) with $\zeta=0$:
\be {f_e}/{\beta_v} - f_p = (D-1)p = (D-1)[f_p - {f_e}{\beta_v}]. \lb{quadeq} \ee
%The values of the fields 
%on the two sides of the shock are related by the Rankine-Hugoniot conditions:
%\be\label{Rankine-Hugoniot}
%\Delta \left[ p \gamma_v^2\beta_v\right] = 0 , \ \  \Delta \left[p(1+ D\gamma_v^2\beta_v^2)\right] = 0 
%\ee
%which follow from (\ref{feEqn}) and (\ref{fpEqn}) 
This yields a quadratic polynomial in $\beta_v$ with coefficients depending upon $D$ and $R:=f_p/f_e.$
The condition for two distinct real roots is $|R|>2(D-1)^{1/2}/D.$ The product of the roots is given by 
\begin{eqnarray}\label{velocityJump}
\beta_0\beta_1&=& 1/(D-1):= \beta_s^2,
%,
\end{eqnarray}
where $\beta_s=c_s/c$ and $c_s=c/\sqrt{D-1}$ is the sound speed.
% and the sum is given by 
%\begin{eqnarray}\label{velocitymean}
%\beta_0+\beta_1 &=& \frac{DR}{D-1}.  
%T_0/T_1&=& (\beta_1\gamma_1^2/\beta_0\gamma_0^2)^{1/d},
%\end{eqnarray}
The condition $h=\epsilon+p=D\cdot p>0$ requires that both sides of (\ref{quadeq}) be positive. 
Using the quadratic formula for the roots, it is easy to check that this holds if and only if $|R|<1.$ The 
simultaneous conditions
\be  1>|R|> 2(D-1)^{1/2}/D \ee
require $D>2$ in order for inviscid shock solutions to exist. A relation between pressures $p_0,p_1$
or temperatures $T_0,T_1$ on both sides of the shock can be obtained by using (\ref{feEqn})
for $\zeta=0,$ which gives
\be p_0/p_1=(T_0/T_1)^D =(\beta_1\gamma_1^2/\beta_0\gamma_0^2). 
 \label{tempJump} \ee
Equations (\ref{velocityJump}) and (\ref{tempJump}) imply that the fluid on one side of the shock 
has supersonic velocity and lower temperature, whereas the other side is subsonic with higher temperature.  
As noted in \cite{liu2011shocks}, positive entropy production requires that colder, supersonic fluid 
flows into the shock front and hotter, subsonic fluid flows out.  
 %Using (\ref{velocityJump}) one can show easily from (\ref{PEqn}) that:
%$$
%\Delta p = - f_e \Delta \beta_v. 
%$$

We derive here all of the source terms which appear in the internal energy and entropy balances for these  
shock solutions, both those in the fine-grained (dissipation-range) balances as $\zeta\rightarrow 0$ and those 
in the coarse-grained (inertial-range) balances as $\ell\rightarrow 0.$  It should be pointed out that first-order 
dissipative relativistic fluid models of the type considered are acausal and have unstable solutions even 
at global equilibrium \cite{hiscock1985generic}.  Thus, the viscous shock solutions 
obtained by \cite{liu2011shocks} are expected to be unstable to small perturbations. However, they 
are exact stationary solutions that as $\zeta\rightarrow 0$ converge in $L^p$ norms for any $p\in [1,\infty)$ 
to stationary shock solutions of relativistic Euler equations, and thus provide 
an example for our general mathematical framework. We emphasize that the viscous model solutions are 
employed only to evaluate dissipation-range  quantities, whereas all of our inertial-range limit results hold 
with complete generality for all relativistic Euler shocks with the equation of state (\ref{pressEqn}).  
Inviscid solution fields are all discontinuous step-functions
\be  f(x) = \left\{\begin{array}{ll}
                      f_0 & x<0 \cr
                      f_1 & x>0 
                      \end{array} \right. = f_0 + (\Delta f) \theta(x). \lb{eqA5} \vspace{2mm}
\ee
where $\Delta f=f_1-f_0$ and $\theta(x)$ is the Heaviside step function. We shall also use the 
notation $f_{av}=\frac{1}{2}(f_0+f_1)$ for the average value on both sides of the shock. 
 A fact that we shall use frequently for ideal step-function fields is 
\be  \bar{f}(x) = f_0 + (\Delta f)\bar{\theta}(x), \quad \bar{g}(x) = g_0 + (\Delta g)\bar{\theta}(x) \lb{eqA18} \ee
and thus
\be  \bar{g} = g_0 + \frac{\Delta g}{\Delta f}(\bar{f}-f_0), \quad 
\partial_x \bar{g} =   \frac{\Delta g}{\Delta f}\partial_x \bar{f}. \lb{eqA19} \ee
Furthermore, 
\be  \partial_x\bar{f}(x) = (\Delta f)\bar{\delta}(x). \lb{eqA20} \ee
%Similar results can be obtained from
%\be  \bar{f}(x) = f_{av} + \frac{1}{2}(\Delta f)\overline{{\rm sign}}(x), \quad 
%\bar{g}(x) = g_{av} + \frac{1}{2}(\Delta g)\overline{{\rm sign}}(x) \lb{eqA21} \ee
%These relations are very helpful to derive inertial-range expressions for the shock solution. 
The coarse-graining that is employed here is purely spatial, with a kernel $G.$ 
Because the solutions are stationary in the rest-frame of the shock, there is no need for 
temporal coarse-graining. 

\subsection{Energy Balance}

\subsubsection{Dissipation Range}
   
It can be easily shown for stationary shocks of these RCF's that $\mathcal{Q}_{diss}$ and $p *\theta$ exist
as distributions separately, not just in combination. The fine-grained energy balance equation \eqref{eq9} 
in the $\zeta\to 0$ limit thus reads:
 \begin{eqnarray}\label{stationaryEbalance}
\partial_x (\epsilon \gamma_v\beta_v) =\mathcal{Q}_{diss} -p *\theta .
\end{eqnarray}
We now calculate the two distributions $\mathcal{Q}_{diss}$ and $p *\theta$ appearing above as sources/sinks of the energy density.
\\

\textit{Viscous Pressure-work $p *\theta$:} Direct differentiation yields the dilatation factor:
  \be
  \theta :=\partial_x (\gamma_v \beta_v)=   \gamma_v^3 \partial_x \beta_v,
  \ee  
  and making use of the Bernoulli relation (\ref{PEqn}) for the pressure, one obtains:
\be\label{pressure_eq1}
(D-1)p \theta %= (D-1)p  \partial_x (\gamma_v \beta_v) =  (D-1)p \gamma_v^3 \partial_x \beta_v
=  \left( \frac{f_e}{\beta_v} - f_p\right)\gamma_v^3 \partial_x \beta_v.
\ee
It is straightforward to check that the right-hand-side of (\ref{pressure_eq1}) can be expressed as a total $x$-derivative:
\begin{eqnarray}\nonumber
&(D-1)& p \theta  %= \frac{d}{dx} \left[f_e \ln\left(\frac{\beta_v}{1+\sqrt{1-\beta_v^2}}\right)+ \gamma_v (f_e - \beta_v f_p) \right]\\
%&=&
= \frac{d}{dx} \left[f_e  \ln\left(\frac{\beta_v}{1+\sqrt{1-\beta_v^2}}\right) + (D-1)\gamma_v\beta_v p  \right]. \label{pstarTheta1}
 \end{eqnarray}
The distributional limit as $\zeta\to 0$ is thus found to be 
%\begin{eqnarray}\nonumber
% \Dlim_{\zeta \rightarrow 0}\left[ \partial_x f, \ \partial_x\ln f\right]  = \left[ \Delta f, \ \ln(f_1/f_0)\right]\delta(x)  
% \end{eqnarray}
%and one finds
\begin{eqnarray}\label{pstarTheta} 
p*\theta &=& \Dlim_{\zeta\to 0} p\theta\\
&=&\Bigg\{\frac{f_e}{D-1} \ln\left(\frac{\beta_1}{\beta_0}\frac{1+\sqrt{1-\beta_{0}^2}}{1+\sqrt{1-\beta_{1}^2}}\right)  + \Delta[\gamma_v\beta_vp]\Bigg\} \delta(x). \nonumber
\end{eqnarray}

\textit{Viscous Dissipation $\mathcal{Q}_{diss}$:} The simplest approach to derive  $\mathcal{Q}_{diss}$
is to use the fine-grained energy balance 
\be \zeta\theta^2-p\theta = \partial_x(\epsilon\gamma_v\beta_v) \ee 
to obtain as $\zeta\to 0$ that
%Multiplying Eq.~(\ref{feEqn}) by $\theta$, we find that
%\be
%\zeta \theta^2 = D  p \theta - \frac{f_e \theta}{\gamma_v^2 \beta_v}
%\ee
%Above we calculated $p*\theta$.  Here we note that:
%\be \label{secondterm}
% \frac{ \theta}{\gamma_v^2 \beta_v}= \frac{\partial_x (\gamma_v \beta_v)}{\gamma_v^2 \beta_v}  =\frac{d}{dx}\ln\left(\frac{\beta_v}{1+\sqrt{1-\beta_v^2}}\right).
%\ee
%From (\ref{pstarTheta1}) and (\ref{secondterm}),  we see that the log term cancels in the difference:
\begin{eqnarray}
\mathcal{Q}_{diss} - p* \theta  %= (D-1)p \theta - \frac{f_e \theta}{\gamma_v^2 \beta_v}=  (D-1)\frac{d}{dx} [\gamma_v \beta_v p]\\
&=&  %(D-1)\Delta[\gamma_v\beta_v p]\delta(x) =  
\Delta[\gamma_v\beta_v \epsilon ]\delta(x).  \label{qdiff}
\end{eqnarray}
%This result could have been expected already from the distributional energy balance (\ref{stationaryEbalance}).  
From Eqs.  (\ref{pstarTheta1}) and (\ref{qdiff}), we get:
\begin{eqnarray}\nonumber
&& \mathcal{Q}_{diss} =\Dlim_{\zeta\to 0} \zeta \theta^2 \\ \nonumber
%&=&\lim_{\zeta\to 0}  \frac{d}{dx} \Bigg[\frac{f_e}{D-1}\ln\left(\frac{\beta_v}{1+\sqrt{1-\beta_v^2}}\right)    + D\gamma_v \beta_v p\Bigg]\\
&=& \Bigg\{\frac{f_e}{D-1}  \ln\left(\frac{\beta_1}{\beta_0}\frac{1+\sqrt{1-\beta_{0}^2}}{1+\sqrt{1-\beta_{1}^2}}\right)  +D \Delta[\gamma_v\beta_vp]  \Bigg\} \delta(x). \nonumber\\
\label{Qvisc}
\end{eqnarray}

\subsubsection{Inertial Range}
The resolved energy in the limit $\zeta\to 0$ satisfies:
\be\label{shockLSenergy}
\partial_x (\overline{\gamma_v \beta_v} \overline{\epsilon}) = \mathcal{Q}_\ell^{flux}- \overline{p}\overline{\theta}.
\ee
We now calculate distributional limit as $\ell\to 0$ of the two terms appearing above as sources/sinks.  
%for the resolved energy density.  
%We will show that they have distributional limits $\mathcal{Q}_{flux}$ and $p\circ \theta$ separately. 
%Moreover, we will directly see that 
%In combination they balance with the source/sink terms in the fine-grained balance 
%(\ref{stationaryEbalance}), in agreement with our general theory. \\

\textit{Inertial Pressure-Work $p\circ \theta$:}  Since $\gamma_v$, $\beta_v$ and $p$ are all step functions in the ideal limit,
(\ref{eqA19}) gives 
\begin{eqnarray}\nonumber
\overline{p} \partial_x \overline{\gamma_v \beta_v} &=& \frac{\Delta [\gamma_v \beta_v]}{\Delta p } \partial_x \left(\frac{1}{2} \overline{p}^2 \right).
\end{eqnarray}
%Since, in the limit we have:
%$$
% \Dlim_{\ell \rightarrow 0}\partial_x\left(\frac{1}{2} \overline{f}^2 \right)= \frac{1}{2} (f_1^2 - f_0^2)\delta(x),
%$$ 
It follows that
\begin{eqnarray}
p\circ \theta &=& \Dlim_{\ell \to 0} \overline{p}\partial_x \overline{\gamma_v \beta_v}
%=  \frac{1}{2(D-1)}\left\{ \frac{\Delta(\gamma_v\beta)}{\Delta e} \left(e_1^2 -e_0^2\right)\right\} \delta(x) 
=p_{av}  \Delta[\gamma_v\beta_v] \delta(x).\label{pcircTheta}
\end{eqnarray}
Note that, as required, this result is completely independent of the choice of the filter kernel $G$.
\\

\textit{Energy Flux $\mathcal{Q}_{flux}$}: By definition \eqref{eq63} \begin{eqnarray}\nonumber
\mathcal{Q}_{flux} &=& -\Dlim_{\ell \to 0} \overline{(h u^\mu u^\nu)} \nabla_\mu  \overline{u}_\nu \\ \nonumber
&=&   \Dlim_{\ell \to 0} \Bigg( \overline{(h\gamma_v^2 \beta_v)} \ \partial_x \overline{\gamma_v} - \overline{(h\gamma_v^2 \beta_v)\beta_v} \ \partial_x \overline{\gamma_v \beta_v}   \Bigg).\nonumber
\end{eqnarray}
Enthalpy can be replaced with pressure using $h=D\cdot p$. 
The balance (\ref{feEqn}) with $\zeta=0$ for both terms then gives in the limit as $\ell\to 0$
\begin{eqnarray} \nonumber
D \overline{(p\gamma_v^2 \beta_v)} \ \partial_x \overline{\gamma_v}&=& f_e\partial_x \overline{\gamma_v} 
\Dto f_e\Delta \gamma_v\delta(x)\\ \nonumber
D \overline{(p\gamma_v^2 \beta_v)\beta_v} \ \partial_x \overline{\gamma_v \beta_v} &=& 
f_e \overline{\beta_v}\   \partial_x\overline{\gamma_v \beta_v }\\\nonumber
&=&  f_e  \frac{\Delta(\overline{\gamma_v \beta_v})}{\Delta \beta_v} \partial_x(\overline{\beta_v}^2)     \\\nonumber
&\Dto&  f_e \Delta [\gamma_v \beta_v]\beta_v^{av}\delta(x),
\end{eqnarray}
where (\ref{eqA19}) was used for the second term.  Together, these yield that:
\be\label{qflux1}
\mathcal{Q}_{flux}  =f_e\Big\{ \Delta \gamma_v  -\beta_v^{av}  \Delta [\gamma_v \beta_v]\Big\} \delta(x) .
\ee
We see again that the limiting inertial range result is independent of the choice of filter kernel $G$.  
To compare this term with those previously calculated, we note that for any ideal shock solution 
(\ref{PEqn}) implies 
\be \Delta(\epsilon\gamma_v\beta_v) = f_e \Delta\gamma_v - f_p\Delta(\gamma_v\beta_v) \ee
and (\ref{PEqn2}) with $\zeta=0$ implies that 
\be p_{av}=f_p-f_e\beta_{av}. \ee
These relations can be used to rewrite the formula (\ref{qflux1}) for $\mathcal{Q}_{flux}$ as:
\be\label{qflux2}
\mathcal{Q}_{flux}  = \Big\{ \Delta[\gamma_v\beta_v \epsilon] + p_{av} \Delta[\gamma_v\beta_v]\Big\}\delta(x).
\ee
Eqns. (\ref{pcircTheta}),(\ref{qflux2}) immediately show that 
\be\label{qfluxpcirctheta}
\mathcal{Q}_{flux} -p\circ \theta  =  \Delta[\gamma_v\beta_v \epsilon]\delta(x), 
\ee
as required by the limit of the balance (\ref{shockLSenergy}). 

The relation (\ref{qflux1}) has a further interesting implication that $\calQ_{flux}<0$ for 
relativistic Euler shocks with the equation of state (\ref{pressEqn}).  Using the relation
(\ref{velocityJump}) for the product $\beta_0\beta_1,$ it is easy to show that 
\be J(\beta_0,D):= \beta_{av} \frac{\Delta(\gamma_v\beta_v)}{\Delta \gamma_v}=
\frac{1}{2}\left(\frac{1}{\gamma_0\gamma_1}+ \frac{D}{D-1}\right),\ee 
which may be regarded as a function of just one of the two velocities (say, $\beta_0$) and $D.$
Using the above definition and (\ref{qflux1}), 
\be\label{qflux3}
\mathcal{Q}_{flux}  =f_e \Delta \gamma_v \left[1 - J(\beta_0,D)\right] \delta(x) .
\ee
As noted earlier, positive entropy production at the shock requires that $\Delta \gamma_v<0,$ 
so that $\mathcal{Q}_{flux}<0$ if the second factor in (\ref{qflux3})
is positive over the range $\beta_s<\beta_0<1.$
Direct calculation of the derivative gives
\be \frac{\partial}{\partial\beta_0}J(\beta_0,D)= -(\beta_0^4-\beta_s^4)\frac{\gamma_0\gamma_1}{\beta_0^3}<0, \ee 
while 
\be J(\beta_s,D) = 1, \quad J(1,D) = \frac{D}{2(D-1)}>\frac{1}{2}. \ee 
Thus $1/2<J(\beta_0,D)<1$ over the permitted range of $\beta_0,$ so that the second factor in (\ref{qflux3}) 
remains positive and  $\mathcal{Q}_{flux}<0$.  This a more extreme version of 
what occurs for shocks in a nonrelativistic, compressible Navier-Stokes fluid, where $Q_{flux}=0$
(Appendix A of paper I). In both cases, irreversible shock-heating is not due to energy cascade,
and in the relativistic case inverse energy cascade even contributes cooling rather than heating. 

\textit{Pressure-Dilatation Defect:} 
By subtracting (\ref{pcircTheta}) from (\ref{pstarTheta}), we find:
\begin{eqnarray}\nonumber
\tau(p,\theta) &\equiv& p*\theta- p\circ\theta\\ \nonumber
&=&\Bigg\{\Delta[\gamma_v\beta_vp]   -  p_{av}\Delta [\gamma_v \beta_v] +  \\
&& \frac{ f_e}{D-1} \ln\left(\frac{\beta_1}{\beta_0}\frac{1+\sqrt{1-\beta_{0}^2}}{1+\sqrt{1-\beta_{1}^2}}\right)\Bigg\} \delta(x).
\end{eqnarray}
Together with (\ref{qdiff}),(\ref{qflux2}) this yields 
\be \mathcal{Q}_{diss} = \mathcal{Q}_{flux} + \tau(p,\theta). \ee 
The latter equality can also be obtained by comparing the relations (\ref{qdiff}) and (\ref{qfluxpcirctheta}),  
corroborating the general result (\ref{eq77}). Because $\mathcal{Q}_{diss}>0$ whereas $ \mathcal{Q}_{flux}<0,$
it follows that $\tau(p,\theta)>0.$ Just as for the non-relativistic shocks discussed in paper I, the pressure-dilation 
defect is responsible for the net irreversible heating at the shock. 

\subsection{Entropy Balance}

\subsubsection{Dissipation Range}

The fine-grained entropy balance for  stationary solutions is given simply by:
\be
\partial_x (s \gamma_v\beta_v )   = \frac{\zeta \theta^2}{T} 
\ee

\textit{Viscous Entropy Production:}  It follow immediately from the above that,  for discontinuous shock solutions, 
\be\label{sigDiss}
\Sigma_{diss} :=\Dlim_{\zeta \to 0} \frac{\zeta \theta^2}{T} = \Delta [\gamma_v\beta_v s]\delta(x).
\ee
The entropy production anomaly is thus completely independent of the details of the molecular dissipation 
and, obviously, $\Sigma_{diss}\geq 0.$ As already noted in \cite{liu2011shocks}, this positivity is equivalent 
to the condition that 
\be 1<\frac{s_1\gamma_1\beta_1}{s_0\gamma_0\beta_0}
=\left(\frac{\beta_1\gamma_0^{D-2}}{\beta_0\gamma_1^{D-2}}\right)^{1/D}, \ee 
where (\ref{tempJump}) has been used to obtain the second expression. This ratio 
is 1 for $\beta_0=\beta_1=\beta_s$ and, considered as a function of $\beta_0$ 
and $D$, it is shown by a straightforward calculation to have positive $\beta_0$-derivative 
for $\beta_0\neq \beta_s$. This implies that $\beta_0>\beta_s>\beta_1$ is required for positive entropy 
production, as earlier claimed. 

\subsubsection{Inertial Range}

The resolved entropy equation for stationary solutions of the RCF models is:
\begin{eqnarray} \nonumber
\partial_x ( \underline{s} \overline{\gamma_v\beta_v}&+&\underline{\beta} \tau(\epsilon, \gamma_v\beta_v))   =  \Sigma_\ell^{flux}\\
&& \quad\quad + \underline{\beta} ( \overline{\mathcal{Q}}_{diss} -\overline{\tau}(p,\theta))  \label{entropyBalance}
\end{eqnarray}
where $\Sigma_\ell^{flux}=\partial_x \underline{\beta} \overline{\tau}(\epsilon,\gamma_v\beta_v).$
This entropy evolution equation is considerably simpler than the general Eq. (\ref{eq86}), since $\lambda=0$ 
and because the pressure is proportional to the energy density so that $I_\ell^{flux} \equiv 0$. 

\textit{Inertial-Range Viscous Heating $\beta \circ {\mathcal{Q}}_{diss}$:}  
From (\ref{Qvisc}), $\mathcal{Q}_{diss}= q_*\delta(x)$ so that 
\be
 \overline{\mathcal{Q}}_{diss}= q_*  \overline{\delta}(x).
 \ee
From the formula  (\ref{energyForm}), we see that the inverse temperature $\beta=1/T$ satisfies  
$\beta =\alpha^{1/D} p^{-1/D}$ and thus
\be\label{betaEqn}
\underline{\beta}= \alpha^{1/D}\left(\frac{\overline{\epsilon}}{D+1}\right)^{-1/D}= \alpha^{1/D}\overline{p}^{-1/D}.
\ee
Using (\ref{eqA20}) to write
$ \overline{\delta} = \frac{\partial_x \overline{p}}{\Delta p},$ we get 
\be
\underline{\beta} \overline{\mathcal{Q}}_{diss}%=\alpha^{-1/D} q_* \overline{p}^{-1/D}  \frac{\partial_x \overline{p}}{\Delta p}
= \alpha^{1/D}\frac{D q_*}{D-1} \frac{1}{\Delta p} \frac{d}{dx} [\overline{p}^{(D-1)/D}]
= \frac{q_*}{\Delta\epsilon} \frac{d\underline{s}}{dx}
 \ee
 and therefore, as $\ell\to 0$:
\be
\beta \circ  {\mathcal{Q}}_{diss}= q_*\frac{\Delta s}{\Delta \epsilon}  \delta(x). 
 \ee
%where in the last equality we used (\ref{entrEqn}) which implies: 
% \be\label{entropyform}
%\alpha^{-1/D} \Delta[p^{(D-1)/D}] = \frac{1}{D} \Delta s.
% \ee

\textit{Pressure-Dilatation Defect $\beta \circ \tau(p,\theta)$:}  Our earlier result Eq. (\ref{pstarTheta}) that ${p*\theta} = q_{PV} \delta(x)$, yields, by the same argument:
\be
\Dlim_{\ell\to 0} \underline{\beta} \overline{p*\theta} = q_{PV}\frac{\Delta s}{\Delta \epsilon} \delta(x).
\ee
On the other hand, using (\ref{betaEqn}), we have
\be
\underline{\beta}\overline{p}\overline{\theta} =\alpha^{1/D} \overline{p}^{\frac{D-1}{D}}\partial_x(\overline{\gamma_v \beta_v })  =  \frac{D\alpha^{1/D}}{2D-1}\frac{\Delta[\gamma_v \beta_v]}{\Delta p}\partial_x\left[\overline{p}^{\frac{2D-1}{D}}\right].
\ee
Thus,
\be
\Dlim_{\ell\to 0} \underline{\beta}\overline{p}\overline{\theta} 
=  \frac{D\alpha^{1/D}}{2D-1}\frac{\Delta[\gamma_v \beta_v]}{\Delta p} \Delta\left[{p}^{\frac{2D-1}{D}}\right]\delta(x)
\ee
The following relations are useful and follow directly from (\ref{energyForm}) and (\ref{entrEqn}):
\begin{eqnarray}\label{pSimplif1}
&& \Delta\left(\frac{1}{T}\right)=\alpha^{1/D}\Delta[p^{-1/D}] = \frac{(D-1)(s_1\epsilon_0-s_0\epsilon_1)}{D\epsilon_0\epsilon_1}
\\ 
\label{pSimplif3}
&& \Delta\left(\frac{p^2}{T}\right)=\alpha^{1/D}\Delta[p^{\frac{2D-1}{D}}] = \frac{\Delta[s \epsilon]}{D(D-1)}.
\end{eqnarray}
With these, we have that $I_{mech}=\beta\circ\tau(p,\theta)$ is given by 
\begin{eqnarray}\nonumber
\beta \circ \tau(p,\theta)
&=&\frac{1}{\Delta \epsilon}   \left[  q_{PV}  {\Delta s}- \frac{\Delta[\gamma_v \beta_v] \Delta[s \epsilon] }{2D-1} \right]\delta(x).\\
\end{eqnarray}

\textit{Combined Contribution  $\beta \circ \mathcal{Q}_{diss} - \beta \circ \tau(p,\theta)$:}  From Eq. (\ref{qdiff}), we obtain that:
\be
q_*-q_{PV}=\Delta[\gamma_v\beta_v\epsilon].
\ee
 Thus, the combined contribution of these terms is simply:
\begin{eqnarray}\nonumber
\beta &\circ& \mathcal{Q}_{diss} -\beta \circ \tau(p,\theta)\\ \nonumber
 &=&\frac{1}{\Delta \epsilon}\left[ \Delta s\Delta[\gamma_v\beta_v\epsilon]  + \frac{\Delta[\gamma_v \beta_v] \Delta[s \epsilon] }{2D-1} \right]\delta(x).\\  \label{simpCombined}
%&=&\Bigg\{ \frac{2d}{2D-1} \frac{\Delta s\Delta [\gamma_v\beta_v\epsilon]}{\Delta \epsilon}  - \frac{s_1\gamma_0\beta_0- s_0\gamma_1\beta_1}{2D-1}\Bigg\}\delta(x)\\
\end{eqnarray}

\textit{Negentropy Flux $\Sigma_\ell^{flux}$:} First, we consider the contribution $(\partial_x \underline{\beta}) \overline{\epsilon\gamma_v\beta_v}$.  From Eqn. (\ref{betaEqn}) we have:
\be
\partial_x  \underline{\beta} = -\frac{\alpha^{1/D}}{D} \overline{p}^{-\frac{D+1}{D}} \partial_x \overline{p}.
\ee
Using (\ref{eqA18}) to write 
\be \overline{\epsilon\gamma_v\beta_v} = \epsilon_0\gamma_0\beta_0
+ \frac{\Delta(\epsilon\gamma_v\beta_v)}{\Delta p}(\overline{p}-p_0), \ee 
a straightforward calculation shows
\begin{eqnarray}\label{Fluxterm1}
 (&\partial_x & \underline{\beta})\overline{\epsilon\gamma_v\beta_v} \Dto-\frac{1}{D} \frac{ \Delta s \Delta [\gamma_v\beta_v\epsilon] }{\Delta \epsilon} \delta(x)\\
 &-& \left(\frac{\Delta [\gamma_v\beta_v\epsilon] - \gamma_0\beta_0 \Delta \epsilon }{ \Delta \epsilon } \right)\epsilon_0  \alpha^{-1/D}\Delta[p^{-1/D}]\delta(x)\nonumber
\end{eqnarray}

The other term is computed likewise using:
\begin{eqnarray}
(&\partial_x&  \underline{\beta})\overline{\epsilon}\overline{\gamma_v\beta_v}\\ \nonumber
&=&  -\alpha^{1/D}\frac{D-1}{D} \overline{p}^{-\frac{1}{D} } \partial_x \overline{p} \left(\gamma_0\beta_0 + \frac{\Delta[\gamma\beta]}{\Delta p}(\overline{p}-p_0)\right)
\end{eqnarray}
whence, after some calculation, one has:
\begin{eqnarray}\nonumber
(\partial_x  \underline{\beta})\overline{\epsilon}\overline{\gamma_v\beta_v}&\Dto& -\alpha^{1/D}\frac{(D-1)^2}{2D-1}\frac{\Delta [\gamma_v\beta_v]}{\Delta \epsilon}\Delta[p^{(2D-1)/D}]\delta(x)\\ \nonumber
 && \!\!\!\! \! \! \!\! \!  \!\!\! + \frac{1}{D} \left(\frac{\Delta [\gamma_v\beta_v] \epsilon_0 - \gamma_0\beta_0 \Delta \epsilon }{ \Delta \epsilon } \right) { \Delta s}\delta(x).\\ \label{Fluxterm2}
\end{eqnarray}
Therefore, (\ref{Fluxterm1}) and (\ref{Fluxterm2}) in combination show:
\begin{eqnarray}\nonumber
\Sigma_{flux}&:=&\Dlim_{\ell \to 0} \partial_x \underline{\beta} \overline{\tau}(\epsilon,\gamma_v\beta_v)
\\ \nonumber
&=& \alpha^{1/D}\frac{(D-1)^2}{2D-1}\frac{\Delta [\gamma_v\beta_v]}{\Delta \epsilon}\Delta[p^{(2D-1)/D}]\delta(x)\\
 &-& \frac{1}{D} \left(\Delta [\gamma_v\beta_v] \epsilon_0 - \gamma_0\beta_0 \Delta \epsilon+ \Delta [\gamma_v\beta_v\epsilon] \right) \frac{ \Delta s}{\Delta \epsilon}\delta(x)\nonumber\\ \nonumber
 &-& \alpha^{1/D}\left(\frac{\Delta [\gamma_v\beta_v\epsilon]  - \gamma_0\beta_0 \Delta \epsilon }{ \Delta \epsilon } \right)\epsilon_0 \Delta[p^{-1/D}]\delta(x)\\
\end{eqnarray}
The relations (\ref{pSimplif1}), (\ref{pSimplif3}) can then be employed to simplify the expression for the flux to:
\begin{eqnarray}\nonumber
\Sigma_{flux}&=&\frac{1}{\Delta \epsilon}\Bigg\{(\epsilon_1s_0-\epsilon_0s_1)\Delta [\gamma_v\beta_v] -\frac{\Delta [\gamma_v\beta_v]\Delta [s\epsilon]}{2D-1} \Bigg\}\delta(x).\\
\label{SimpSdiss}
\end{eqnarray}
Adding together the formulas (\ref{simpCombined}) and (\ref{SimpSdiss}), one has, after minor manipulation, that:
\be
\beta\circ \mathcal{Q}_{diss} -\beta\circ {\tau}(p,\theta) + \Sigma_{flux} = \Delta[\gamma_v\beta_vs]
\ee
in agreement with (\ref{entropyBalance}) and the dissipation-range result (\ref{sigDiss}), 
as demanded by the general equality Eq.~(\ref{eq98}).  

A further implication of the formula (\ref{SimpSdiss})
for entropy flux is that $\Sigma_{flux}>0$ at these relativistic Euler shocks. Although not presented here,
arguments like those applied to $\mathcal{Q}_{flux}$ show this and are confirmed by numerically 
plotting (\ref{SimpSdiss}) as a function of $R$ for each $D>2.$ It is interesting that $\Sigma_{flux}>0$ was also 
found for planar shock solutions of non-relativistic compressible Euler equations in paper I. In both cases, there 
is a forward cascade of negentropy at the shock, even though the energy flux is vanishing or negative.

\bibliographystyle{apsrev4-1}
\bibliography{bibliography.bib}

\end{document}